\pdfoutput=1

\documentclass[aps,prd,preprint,superscriptaddress,showpacs]{revtex4}
\usepackage{graphics}
\usepackage{epstopdf}
\usepackage{graphicx}
\usepackage{dcolumn}
\usepackage{bm}
\usepackage{amsmath}
\usepackage{color}
\usepackage{enumerate}

\newcommand{\be}{\begin{equation}}
\newcommand{\ee}{\end{equation}}
\newcommand{\ba}{\begin{eqnarray}}
\newcommand{\ea}{\end{eqnarray}}
\newcommand{\bfig}{\begin{figure}[t]\begin{centering}}
\newcommand{\efig}{\end{centering}\end{figure}}

\def\dm03{\hbox{$\Delta m^2_{03}$}}

\begin{document}

\title{Super-PINGU for measurement of the leptonic CP-phase with
  atmospheric neutrinos}

\author{Soebur Razzaque}
\email{srazzaque@uj.ac.za}
\affiliation{Department of Physics, University of Johannesburg,
PO Box 524, Auckland Park 2006, South Africa}

\author{A.\ Yu.\ Smirnov} 
\email{smirnov@mpi-hd.mpg.de} 
\affiliation{Max-Planck-Institute for Nuclear Physics, Saupfercheckweg
  1, D-69117 Heidelberg, Germany}
\affiliation{International Centre for Theoretical Physics, Strada
  Costiera 11, I-34100 Trieste, Italy}


\begin{abstract}

  We explore a possibility to measure the CP-violating phase $\delta$
  using multi-megaton scale ice or water Cherenkov detectors with low,
  $(0.2 - 1)$ GeV, energy threshold assuming that the neutrino mass
  hierarchy is identified.  We elaborate the relevant theoretical and
  phenomenological aspects of this possibility.  The distributions of
  the $\nu_\mu$ (track) and $\nu_e$ (cascade) events in the neutrino
  energy and zenith angle $(E_\nu - \theta_z)$ plane have been
  computed for different values of $\delta$.  We study properties and
  distinguishability of the distributions before and after smearing
  over the neutrino energy and zenith angle.  The CP-violation effects
  are not washed out by smearing, and furthermore, the sensitivity to
  $\delta$ increases with decrease of the energy threshold.  The
  $\nu_e$ events contribute to the CP-sensitivity as much as the
  $\nu_\mu$ events.  While sensitivity of PINGU to $\delta$ is low, we
  find that future possible upgrade, Super-PINGU, with few megaton
  effective volume at ($0.5 - 1$) GeV and e.g.  after 4 years of
  exposure will be able to disentangle values of $\delta = \pi/2,~
  \pi,~ 3\pi/2$ from $\delta = 0$ with ``distinguishability'' ($\sim$
  significance in $\sigma$'s) $S_{\sigma}^{tot} = (3 - 8),~ (6 - 14),~
  (3 - 8)$ correspondingly.  Here the intervals of $S_{\sigma}^{tot}$
  are due to various uncertainties of detection of the low energy
  events, especially the flavor identification, systematics, {\it
    etc.}.  Super-PINGU can be used simultaneously for the proton
  decay searches.

\end{abstract}

\maketitle

\section{Introduction}

Discovery of the leptonic CP violation and measurement of the Dirac CP
phase are among the main objectives in neutrino physics and, in
general, in particle physics.  They may have fundamental implications
for theory and important consequences for phenomenology of atmospheric
and accelerator neutrinos, high energy cosmic neutrinos, {\it etc}.
\cite{reviews}.
 
The present experimental results have very low sensitivity to $\delta$
giving only weak indications of the preferable interval of its values.
Thus, the T2K and reactor data favor the interval $\delta = (1 - 2)\,
\pi$ with central value $\delta = 1.5\,\pi$ \cite{t2k}.  Analysis of
the SuperKamiokande atmospheric neutrino data gives preferable range
$(1.2 \pm 0.5)\pi$~\cite{himmel}.  The global fit of all oscillation
data, e.g. from \cite{fogli}, agrees with these results: $\delta
\approx 1.39^{+0.38}_{- 0.27}\, \pi$ at $1\sigma$ level and no
restriction appears at $3\sigma$ level.  The values around $\delta
\sim \pi/2$ are disfavored.  Similar results with the best fit value
$\delta = 1.34\, \pi$ (NH) have been obtained in \cite{valle} and with
$\delta = {251}^{\circ}$ in \cite{nufit}.

A possibility to measure $\delta$ is generally associated with
accelerator long base-line (LBL) neutrino experiments.  There is
certain potential to improve our knowledge of $\delta$ with further
operation of T2K and NOvA~\cite{ghosh}.  Proposals of more remote
experiments, which will measure $\delta$ with reasonable accuracy,
include LBNE \cite{lbne}, J-PARC - HyperKamiokande \cite{hyperk}, ESS
\cite{ess} and LBNO \cite{lbno}.  Further developments can be related
to the low energy neutrino and muon factories, beta beams, {\it etc.},
see \cite{reviews}.

Another possibility to determine $\delta$ is to use the atmospheric
neutrino fluxes and large underground/underwater detectors.
Sensitivity of future atmospheric neutrino studies by HyperKamiokande
(HK) has been estimated in~\cite{hyperk}: During 10 years of running
with fiducial volume $0.57$ Mton the HK will be able to discriminate
the values of phases $\delta = 40^{\circ},\, 140^{\circ},\,
220^{\circ},\, 320^{\circ}$ at about $(1 - 1.5)\sigma$ CL.  ICAL at
INO alone will have very low sensitivity, but combined with data from
T2K and NOvA, it will reduce degeneracy of parameters, and thus,
increase the global sensitivity \cite{ghosh2}.

Various theoretical and phenomenological aspects of the CP-violation
in atmospheric neutrinos have been explored in a number of
publications before~ \cite{Peres:2003wd, Kimura1,
  GonzalezGarcia:2004wg, our2, our3, Mena:2008rh, ARS, latimer,
  Agarwalla:2012uj, ohlsson, ge1, ge2}.  In particular, pattern of the
neutrino oscillograms (lines of equal probabilities in the $E_\nu -
\cos\theta_z$ plane) with CP violation has been studied in details in
\cite{our3}.  It was realized that structure of the oscillograms is
determined to a large extent by the grid of the magic lines of three
different types \cite{Barger:2001yr, Huber:2003ak, Smirnov:2006sm,
  our3} (solar, atmospheric and interference phase lines).  Although
at the probability level the effects of the CP-violation can be of
order 1, there are a number of factors which substantially reduce the
effects at the level of observable events \cite{ARS}.

Capacities of new generation of the atmospheric neutrino detectors
(PINGU, ORCA) have been explored recently~\cite{pingu, ARS, pingu2,
  orca, Winter:2013ema}.  It was found \cite{ARS, pingu2} that these
detectors with $E_{th} \sim 3$ GeV have good sensitivity to the
neutrino mass hierarchy and the parameters of the 2-3 sector (the 2-3
mixing and mass splitting). However, the CP-violation effects turn out
to be sub-leading.  This helps in establishing the hierarchy without
serious degeneracy with $\delta$ in contrast to the accelerator
experiments, but the information on the CP-phase will be rather poor.

The goal of this paper is twofold: (i) detailed study of the
CP-violation effects in atmospheric neutrinos, and (ii) tentative
estimation of sensitivity to the CP-phase of future large detectors,
assuming that the neutrino mass hierarchy is identified.  We will show
that in spite of averaging of oscillation pattern over the neutrino
energy and direction, the CP- violation effects are not washed out,
and furthermore, increase with lowering the energy threshold $E_{th}$.
This opens up a possibility to measure $\delta$ using multi-megaton
scale ice or water Cherenkov detectors with $E_{th} = (0.2 - 0.5)$
GeV. We study dependence of the energy and zenith angle distributions
of events produced by $\nu_e$ and $\nu_\mu$ on the CP phase.  We
estimate distinguishability of different values of $\delta$.
According to the present proposal \cite{pingu2} PINGU will have low
sensitivity to $\delta$ and only further upgrades, which we will call
Super-PINGU, can measure $\delta$ with potentially competitive
accuracy.  We discuss requirements for such detectors.  We identify
problems and challenges of these CP measurements, and propose ways to
resolve or mitigate the problems. We formulate conditions, in
particular on accuracies of knowledge of external parameters and level
of flavor misidentification, to achieve the goal.

The paper is organized as follows. In Sec.\ II we summarize relevant
information on the oscillation probabilities and their dependence on
CP-phase. We present analytical formulas for the probabilities in
quasi-constant density approximation. The grid of the magic lines will
be described and we will show how the grid determines structure of
oscillograms.  In Sec. \ III we consider a possible upgrade of PINGU,
called Super-PINGU, which will be able to measure $\delta$ and outline
a procedure of computation of numbers of events.  In Sec.\ IV we
compute the distributions as well as relative differences of
distributions of the $\nu_\mu$ events in the $E_\nu - \cos \theta_z$
plane (the relative CP-differences) for different values of $\delta$.
We study dependence of these distributions on $\delta$ before and
after smearing over the neutrino energy and direction.  In Sec.\ V we
perform similar studies of the cascade (mainly $\nu_e$) events.  Sec.\
VI contains estimations of the total sensitivity of Super-PINGU to
$\delta$ and discussion of our results. We conclude in Sec. VII.

\section{Oscillation probabilities, CP-domains}

\subsection{Oscillation amplitudes and probabilities}

We will study the CP-violation phase $\delta$ defined in the standard
parametrization of the PMNS mixing matrix, $U_{PMNS} = U_{23}
I_{\delta} U_{13} I_{\delta}^*U_{12}$, where $U_{ij}$ is the matrix of
rotation in the $ij$-plane and $I_{\delta} \equiv {\rm diag} (1, 1,
e^{i\delta})$.  We consider evolution of the neutrino states $\nu_f
\equiv (\nu_e, \nu_\mu, \nu_\tau)^T$ in the propagation basis,
$\nu_{prop} = (\nu_e, \tilde{\nu}_2, \tilde{\nu}_3)^T$ determined by
the relation $\nu_f = U_{23} I_{\delta} \nu_{prop}$. In this basis the
CP dependence is dropped out from the evolution and appears via
projection of the propagation states $\nu_{prop}$ back onto the flavor
states at the production and detection. Due to this, dependence of the
probabilities and numbers of events on $\delta$ is simple and
explicit.  Therefore the results will be presented in terms of
amplitudes in this basis (see \cite{our3} for details), where the
matrix of amplitudes is defined as
\be
|| A_{\alpha \beta}|| = 
\left(
\begin{array}{ccc}
A_{ee} &  A_{e\tilde{2}} & A_{e\tilde{3}}\\
...  & A_{\tilde{2}\tilde{2}} &   A_{\tilde{2}\tilde{3}}\\
... & ... & A_{\tilde{3}\tilde{3}} 
\end{array}
\right).  
\nonumber
\ee
Here we have taken into account the equalities $A_{e \tilde{i}} =
A_{\tilde{i} e}$ and $A_{\tilde{2}\tilde{3}} = A_{\tilde{3}\tilde{2}}$
valid for symmetric density profile and in absence of the fundamental
CP and T violation in the propagation basis.  In the low energy
domain, $E \lesssim (2 - 3)$ GeV, {\it i.e.} below the 1-3 resonance,
one can further decrease the number of amplitudes involved down to 3
(see \cite{Peres:2003wd} and comment \footnote{The basis used in the
  paper \cite{Peres:2003wd} differs from the basis considered here by
  the additional 1-3 rotation on the 1-3 mixing in matter. This basis
  is useful for description of oscillations at low energies (in the
  sub-GeV range) since it allows to make certain approximations which
  simplify description. Namely, neglecting changes of 1-3 mixing in
  matter with distance one can reduce 3-neutrino evolution problem to
  2-neutrino evolution problem. Correspondingly all the probabilities
  can be expressed in terms of just three real functions $P_2$, $R$
  and $I$. The main dependence on 1-3 mixing as well as on $\delta$ is
  explicit here.  The formulas in \cite{Peres:2003wd} are approximate,
  and in general they are not valid at high energies (in multi-GeV
  range). Since the highest sensitivity to CP is at low energies these
  formulas give accurate description of CP-effects.}).

The oscillation probabilities $P_{\alpha\beta} \equiv
|A_{\alpha\beta}|^2$ can be written as
\be
P_{\alpha\beta} \equiv P_{\alpha\beta}^{ind} + P_{\alpha\beta}^\delta\, , 
\label{eq:genProb}
\ee
where $P_{\alpha\beta}^{ind}$ and $P_{\alpha\beta}^\delta$ are the
$\delta$-independent and $\delta$-dependent parts of the probability
$P_{\alpha\beta}$, respectively.  Notice that $P_{\alpha \beta}^{ind}
\neq P_{\alpha \beta}(\delta=0)$, since $P_{\alpha\beta}^{\delta}$
contains terms which are proportional to $\cos \delta$, generally even
on $\delta$, and these terms do not disappear when $\delta =0$. Then
the total probability is $P_{\alpha \beta}(\delta=0) = P_{\alpha
  \beta}^{ind} + P_{\alpha \beta}^0$.  The probabilities
$P_{\alpha\beta}^{ind}$ equal \cite{our3}
\ba 
P_{e \mu}^{ind} & = & c_{23}^2 |A_{e\tilde{2}}|^2  + s_{23}^2 |A_{e\tilde{3}}|^2, \\  
P_{\mu \mu}^{ind} & = & \left|c_{23}^2 A_{\tilde{2}\tilde{2}}  + 
s_{23}^2 A_{\tilde{3}\tilde{3}} \right|^2. 
\nonumber
\ea
 
The amplitude $A_{\tilde{2}\tilde{3}}$ is doubly suppressed by small
quantities $\Delta m_{21}^2/\Delta m_{31}^2$ and $s_{13}$ \cite{our3}.
Therefore terms that are quadratic in $A_{\tilde{2}\tilde{3}}$ can be
neglected in the first approximation of our analytical study. For the
$\delta-$dependent parts we have then~\cite{our3} $P_{ee}^\delta = 0$,
\be
P_{e\mu}^\delta  =  
\sin 2\theta_{23} {\rm Re} 
\left[e^{i\delta} A_{e\tilde{2}}^* A_{e\tilde{3}} \right]
= 
\sin 2\theta_{23}|A_{e\tilde{2}}A_{e\tilde{3}}|\cos(\phi+\delta)\,,
\label{eq:Pemudelta} 
\ee
where $\phi\equiv {\rm arg}(A_{e\tilde{2}}^*A_{e\tilde{3}}^{})$, and    
\be
P_{\mu\mu}^\delta   =  
- \sin 2\theta_{23} {\rm Re} \left[ A_{e\tilde{2}}^* A_{e\tilde{3}}\right] 
\cos \delta  + D_{23} =  -\sin 2\theta_{23} 
|A_{e\tilde{2}}^{}A_{e\tilde{3}}^{}|\cos\phi \cos\delta + D_{23}.   
\label{eq:Pmumudelta}
\ee
Here   
\be
D_{23} \equiv  \sin 2\theta_{23} \cos \delta 
\cos 2\theta_{23}{\rm Re}[A_{\tilde{2}\tilde{3}}^*(A_{\tilde{3}\tilde{3}}^{}-
A_{\tilde{2}\tilde{2}}^{})]\, . 
\nonumber
\ee
The term $D_{23}$ is small if the 2-3 mixing is close to the maximal
one, and as we said, in addition the amplitude
$A_{\tilde{2}\tilde{3}}$ is small.  Let us emphasize that in
$P_{\mu\mu}^\delta$ the phase dependence, $\cos \delta$, factors out,
whereas in $P_{e\mu}^\delta$ it appears in combination with the
oscillation phase $\phi$.

In matter with symmetric density profile one has for the inverse
channels
$$
P_{\beta \alpha} = P_{\alpha \beta} ({\delta  \rightarrow - \delta}), 
$$
in particular, $P_{\mu e}^\delta = P_{e\mu}^{- \delta}$.  For
antineutrinos the probabilities have the same form as for neutrinos
with substitution:
\be
\delta \rightarrow - \delta, ~~~
\phi_{32}^m \rightarrow \bar{\phi}_{32}^m, ~~~\phi_{21}^m 
\rightarrow \bar{\phi}_{21}^m, ~~~
\theta_{ij}^m \rightarrow \bar{\theta}_{ij }^m ,
\label{antinu}
\ee
where $\bar{\theta}_{ij} = \theta_{ij} (V \rightarrow - V)$ and
$\bar{\phi}_{ij} = \phi_{ij} (V \rightarrow - V)$ are the mixing
angles and phases in matter for antineutrinos, and $V$ is the matter
potential.  In particular,
\be
\bar{P}_{e\mu}^\delta 
=
\sin 2\theta_{23}|\bar{A}_{e\tilde{2}}
\bar{A}_{e\tilde{3}}|\cos(\bar{\phi} - \delta)\, . 
\label{eq:Pemudelta-bar}
\ee

\subsection{Quasi-constant density approximation}

One can further advance in analytical study using explicit expressions
for the amplitudes in the constant (or quasi-constant) density
approximation \cite{our3} (see also \cite{kimura2} and
\cite{Blennow:2013vta}).  According to this approximation, at high
energies for a given trajectory in mantle one can use the mixing
angles computed for the average value of the potential $V =
\bar{V}(\theta_z)$.  For low energies, where adiabaticity condition is
fulfilled, the mixing angle is determined by the surface density.  The
oscillation phases, however, should be computed by integration over
the neutrino trajectory.  For core-crossing trajectories one can use
the three layer model with constant densities in each layer;
corrections are computed in \cite{our3}.

In the case of constant density~\cite{our3}
\ba
A_{e\tilde{2}} & = &- i e ^{i \phi_{21}^m} 
\cos \theta_{13}^m \sin 2 \theta_{12}^m \sin \phi_{21}^m, 
\label{ae2-con}\\
A_{e\tilde{3}} & = & - i e ^{i \phi_{21}^m} \sin 2 \theta_{13}^m 
\left( \sin \phi_{32}^m e ^{-i \phi_{31}^m} + 
\cos^2 \theta_{12}^m \sin \phi_{21}^m \right).
\label{ae3-con}
\ea
The half-phases equal in the high energy range (substantially larger
than the 1-2 resonance, $E_\nu \gtrsim 0.5$ GeV):
\be 
\phi_{32}^m \approx  \frac{\Delta m_{31}^2 L}{4 E_\nu} 
\sqrt{(1 - \epsilon)^2 \mp 2 (1 - \epsilon) \xi \cos 2 \theta_{13} + \xi^2}.  
\label{ph32}
\ee
Here $L = 2R_E \cos \theta_z$ with $R_E$ being the radius of the Earth,  
\be
\xi  \equiv  \frac{2 V E_\nu}{\Delta m_{31}^2},~~~~ 
\epsilon \equiv \sin^2 \theta_{12} \frac{\Delta m_{21}^2}{\Delta m_{31}^2}, 
\nonumber
\ee
and the upper (lower) sign  corresponds to neutrinos (antineutrinos). 
For two other phases we obtain  
\ba
\phi_{21}^m  & \approx &  \frac{\Delta m_{31}^2 L}{8 E_\nu}
[1 + \xi - \epsilon (2\cot^2 \theta_{12} - 1 )] - \frac{1}{2} \phi_{32}^m , 
\label{ph21h}\\
\phi_{31}^m  & \approx &  \frac{\Delta m_{31}^2 L}{8 E_\nu}
[1 + \xi -  \epsilon (2\cot^2 \theta_{12} - 1)] + \frac{1}{2} \phi_{32}^m ,  
\label{phi31}
\ea
where $\phi_{32}^m$ is given in (\ref{ph32}).  In practical cases the
$\epsilon-$terms can be neglected.  For low energies (close to the 1-2
resonance):
\be 
\phi_{21}^m \approx  \frac{\Delta m_{21}^2 L}{4 E_\nu} 
\sqrt{\left(\cos 2 \theta_{12} \mp \frac{2 V E_\nu}{\Delta m_{21}^2} \right)^2 
+ \sin^2  2 \theta_{12}}.   
\label{ph12} 
\ee
Notice that in the energy range above the 1-2 resonance $\cos^2
\theta_{12}^m \approx 0$ and the amplitude $A_{e\tilde{3}}$
(\ref{ae3-con}) is reduced to the two neutrino form, which corresponds
to factorization \cite{our3}.

Inserting expressions for the amplitudes (\ref{ae2-con}) and
(\ref{ae3-con}) into (\ref{eq:Pemudelta}) we obtain
\be
P_{e\mu}^\delta = J_\theta \sin \phi_{21}^m 
\left[ \sin \phi_{32}^m  \cos (\delta - \phi_{31}^m ) +
\cos^2 \theta_{12}^m \sin \phi_{21}^m  \cos \delta  \right], 
\label{pemu1}
\ee
where 
\be
J_\theta \equiv \sin 2\theta_{23} \sin 2 \theta_{12}^m \sin 2 \theta_{13}^m 
\cos \theta_{13}^m
\ee
is the mixing angles factor of the Jarlskog invariant in matter. Using
relation $\phi_{31}^m = \phi_{32}^m + \phi_{21}^m$ we obtain from
(\ref{pemu1})
\be
P_{e\mu}^\delta \approx J_\theta \sin \phi_{21}^m
\left[ \frac{1}{2}  \sin 2 \phi_{32}^m  \cos (\delta - \phi_{21}^m ) +
\sin^2 \phi_{32}^m  \sin (\delta - \phi_{21}^m ) + 
\cos^2 \theta_{12}^m \sin \phi_{21}^m  \cos \delta  \right]. 
\label{pemu2}
\ee   
Similarly, neglecting $D_{23}$ we find for $P_{\mu \mu}^\delta$  
\be
P_{\mu \mu}^\delta =  - \cos \delta  J_\theta 
\sin \phi_{21}^m
\left[ \sin \phi_{32}^m  \cos \phi_{31}^m  +
\cos^2 \theta_{12}^m \sin \phi_{21}^m  \right],
\label{pmumu2a}
\ee
or excluding $\phi_{31}^m$: $\phi_{31}^m = \phi_{32}^m + \phi_{21}^m$
\be
P_{\mu \mu}^\delta =  - \cos \delta  J_\theta \sin \phi_{21}^m
\left[ \frac{1}{2}  \sin 2 \phi_{32}^m  \cos \phi_{21}^m  -
\sin^2 \phi_{32}^m  \sin \phi_{21}^m +
\cos^2 \theta_{12}^m \sin \phi_{21}^m  \right],  
\label{pmumu2}
\ee
where $\delta$ dependence factors out.

For antineutrinos we have the same expressions (\ref{pemu2}) and
(\ref{pmumu2}) with substitution (\ref{antinu}) and $J_\theta
\rightarrow \bar{J}_\theta = J_\theta (\theta_{ij} \rightarrow
\bar{\theta}_{ij})$.

We will use the analytic expressions (\ref{pemu2}) and (\ref{pmumu2})
and the corresponding expressions for antineutrinos for interpretation
of numerical results.

\subsection{Numerical results}

We have computed the probabilities $P_{\alpha \beta} = P_{\alpha
  \beta} (E_\nu,\theta_z)$ by performing numerical integration of the
evolution equation for the complete $3\nu-$system.  We used the PREM
density profile of the Earth \cite{prem} and the values of the
neutrino parameters $\Delta m^2_{32} = 2.35 \cdot 10^{-3}$ eV$^2$,
$\Delta m^2_{21} = 7.6 \cdot 10^{-5}$ eV$^2$, $\sin^2 \theta_{23} =
0.42$, $\sin^2 \theta_{12} = 0.312$ and $\sin^2 \theta_{13} = 0.025$,
which are close to the current best fit values~\cite{fogli}. We assume
the normal neutrino mass hierarchy in the most part of the paper.

\begin{figure}[t]
\includegraphics[width=4.5in]{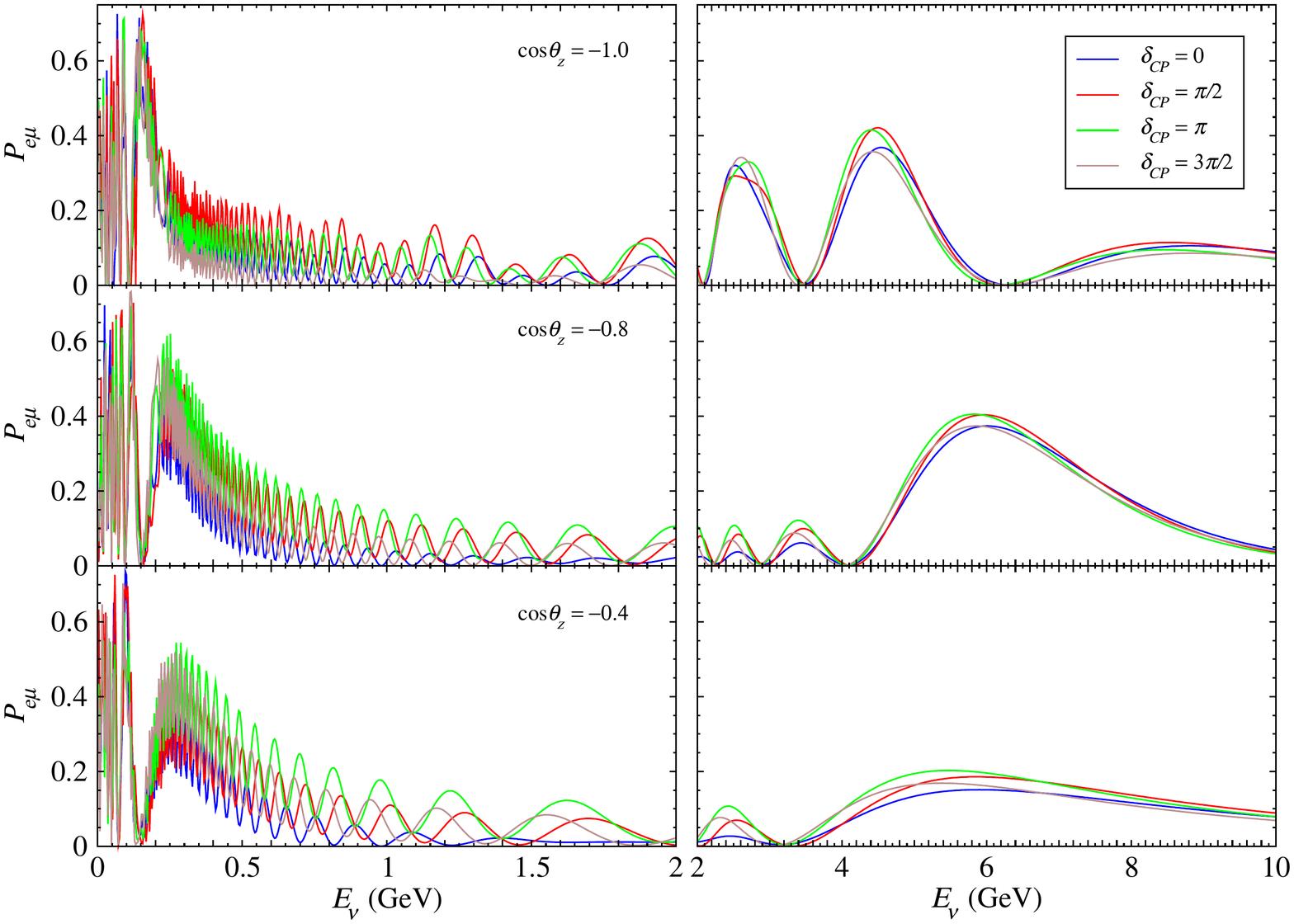}
\includegraphics[width=4.5in]{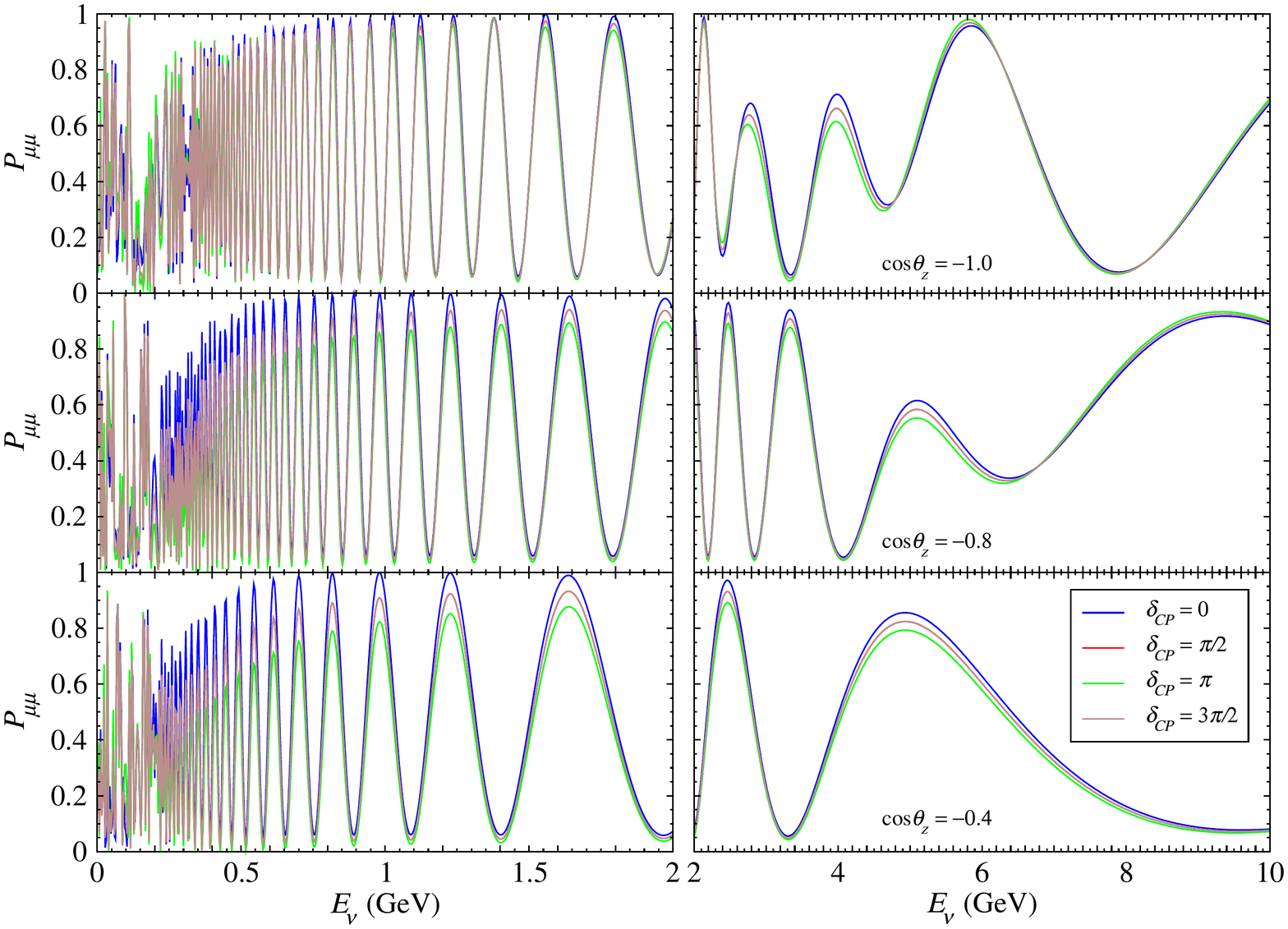}
\caption{
  Probabilities of the $\nu_e \to \nu_\mu$ (top panels) and $\nu_\mu
  \to \nu_\mu$ (bottom panels) oscillations as functions of the
  neutrino energy for different values of $\delta$ and the zenith
  angle.  The probability $P_{\mu\mu}$ is the same for $\delta =
  \pi/2$ and $3\pi/2$ in the bottom panels.  Normal mass hierarchy is
  assumed and the neutrino parameters from the global fits are used
  (see main text).}
\label{fig:emu_prob}
\end{figure}

\begin{figure}[t]
\includegraphics[width=4.5in]{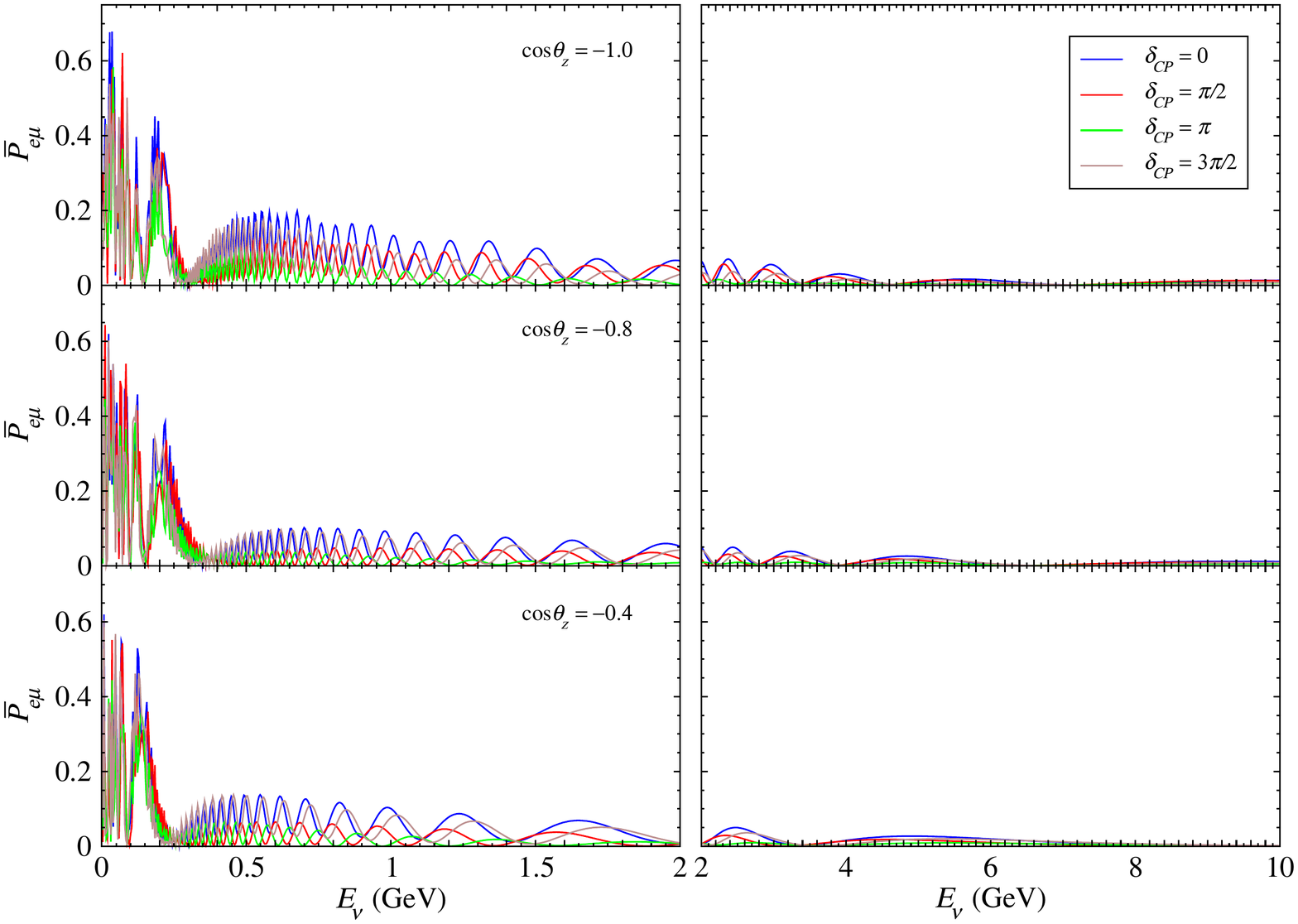}
\includegraphics[width=4.5in]{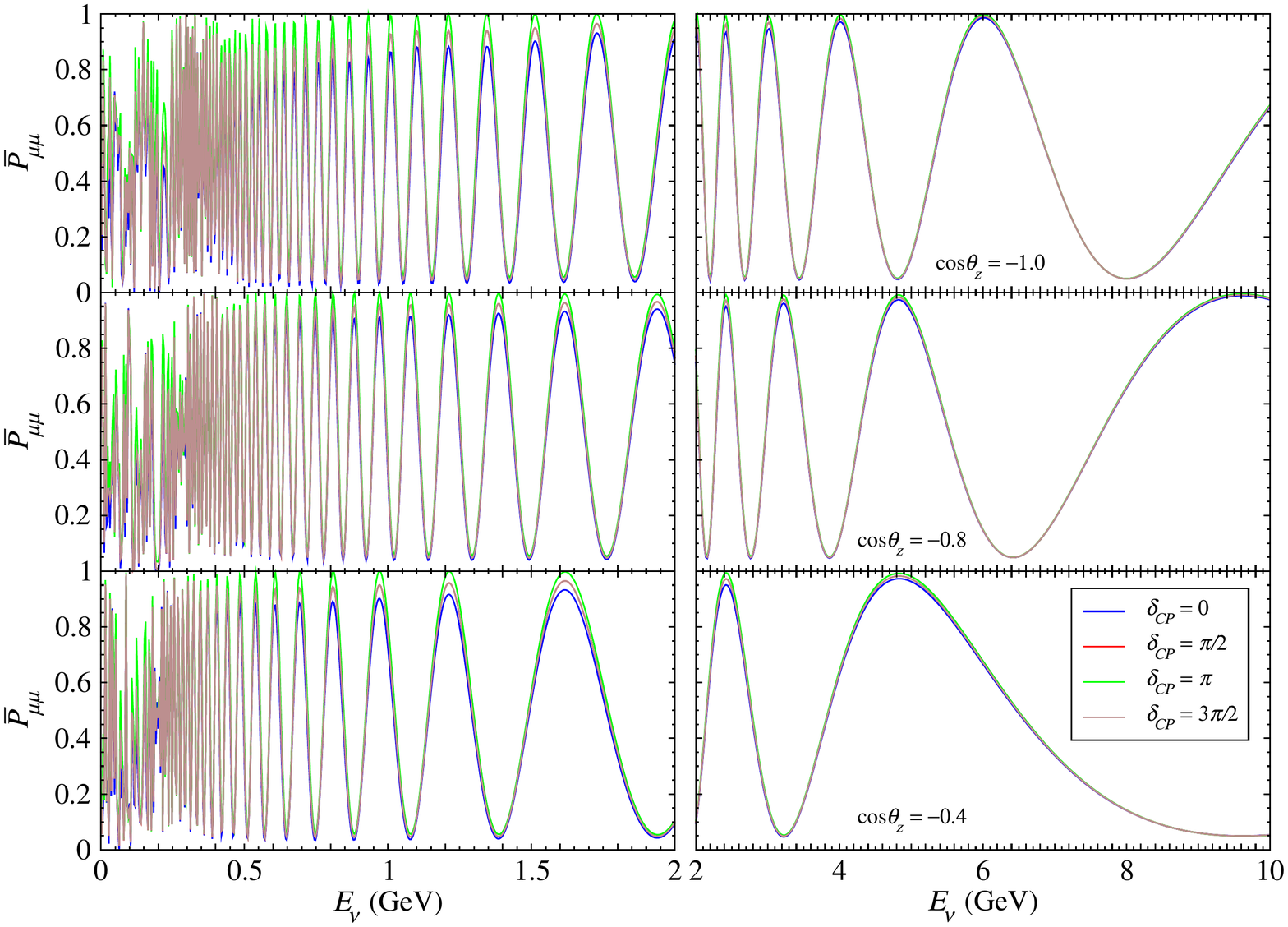}
\caption{Same as Fig.~\ref{fig:emu_prob}, but for antineutrinos.}
\label{fig:anti_emu_prob}
\end{figure}

In Fig.~\ref{fig:emu_prob} we show the oscillation probabilities
$\nu_e \rightarrow \nu_\mu$ and $\nu_\mu \rightarrow \nu_\mu$ as
functions of the neutrino energy for different values of CP-phase and
zenith angles. In the low energy range where sensitivity to $\delta$
is high and consider $P_{e\mu}$.  In Fig.~\ref{fig:emu_prob} the
resonantly enhanced probability due to the 1-2 mixing and mass
splitting is modulated by fast oscillations driven by the 1-3 mass and
mixing.  The 1-2 resonance energy in the mantle is at $E_{12}^R
\approx 0.12$ GeV.  For core crossing trajectories (upper panels) the
parametric effects distort the dependence of probability on energy.

The key feature which opens up a possibility to measure $\delta$ is
the presence of systematic shift of the oscillatory curves
(probabilities) at low energies, $ \lesssim 2$ GeV, with increase of
the phase. The shift occurs in the same way in wide energy interval
$E_{\nu} = (0.2 - 2)$ GeV, and essentially for all trajectories which
cross the mantle only. This systematic shift can be understood using
analytical expressions for the probabilities. Averaging
$P_{e\mu}^\delta$ (\ref{pemu2}) over fast oscillations driven by the
1-3 splitting we find
\be
\langle {P}_{e\mu}^\delta \rangle = \frac{J_\theta}{2}  
\left[
\cos \delta \cos 2 \theta_{12}^m \sin^2 \phi_{21}^m  
+
\frac{1}{2}\sin \delta \sin 2 \phi_{21}^m  
\right]. 
\label{pemu-av}
\ee       
The first term does not change the sign with $\phi_{21}^m$, whereas
the second one does. Notice that above the 1-2 resonance $\cos 2
\theta_{12}^m \approx - 1$, and so
\be
\langle {P}_{e\mu}^\delta \rangle \approx \frac{J_\theta}{2}
\left[- \cos \delta \sin^2 \phi_{21}^m
+ \frac{1}{2}\sin \delta \sin 2 \phi_{21}^m  \right].
\ee
The difference of probabilities for a given value $\delta$ and $\delta
= 0$ equals:
\be
\langle P_{e\mu}^\delta \rangle - \langle P_{e\mu}^{0} \rangle 
= \frac{J_\theta}{2}  
\left[ (1 -  \cos \delta)  \sin^2 \phi_{21}^m + 
\frac{1}{2}\sin \delta \sin 2 \phi_{21}^m  \right]. 
\label{pemu-av1}
\ee
The first term is positive for all $\phi_{21}^m$ and $\delta$, and it
is this term that produces a systematic shift of the probabilities.
 
For the values of $\delta$-phase shown in Fig.~\ref{fig:emu_prob}  
we obtain from (\ref{pemu-av})
\ba
\langle P_{e\mu}^{0} \rangle & = & - \langle P_{e\mu}^{\pi} \rangle =  
\frac{J_\theta}{2}  \cos 2 \theta_{12}^m \sin^2 \phi_{21}^m , \\
\langle P_{e\mu}^{\pi/2} \rangle &  = & - \langle P_{e\mu}^{3\pi/2} \rangle
=  \frac{J_\theta}{4}  \sin^2 2 \phi_{21}^m . 
\nonumber
\ea
These equations show that $\langle P_{e\mu}^{0} \rangle$ is the
smallest one.  The probability increases with $\delta$ and reaches
maximum at $\delta = \pi$. For the trajectory with $\cos \theta_z = -
0.4$, the oscillation phase equals $\phi_{21}^m \approx \pi/2$. That
leads to $\langle P_{e\mu}^{\pi/2}\rangle =
\langle{P}_{e\mu}^{3\pi/2}\rangle = 0$, and consequently, to equal
total probabilities. For $\cos \theta_z = - 0.8$ the phase equals
$\phi_{21}^m = 1.42\, \pi$ which gives different values of probability:
$\langle {P}_{e\mu}^{\pi/2} \rangle = -
\langle{P}_{e\mu}^{3\pi/2}\rangle = 0.24 J_\theta/4 = 0.06 J_\theta $,
and furthermore $\langle{P}_{e\mu}^{\pi}\rangle = -
\langle{P}_{e\mu}^{0} \rangle = 0.94 J_\theta/2 = 0.47 J_\theta$.
These results are in agreement with plots shown in
Fig.~\ref{fig:emu_prob}.

Although there is certain phase shift with change of $\delta$, the
sizes of energy intervals where the difference $P(\delta_1) -
P(\delta_2)$ has positive and negative signs are strongly different.
One sign dominates, and therefore there is no averaging over energy.
Maximal relative upward shift of the probability curves compared to
the $\delta = 0$ curve is around $(0.4 - 1)$ GeV.  For the
core-crossing trajectories ($\cos \theta_z < - 0.83$) due to the
parametric effects the transition probability first increases with
increase of $\delta$, it reaches maximum at $\delta \sim \pi/2$ and
then decreases.

The $\nu_\mu - \nu_\mu$ probability, $P_{\mu \mu}^\delta$
(\ref{pmumu2}), averaged over the 1-3 oscillations equals
\be
\langle P_{\mu \mu}^\delta \rangle =  - \frac{J_\theta}{2}  \cos \delta  
\sin^2 \phi_{21}^m \cos 2\theta_{12}^m ,   
\label{pmumu-av}
\ee 
where the $D_{23}$ term is neglected.  Notice immediately that the
CP-effect in the $\nu_\mu - \nu_\mu$ channel has an opposite sign with
respect to that in the $\nu_e - \nu_\mu$ channel (\ref{pemu-av}).
Therefore the presence of both $\nu_e$ and $\nu_\mu$ original fluxes
weakens the total CP-effect, and consequently, the sensitivity to
$\delta$ which is unavoidable.  We will call this {\it the flavor
  suppression}.
 
According to (\ref{pmumu-av}) dependence of the $\nu_\mu - \nu_\mu$
probability on $\delta$ factors out and therefore turns out to be very
simple.  The maximal effect is for $\delta = 0$,
\be
\langle P_{\mu \mu}^{0} \rangle = - \langle P_{\mu \mu}^\pi \rangle  
\approx  \frac{J_\theta}{2}  \sin^2 \phi_{21}^m, 
\nonumber
\ee
and $\langle P_{\mu \mu}^{\pi/2} \rangle = \langle P_{\mu
  \mu}^{3\pi/2} \rangle = 0$, so that the total probabilities are
equal for $\pi/2$ and $3\pi/2$ which in perfect agreement with result
of Fig.~\ref{fig:emu_prob}.

The difference of probabilities for a given value of $\delta$ and zero
phase equals
\be
\langle {P}_{\mu \mu}^\delta \rangle -    \langle {P}_{\mu \mu}^{ 0} \rangle
= \frac{J_\theta}{2} 
(1 - \cos \delta) \sin^2 \phi_{21}^m \cos 2\theta_{12}^m \approx 
- \frac{J_\theta}{2}(1 - \cos \delta) \sin^2 \phi_{21}^m. 
\label{numudiff}
\ee
Only CP-even contribution is present.

The probabilities in antineutrino channels are shown in
Fig.~\ref{fig:anti_emu_prob}.  Their dependencies on $E_\nu$ and $\cos
\theta_z$ can be immediately understood from our analytical treatment.
According to (\ref{antinu}) the averaged probabilities equal
\ba
\langle \bar{P}_{e\mu}^\delta \rangle & = & \frac{\bar{J}_\theta}{2} 
\left[ 
 \cos \delta \cos 2 \bar{\theta}_{12}^m \sin^2 \bar{\phi}_{21}^m 
- \frac{1}{2} \sin \delta \sin 2 \bar{\phi}_{21}^m \right],
\label{anpemu-av}\\
\langle \bar{P}_{\mu \mu}^\delta \rangle & = & - \frac{\bar{J}_\theta}{2} \cos \delta
\sin^2 \bar{\phi}_{21}^m \cos 2 \bar{\theta}_{12}^m . 
\label{anpmumu-av}
\ea
For energies far above the 1-2 resonance, the expressions are further
simplified since $\cos 2 \bar{\theta}_{12}^m \approx 1$ (recall, for
neutrinos $\cos 2\theta_{12}^m \approx - 1$):
\be
\langle \bar{P}_{e\mu}^\delta \rangle  =   \frac{\bar{J}_\theta}{2} 
\left[\cos \delta  \sin^2 \bar{\phi}_{21}^m 
- \frac{1}{2}\sin \delta \sin 2 \bar{\phi}_{21}^m  
\right], ~~~~~
\langle \bar{P}_{\mu \mu}^\delta \rangle  =  
- \frac{\bar{J}_\theta}{2} \cos \delta
\sin^2 \bar{\phi}_{21}^m . 
\label{anpmumu-avlim}
\ee
Comparing this with (\ref{pemu-av1}) and (\ref{pmumu-av}) we find that
for antineutrinos the probabilities have opposite sign with respect to
the probabilities for neutrinos.  Indeed, according to
Fig.~\ref{fig:anti_emu_prob} for mantle trajectories the biggest
amplitude $\bar{P}_{e \mu}$ is for $\delta = 0$ and the smallest one
is for $\delta = \pi$ which is opposite to the $P_{e \mu}$ case. This
means that summation of signals from neutrinos and antineutrinos
reduces the effect of CP-phase, and consequently, the sensitivity to
this phase.  This {\it C-suppression} can be reduced if $\nu$ and
$\bar{\nu}$ signals are separated at least partially (see Sec. IV C).

As follows from Fig.~\ref{fig:anti_emu_prob} for the mantle crossing
trajectories, only the largest CP effect on $\bar{P}_{e \mu}$ is in
the range $E_\nu = (0.4 - 0.7)$ GeV where maximal values equal
$\bar{P}_{e \mu} \approx 0.1$ and $0.15$ for $\cos \theta_z = - 0.8$
and $- 0.4$ correspondingly.  These numbers are about 2 times smaller
than for neutrinos.  The reason is that, in the case of NH for
neutrinos both $\theta_{12}^m$ and $\theta_{13}^m$ are enhanced in
matter whereas for antineutrinos both $\bar{\theta}_{12}^m$ and
$\bar{\theta}_{13}^m$ are suppressed.  The antineutrino probabilities
decrease with increase of energy above 0.8 GeV.  This, as well as
smaller $\bar{\nu}$ cross-sections suppresses number of $\bar{\nu}$
events and therefore reduces cancellation of the CP-effect.

Similar consideration can be performed for the $\nu_\mu - \nu_\mu$
channel for which $ \langle \bar{P}_{\mu \mu}^\delta \rangle = -
{\bar{J}_\theta}{2} \cos \delta \sin^2 \bar{\phi}_{21}^m$.  Notice
that in vacuum $\langle \bar{P}_{\mu \mu}^\delta \rangle = \langle
P_{\mu \mu}^\delta \rangle$, {\it i.e.}\ the probability is even
function of $\delta$. In the matter dominated region we have $\langle
\bar{P}_{\mu \mu}^\delta \rangle \approx - \langle P_{\mu \mu}^\delta
\rangle$ due to change of sign of the potential.  The differences of
the antineutrino probabilities for a given $\delta$ and $\delta= 0$
equals at $\cos 2 \theta_{12}^m \approx 1$
\ba
\langle \bar{P}_{e\mu}^\delta \rangle - 
\langle \bar{P}_{e\mu}^{0} \rangle 
& = & - \frac{\bar{J}_\theta}{2} 
\left[(1 - \cos \delta)  \sin^2 \bar{\phi}_{21}^m 
+  \frac{1}{2}\sin \delta \sin 2 \bar{\phi}_{21}^m  
\right],
\label{anpemu-avlim}\\
\langle \bar{P}_{\mu \mu}^\delta \rangle - 
\langle \bar{P}_{\mu \mu}^{0} \rangle  
& = & \frac{\bar{J}_\theta}{2} (1 -  \cos \delta)
\sin^2 \bar{\phi}_{21}^m .   
\label{anpmumu-avdiff}
\ea
They also have an opposite sign with respect to the differences for
neutrinos (\ref{pemu-av1}) and (\ref{numudiff}), and equal up to
change of mixing angles and phases in matter.

\subsection{Magic lines and CP-domains}

In what follows we will study differences of probabilities as well as
distributions of events in the $E_\nu - \cos \theta_z$ plane for
different values of $\delta$. The patterns of distributions are
determined to a large extent by the grid of the magic lines
\cite{Barger:2001yr, Huber:2003ak, Smirnov:2006sm, our3}.  The lines
fix the borders of the CP-domains -- the regions in the $E_\nu - \cos
\theta_z$ plane of the same sign of the CP-difference.

Let us summarize relevant information about properties of the magic
lines.  Recall that the magic lines are defined as the lines in the
$E_\nu - \theta_z$ plane along which the oscillation probabilities do
not depend on phase $\delta$ in the so called factorization (quasi $2
\nu$) approximation \cite{our3}. Correspondingly, the CP-differences
vanish along these lines.

\begin{enumerate}[(i)]

\item The {\it solar magic lines} are determined by the condition
\be
\phi_S = \phi_{21}^m  = n \pi, ~~~~n = 1, 2, 3, ..., 
\label{solarlines}
\ee
where in neutrino channels $\phi_S$ is given by the expression
(\ref{ph12}) for $\phi_{21}^m$ valid in $3\nu$ framework below 1-3
resonance but extended to all the energies. For antineutrinos in the
NH case $\bar{\phi}_S = \bar{\phi}_{21}^m$ everywhere.  Along these
lines $|A_{e \tilde{2}}| = 0$ below the 1-3 resonance. That would be
the line of zero solar amplitude in the $2\nu$ approximation. The
minimum of $\nu_e - \nu_\mu$ probability at $\sim 0.15$ GeV for $\cos
\theta_z = - 0.4$ in Fig.~\ref{fig:emu_prob} corresponds to the first
magic line with $ \phi_S = \phi_{21}^m = \pi$.  The minimum at $0.17$
GeV for $\cos \theta_z = - 0.8$ (Fig.~\ref{fig:emu_prob}) is on the
second magic line with $\phi_{21}^m = 2\pi$.

Notice that the energy of minimal level splitting (maximal oscillation
length) is given by $E_{12}^R / \cos^2 2 \theta_{12} \approx 0.7$ GeV
which is much bigger than $E_{12}^R = 0.12$ GeV due to large 1-2
mixing. So, below 0.7 GeV the splitting increases and correspondingly
the oscillation length decreases. Therefore the same phase can be
obtained for smaller $|\cos \theta_z|$, and consequently, the solar
magic lines bend toward smaller $|\cos \theta_z|$.  At energies much
above the 1-2 resonance these lines do not depend on energy and are
situated at
\be
\cos \theta_z = - 0.60,~ -0.86, ~ - 0.97,    
\label{solarlll}
\ee
for $\phi_{21}^m = \pi, 2\pi$, and $3\pi$ correspondingly.  

\item The {\it atmospheric magic lines} are determined by the equality
\be
\phi_A = \phi_{23}^m = n \pi, ~~~~n = 1, 2, 3, ...
\label{atmlines}
\ee
Along these lines $|A_{e \tilde{3}}| \approx 0$. It would vanish
exactly in the $2\nu$ approximation, when $\cos^2 \theta_{12}^m
\approx 0$ that is far above the 1-2 resonance.  Zeros of the $\nu_e -
\nu_\mu$ probability at $E_\nu \geq 2$ GeV (see
Fig.~\ref{fig:emu_prob}) which do not depend on $\delta$ are situated
on the atmospheric magic lines. E.g., for $\cos \theta_z = - 0.4$
these points are at $E_\nu = 2$ GeV and $E_\nu = 3.2$ GeV.  For $\cos
\theta_z = - 0.8$, zeros are at $E_\nu = 2.3,~ 2.9,~ 4.1$ GeV.  For
$P_{\mu \mu}$ the solar and atmospheric magic lines coincide with
those for $P_{e \mu}$ in the limit $D_{23} = 0$.

The magic lines determined by (\ref{solarlines}) and (\ref{atmlines})
do not coincide with lines where $|A_{e \tilde{2}}| = 0$ and $|A_{e
  \tilde{3}}| = 0$ in the $3 \nu$ framework.  But they play the role
of asymptotics of the true lines where dependence of probabilities on
$\delta$ disappears. The latter interpolate between different magic
lines.

\item The {\it interference phase} lines are important for distinguishing
different values of the CP-phase: a given value $\delta$ and a
different value $\delta_0$. Along these lines $P_{\alpha \beta}^\delta
- P_{\alpha \beta}^{\delta_0} = 0$. According to (\ref{eq:Pemudelta})
for $P_{e \mu}$ the condition reads
\be
\cos (\phi + \delta) = \cos (\phi + \delta_0), 
\nonumber
\ee
where $\phi \approx - \phi_{31}$ and the latter is the vacuum
oscillation phase.  This condition corresponds to intersection of
probability curves for different values of phases $\delta$ and
$\delta_0$ in Fig.~\ref{fig:emu_prob}.  For $\delta_0 = 0$ the
condition can be written as $\phi_{31} + \delta = - \phi_{31}$ or
\be
\phi_{31}= \frac{\Delta m_{31}^2 L}{4 E_\nu} = - \frac{\delta}{2} + n \pi.  
\label{int-mu}
\ee
For the inverse channel, $\nu_\mu \rightarrow \nu_e$, the sign of
$\delta$ should be changed.  According to (\ref{eq:Pmumudelta})
dependencies of the $\nu_\mu \rightarrow \nu_\mu$ probability on
$\phi$ and $\delta$ factor out in the approximation $D_{23} = 0$, and
the corresponding interference phase line is determined by the
condition $\cos \phi = 0$, or
\be
\phi \approx \phi_{31} = \frac{\pi}{2} + n \pi. 
\nonumber
\ee
The condition can be written as 
\be
E_{\nu} = - A(\phi) \cos\theta_z = 
- \frac{R_E \Delta m_{31}^2}{2 \phi (\delta)} \cos\theta_z,  
\label{eq:magicE}
\ee
where $R_E$ is the Earth radius and in general $\phi (\delta)$ should
lead to the vanishing CP-difference of probabilities.

The exact value of interference phase $\phi$ does not coincide with $-
\phi_{31}$.  In the constant density approximation $\phi$ equals the
phase of the expression in brackets of $A_{e\tilde{3}}$
(\ref{ae3-con}):
\be
\tan \phi = - \frac{\sin \phi_{32}^m  \sin \phi_{31}^m}
{\cos \phi_{31}^m \sin \phi_{32}^m    +
\cos^2 \theta_{12}^m  \sin \phi_{21}^m}.
\label{intph}
\ee
Notice that $\phi$ would be equal $- \phi_{31}^m$, if $\cos
\phi_{31}^m = 0$. The latter is satisfied for high energies $E \gg
E_{12}^R$, where $\cos^2 \theta_{12}^m \approx 0$.  However, if
$\phi_{31}^m \approx \pi/2$ we can not neglect the second term in the
denominator of (\ref{intph}).  Notice that in the limit $\cos^2
\theta_{12}^m = 0$ we obtain from (\ref{pmumu2a})
\be
P_{\mu \mu}^\delta =  - \cos \delta  J_\theta
\sin \phi_{21}^m \sin \phi_{32}^m  \cos \phi_{31}^m,
\label{pmumu2b}
\ee
where one can see immediately all three ``magic'' conditions.

\end{enumerate}

Notice that magic lines could be introduced immediately in the $3\nu$
framework as the lines along which $P_{\alpha \beta}^\delta -
P_{\alpha \beta}^{\delta_0} = 0$.  In this case they would, indeed,
determine the borders of domains with different sign of the
CP-difference of the probabilities. We use the original definitions of
magic lines to match with previous discussions. Still as we said
before, the solar, atmospheric and interference lines nearly coincide
with the exact lines of zero CP-differences in certain energy
intervals or in asymptotics. The corresponding phases are related as
\be
\phi_{21}^m \approx 
\left\{
\begin{array}{l}
\phi_S^m, ~~~ E \ll E_{31}^R\\
\phi_A^0, ~~~  E \geq E_{31}^R 
\end{array}
\right. , ~~~~
\phi_{31}^m \approx
\left\{
\begin{array}{l}
\phi_A^0, ~~~ E \ll E_{31}^R\\
\phi_S, ~~~  E \gg E_{31}^R
\end{array}
\right. , ~~~~
\phi_{32}^m \approx \phi_A^m ,  
\nonumber
\ee
where $\phi_A^0$ is the phase in vacuum. So, the true lines of zero
difference of probabilities interpolate between the magic lines (see
\cite{our3} for details).

\section{PINGU, Super-PINGU and CP}

The key conclusion of the previous section is that integration over
the neutrino energy and direction does not suppress the CP effect
significantly. Furthermore, for all trajectories which cross the
mantle of the Earth only, the CP violation effect is similar: it has
the same sign and the same change with $\delta$. Effect is different
for the core crossing trajectories, $\cos \theta_z < - 0.83$. So,
there could be partial cancellation due to smearing over the zenith
angle.  Another important feature is that the relative CP effect at
the probability level increases with decrease of energy.  In this
connection we will explore a possibility to measure $\delta$ using
multi-megaton scale neutrino detectors with low energy threshold.  As
it was already realized in \cite{ARS}, sensitivity of PINGU to
$\delta$ is low. So, we will consider future possible upgrades of
PINGU.  We will also quantify capacity of PINGU to obtain information
about $\delta$. For definiteness we will speak about PINGU for which
more information is available. Similar upgrades can be considered for
ORCA detector \cite{orca}.

\subsection{PINGU and Super-PINGU}

We calculate event rates for the proposed PINGU detector and for
possible future PINGU upgrade which we call Super-PINGU. The PINGU
detector \cite{pingu2} will have 40 strings additional to the DeepCore
strings with 60 digital optical modules (DOM's) at 5 m spacing in each
string. A compact array like PINGU could detect neutrinos with
energies as low as $(1 - 3)$ GeV. Strict criteria allow over $90\%$
efficiency of event reconstruction for all 3 flavors \cite{pingu2}. We
parametrize the PINGU effective mass as
\begin{equation}
\rho V_{{\rm eff}, \mu} (E_\nu ) =  3.0 \left[ \log (E_\nu /{\rm GeV}) 
\right]^{0.61} \, {\rm Mt}
\label{PINGU_num}
\end{equation}
and
\begin{equation}
\rho V_{{\rm eff}, e} (E_\nu ) =  3.1 \left[ \log (E_\nu /{\rm GeV}) 
\right]^{0.60} \, {\rm Mt},
\label{PINGU_nue}
\end{equation}
respectively for $\nu_\mu$ and $\nu_e$. Here $V_{{\rm eff}, \alpha}$
is the effective volume and $\rho$ is the density of the ice.  These
parametrizations well represent simulated volumes \cite{pingu2} from
$\gtrsim 1$ GeV up to 25 GeV.  We will use an accuracy of the energy
and angle reconstruction for PINGU from \cite{pingu2}.

Along with the PINGU proposal the idea has been discussed to construct
``ultimate'' multi-megaton-scale detector MICA with a threshold about
10 MeV allowing to detect the solar and supernova neutrinos
\cite{mica}.  Clearly reducing the threshold by more that 2 orders of
magnitude is very challenging.  In this connection we would like to
consider a kind of intermediate step - the detector with an effective
energy threshold about (0.1 - 0.2) GeV, {\it i.e.} \ one order of
magnitude below the threshold in the present PINGU proposal. For this,
a denser array of DOM's is required which will lead to increase of the
effective volume of a detector at low energies. For definiteness we
will take the effective volume which corresponds to the PINGU detector
simulations with a total of 126 strings and 60 DOM's per string each
\cite{cowen}. The effective mass can be parameterized as
\begin{equation}
\rho V_{\rm eff} (E_\nu )  =  2.6 \left[ {\rm log} (E_\nu /{\rm GeV}) + 1 
\right]^{1.32} \, {\rm Mt},
\label{new-v}
\end{equation}
for both $\nu_\mu$ and $\nu_e$ events.  We call this version
Super-PINGU.  According to (\ref{PINGU_num}) and (\ref{new-v}) the
effective mass, $\rho_{eff} V_{eff}$, in the range (1 - 2) GeV equals
$0.7$ Mton for PINGU and $\sim 2.8$ Mton for Super-PINGU, {\it i.e.} 4
times larger. For the bin below 1 GeV the corresponding numbers are
0.3 and 2.2 Mton (7 times larger).  This can be compared with MICA,
which may have 220 strings and 140 DOMs per string.  We will
extrapolate to lower energies some PINGU characteristics from the
proposal \cite{pingu2}. 

Going to further upgrade has double effect: 
\begin{itemize}
\item increase of the effective volume, especially in the low energy
  bins, and
\item improvements of reconstruction of the neutrino energy and
  direction as well as the flavor identification of events for all
  energies.
\end{itemize}

Super-PINGU will have three times denser DOM array than PINGU.
Therefore it will collect about 3 times more photons from the same
event (with the same neutrino energy). Recall that, in PINGU the
average distance between DOM's is smaller than the photon scattering
length (50 m).

We describe uncertainties of reconstruction of the neutrino energy and
direction by smearing functions
\be
G_E (E_{\nu}^r , E_\nu),  ~~~G_\theta (\theta_{z}^r , \theta_z), 
\nonumber
\ee
where $E_\nu$ and $\theta_z$ ($E_\nu^r$ and $\theta_z^r$) are the true
(reconstructed) energy and zenith angle of the neutrinos. For PINGU we
use $G_E$ and $G_\theta$ from \cite{pingu2} determined down to
energies $\sim 1$ GeV.  The distributions are normalized in such a way
that
\be
\int  d E_\nu d\theta_z  G_E (E_{\nu}^r , E_\nu)  
G_\theta (\theta_{z}^r , \theta_z)  = 1. 
\nonumber
\ee
Notice that PINGU distributions have longer tails than the Gaussian
functions.

Characteristics of the Super-PINGU reconstruction are expected to be
better.  We estimate parameters of $G_E$ and $G_\theta$ for
Super-PINGU using the DeepCore resolutions and the simulated PINGU
resolutions \cite{dcres}, \cite{gross} in the following way.  For a
given event the number of photons collected is proportional to the
density of DOMs, that is $N_{\rm DOM}$ for fixed total volume of the
detector.  Therefore the relative statistical error in determination
of characteristics is proportional to $1/\sqrt{N_{\rm DOM}}$, so we
can assume that
\be 
\sigma_\theta  \propto  
\frac{1}{\sqrt{N_{\rm DOM}}},~~~~
\sigma_E \propto \frac{1}{\sqrt{N_{\rm DOM}}}.  
\label{resDOM}
\ee 
Estimations of resolutions of the DeepCore and PINGU confirm
(\ref{resDOM}). Indeed, DeepCore has about $N_{\rm DOM}^{\rm DC} =
530$ DOM's, while PINGU (40 strings with 60 DOM's per string) will
have $N_{\rm DOM}^{\rm PINGU} = 2400$ DOM's (also with higher quantum
efficiency), that is, $N_{\rm DOM}^{\rm PINGU}/ N_{\rm DOM}^{\rm DC} =
4.5$.  Since the density of DOM's in PINGU is about 4.5 times larger,
amount of light detected from the same event will be about 4.5 times
larger.  According to \cite{dcres}, \cite{gross} and \cite{pingu2} for
the $\nu_{\mu}$ events the ratio of resolutions (median errors)
\be
\frac{\sigma_\theta^{\rm PINGU}}{\sigma_\theta^{\rm DC}} \approx 0.5. 
\label{eq:median}
\ee
The ratio equals 0.66 at $E_\nu = 5$ GeV, however estimation of DC
parameters become not very reliable at low energies.  For the $\nu_e$
events the improvement is even better: The ratio of median errors
(\ref{eq:median}) is ($0.38 - 0.43$) in the interval $E_\nu = (10 -
20)$ GeV and it becomes 0.6 at 5 GeV.

For neutrino energy reconstruction (median energy resolution) of the
$\nu_\mu$ events we have
\be
\frac{\sigma_E^{\rm PINGU}}{\sigma_E^{\rm DC}} \approx 0.58 - 0.61  
\label{eq:rat}
\ee
in the interval $E_\nu = (10 - 20)$ GeV. It decreases down to 0.52 at
$E_\nu = 5$ GeV. Similar improvement is expected for the $\nu_e$
events.

The Super-PINGU will have 3 times larger number (and therefore density)
of DOM's, than PINGU. Therefore according to (\ref{resDOM}) the
resolutions will be further improved by factor $1/\sqrt{3} \approx
0.58$.  So for Super-PINGU we use the resolution functions from Fig.~7
and 8 of \cite{pingu2}, scaling their widths as
\be 
\sigma_\theta^{\rm SuperPINGU} = \frac{1}{\sqrt{3}} 
\sigma_\theta^{\rm PINGU}, ~~~ 
\sigma_E^{\rm SuperPINGU} = \frac{1}{\sqrt{3}} \sigma_E^{\rm PINGU}. 
\label{width2}
\ee 
We extrapolate these functions down to $E_\nu = 0.5$ GeV and for
simplicity neglect possible dependences of the factors in Eq.
(\ref{width2}) on energy. (Notice that according to \cite{pingu2} the
median value of angle is very similar for cascades and tracks.)
  
This estimation of improvement can be considered as conservative.
Indeed, the DeepCore characteristics have been obtained after
stringent kinematical cuts which allows one to select a sample of high
quality events.  That reduces efficiency of reconstruction (fraction
of reconstructed events) down to $(10 - 20) \%$, whereas PINGU
characteristics have been obtained with $(60 - 70) \%$ efficiency.
With stronger cuts in PINGU the reconstruction characteristics could
be even better. Also, developments of electronics may lead to further
improvements.  Clearly, configuration of Super-PINGU should be
optimized taking into account also the cost of construction.  For
large density of strings the issue of the ice stability may become
important.  One can reduce number of strings by increasing number of
DOMs per string (decreasing vertical spacing).  Since typical size of
an event is about 100 m, for distances between strings (17 m) the
total number of DOMs in the unit volume matters and geometry plays
only secondary role.  Another option is to consider underwater
detector, i.e., an upgrade of ORCA.

\subsection{Distributions of events in the  neutrino energy and zenith angle plane}

To evaluate sensitivity of Super-PINGU to $\delta$ we will compute the
$(E_\nu - \cos \theta_z)$ distributions of events of different types
and explore their dependence on $\delta$.  The numbers of events
$N_\alpha$, produced by neutrinos $\nu_\alpha$ ($\alpha = e, \mu$)
with energies and zenith angles in small bins $\Delta (E_\nu)$ and
$\Delta(\cos \theta_z)$ marked by subscript $ji$ equal
\be 
N_{ij, \alpha} = 2 \pi N_A \rho T 
\int_{\Delta_i\cos\theta_z} d\cos\theta_z \int_{\Delta_jE_\nu} dE_\nu~ 
V_{\rm eff, \alpha} (E_\nu) d_\alpha (E_\nu, \theta_z). 
\label{eq:nev} 
\ee 
Here $T$ is the exposure time, $N_A$ is the Avogadro's number.  The
density of events of type $\alpha$, $d_\alpha$, (the number of events
per unit time per target nucleon) is given by
\be
d_\alpha(E_\nu, \theta_z)    = d_\alpha^{\nu} + d_\alpha^{\bar{\nu}} =  
\left[\sigma_\alpha \Phi_\alpha  + {\bar \sigma}_\alpha 
{\bar \Phi}_\alpha  \right], 
\label{eq:den1}
\ee
where $\Phi_\alpha$ and ${\bar \Phi}_\alpha$ are the fluxes of
neutrinos and antineutrinos at the detector which produce events of
the type $\alpha$, and $\sigma_\alpha$ and ${\bar \sigma}_\alpha$ are
the corresponding cross-sections. In turn, the fluxes at the detector
equal
\be
\Phi_\alpha = \Phi_\mu^0 P_{\mu \alpha} + \Phi_e^0 P_{e \alpha}, 
\nonumber
\ee 
$\Phi_\mu^0 = \Phi_\mu^0 (E_\nu,\theta_z) $ and $\Phi_e^0 =
\Phi_e^0(E_\nu,\theta_z)$ are the original muon and electron neutrino
fluxes at the production.

With decrease of energy, resonance processes (pion production) and
quasi-elastic processes will contribute, and the latter dominates
below 1 GeV.  In our estimations we use the total neutrino-nucleon
cross-sections down to (0.2 - 0.3) GeV as they are parametrized in
\cite{pingu2}, We assume that different contributing processes would
produce visible effect at the detector with the same efficiency.  For
antineutrinos there is no data below 1 GeV and we use extrapolation
given in \cite{pingu2}.  Clearly in future these computations should
be refined.

We use the atmospheric neutrino fluxes, $\Phi_\mu^0$ and $\Phi_e^0$
(and corresponding fluxes of antineutrinos) from
Refs.~\cite{Honda:1995hz, Athar:2012it}.  At low energies the
geomagnetic effects become important which break azimuthal symmetry.

After smearing in the ($E^r_\nu - \cos \theta_{z}^r$) plane, we
obtained the unbinned distribution of events as
\be
N_{\alpha}(E^r, \cos \theta^r)  =  
{2\pi N_A T \rho}  \int  d\cos\theta_z  \int dE_\nu~
G_E (E_{\nu}^r , E_\nu)~G_\theta (\theta_{z}^r , \theta_z) 
 ~V_{\rm eff} (E_\nu)~ d_\alpha (E_\nu, \cos\theta_z),   
\label{eq:nres1}
\ee
$\alpha = e, \mu$, and then  binned them according to   
\be
N_{ij, \alpha}   =  \int_{\Delta_i (\cos\theta_z^r)} 
d\cos\theta_z^r \int_{\Delta_j (E_\nu^r)} dE_\nu^r
~N_\alpha (E^r, \cos \theta_z^r), 
\label{eq:nres2}
\ee
with $\Delta (E^r_\nu) = 1$ GeV and $\Delta (\cos\theta_z^r) =0.05$.
Again, we can split number of events onto $\delta - $ dependent and
$\delta - $independent parts: $N_{ij}(\delta) = N_{ij}^{ind} +
N_{ij}^{\delta}$.

\subsection{CP-asymmetry and distinguishability}

As in \cite{ARS}, we will employ the distinguishability $S_\sigma$ as
a quick estimator of sensitivity of measurements.  For a given type of
events (we omit the index $\alpha$) and each $ij$-bin we define the
relative CP-difference as
\be
S_{ij}(f) = \frac{N_{ij}^{\delta} - N_{ij}^{0}}{\sigma_{ij}},
\label{sij}
\ee 
where $N_{ij}^{\delta}$ and $N_{ij}^{0}$ are the numbers of events
computed for a given value of $\delta$ and for $\delta = 0$
correspondingly, and
\be 
\sigma^2_{ij} = N_{ij}(\delta = 0) + [f N_{ij}(\delta = 0)]^2 
= N_{ij}^{ind} + N_{ij}^{0} + f^2 (N_{ij}^{ind} + N_{ij}^{0})^2 
\nonumber 
\ee 
is the total ``error'' in the $ij$-bin.  If $N_{ij}(\delta = 0)$ is
interpreted as a result of measurement, the first term in the equation
above would correspond to the statistical error and the second one to
the uncorrelated systematic errors.  As in \cite{ARS} we assume that
the latter is proportional to the number of events: $f N_{ij}(\delta =
0) $.  In general $f$ is a function of neutrino energy and zenith
angle.  The uncorrelated errors could be due to local impurities
(dust) in the ice, uncontrolled efficiency of individual DOMs,
uncertainties in neutrino fluxes and cross-sections (on top of overal
normalization and tilt uncertainties). The level of these
uncertainties is not known. So, only what we can do is to explore how
sensitivity changes depending on the level of uncertainties.  This
allows us to conclude about tolerable level of $f$.

Notice that, since here contribution from the systematic error is
proportional to $(N^0_{ij})^2$, for the same $f$ the role of this
error decreases with decreasing size of the bin.  Since here we use 2
times larger both in $E_\nu$ and $\cos \theta_z$ bins after smearing
than in \cite{ARS}, to keep the same level of systematic errors with
respect to statistical error we need to use 2 times smaller $f$. For
illustration we will take values $f = 2.5 \%$ and $5\%$ which
correspond to the cases when uncorrelated systematic error smaller and
comparable with the statistical error.  Notice that $f = 2.5\%$ gives
the closest approximation of our results to the results of PINGU
simulations \cite{pingu2} in the case of sensitivity to the neutrino
mass hierarchy.

If $N_{ij}^{\delta}$ is considered as the fit value, the moduli
$|S_{ij}|$ would give the standard deviation and so the statistical
significance.  However, in contrast to real situation the ``measured''
value $N^{0}_{ij}$ does not fluctuate. Therefore we will not interpret
it as number of sigmas, but just use $|S|$ as independent
characterization - the distinguishability.

Considering the effect in each bin as an independent measurement
(which is possible after smearing), we can define the total
distinguishability as
\be 
S_{\sigma} = \sqrt{\sum_{ij} 
S_{ij}^2} = \sqrt{\sum_{ij} \frac{(N_{ij}^{\delta} - 
N_{ij}^{0})^2}{\sigma^2_{ij}}}, 
\ee 
where the sum is over all the bins.  

Although the correlated systematic errors, e.g., those of the overall
flux normalization and the tilt of the spectrum, do not reproduce the
pattern of the distribution for different values of $\delta$, still
they can reduce significance substantially.  Effects of the correlated
errors will be considered in Sec. VI B.

We will avoid precise statistical interpretation of distinguishability
and just consider that it gives some idea about significance and
sensitivity. Still, in various cases $|S|$ turns out to be close to
the significance as follows from comparison of our previous
estimations with results of complete MC simulations \cite{ARS,
  pingu2}.  Furthermore it reproduces rather precisely dependences of
sensitivities on characteristics of detectors and neutrino parameters.

Apart from the total distinguishability, the sensitivity can be
characterized also by maximal positive and negative CP-differences in
individual bins in a given range of energies and zenith angles.

Of course, the $\chi^2$ or maximal likelihood analyses would give
higher quality, more reliable and precise estimation of the
sensitivity. For this, however, one needs to perform Monte Carlo (MC)
simulation of events at Super-PINGU. We are certainly not in position
to make these simulations and this is beyond the scope of this paper.

\section{Distributions of the $\nu_\mu$ events}

\subsection{The density of $\nu_\mu$ events}

The $\nu_\mu$ (track) events produced mainly by the charged current
$\nu_\mu$ interactions, $\nu_\mu + N \rightarrow \mu + X$,
$\bar{\nu}_\mu + N \rightarrow \mu^+ + X$, are observed as muon tracks
accompanied by hadronic cascades.  For these events the energy of the
muon $E_\mu$ and the direction of its trajectory characterized by the
angles $\theta_\mu$ and $\phi_\mu$ as well as the total energy of the
hadronic cascade (for deep-inelastic scattering) $E_h$ can be measured.
Using this information one can reconstruct the neutrino energy as
\be 
E_\nu^{r} \approx  E_\mu + E_h - m_N\,,
\nonumber
\ee
where $m_N$ is the nucleon mass.  Also the direction of cascade can be
determined to some extent.  So one can reconstruct the neutrino
direction.

At low energies processes with one pion resonance production is
important and below 1 GeV the quasi-elastic scattering dominates.  For
these events procedure of reconstruction of the neutrino energy and
direction becomes different.  So, the detection of the low energy
events should be considered separately, and such a study is beyond the
scope of this paper.  There are also some contributions from
$\nu_\tau$ which produce $\tau$ leptons with subsequent decay into
muons.  In this initial study for estimations we extrapolate
characteristics of reconstruction functions determined at high
energies down to low energies.

The $\delta$-dependent part of the number density of the $\nu_\mu$
events in a single bin equals
\be
d_\mu^\delta\equiv 
\sigma^{CC} \Phi_\mu^0 \left[\left(P_{\mu\mu}^\delta+\frac{1}{r} 
P_{e\mu}^\delta\right) 
+\kappa_\mu \left(\bar{P}_{\mu\mu}^\delta+\frac{1}{\bar{r}} 
\bar{P}_{e\mu}^\delta\right)\right].
\label{eq:Nmudelta1}
\ee
where 
\be
\kappa_\mu \equiv
\frac{{\bar \sigma}^{CC} \bar{ \Phi}_\mu^0}
{\sigma^{CC} \Phi_\mu^0},  ~~~~
r \equiv \frac{\Phi_\mu^0}{\Phi_e^0},  ~~~~
\bar{r} \equiv \frac{\bar{\Phi}_\mu^0}{\bar{\Phi}_e^0}.   
\nonumber
\ee
The ratios $r$ and $\bar{r}$ depend both on the neutrino energy and
zenith angle, e.g., in the range $E_\nu=(2 - 25)$ GeV and for
$\cos\theta_z = -0.8$ the ratio can be roughly parameterized as $r =
1.2 \cdot (E_\nu/ 1~ {\rm GeV})^{0.65}$.  Below 2 GeV one has $r
\approx 2$.

From Eqs.~(\ref{eq:Pemudelta}) and (\ref{eq:Pmumudelta}) we find for
neutrino contribution
\ba
d_\mu^\delta - d_\mu^{0}  & = &  
\sigma^{CC} \frac{1}{r} \Phi_\mu^0 \sin 
2\theta_{23}|A_{e\tilde{2}} A_{e\tilde{3}}|
\left[(r - 1)\cos\phi(1 - \cos\delta) - \sin\phi 
\sin\delta \right]
\nonumber\\
& + & \sigma^{CC} \Phi_\mu^0 D_{23} \left(1 - \frac{1}{\cos\delta}\right).
\label{eq:Nmudelta2}
\ea
This shows that in the case of $D_{23} = 0$ the difference
$d_\mu^\delta - d_\mu^{0}$ should vanish whenever $A_{e\tilde{2}} = 0$
or $A_{e\tilde{3}} = 0$, {\it i.e.} along the solar and atmospheric
magic lines for probabilities considered above.  The antineutrino
contribution can be written similarly.  It is suppressed in comparison
to the neutrino contribution by factor $\sim 0.2$ due to smaller
probabilities (factor of 2 at low energies, see
Fig.~\ref{fig:anti_emu_prob}) and smaller cross-section.
 
Notice that the main sensitivity to the mass hierarchy searches comes
from $P_{e \tilde{3}}$ which is screened at low energies where $r =
2$. In contrast, no screening of the CP-dependent terms occurs.  The
phase $\delta$ affects relative contribution of the $\nu_e - \nu_\mu$
and $\nu_\mu - \nu_\mu$ channels.

\begin{figure}[t]
\includegraphics[width=3in]{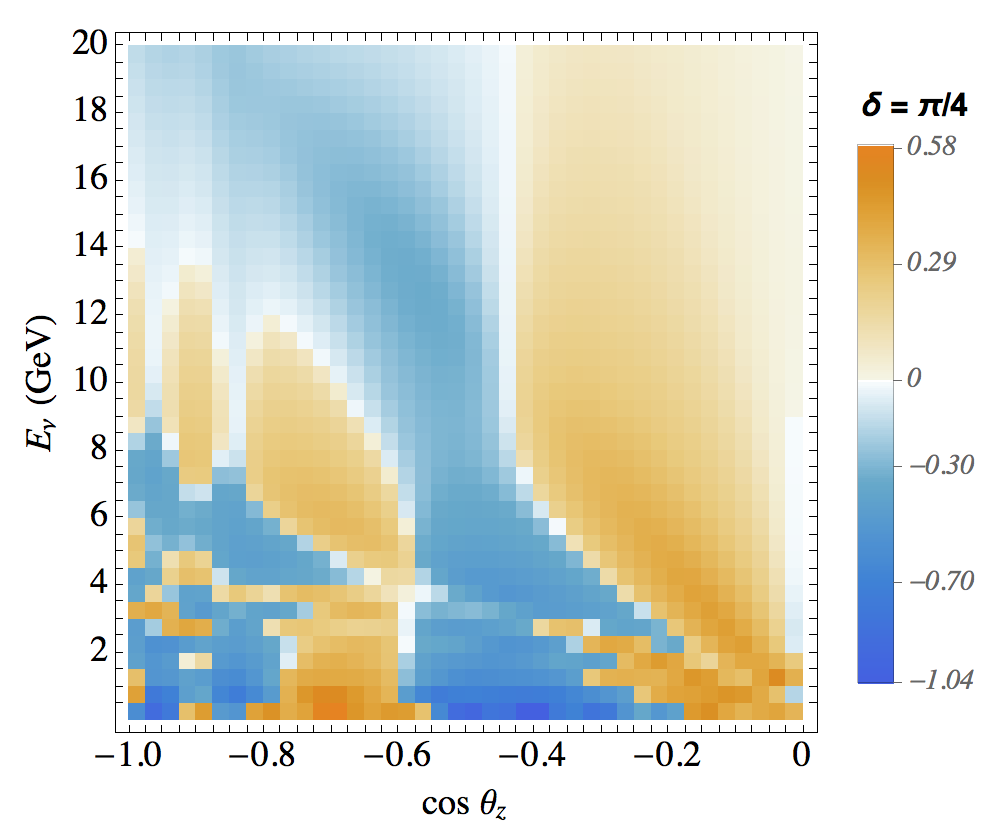} \hskip 0.25in
\includegraphics[width=3in]{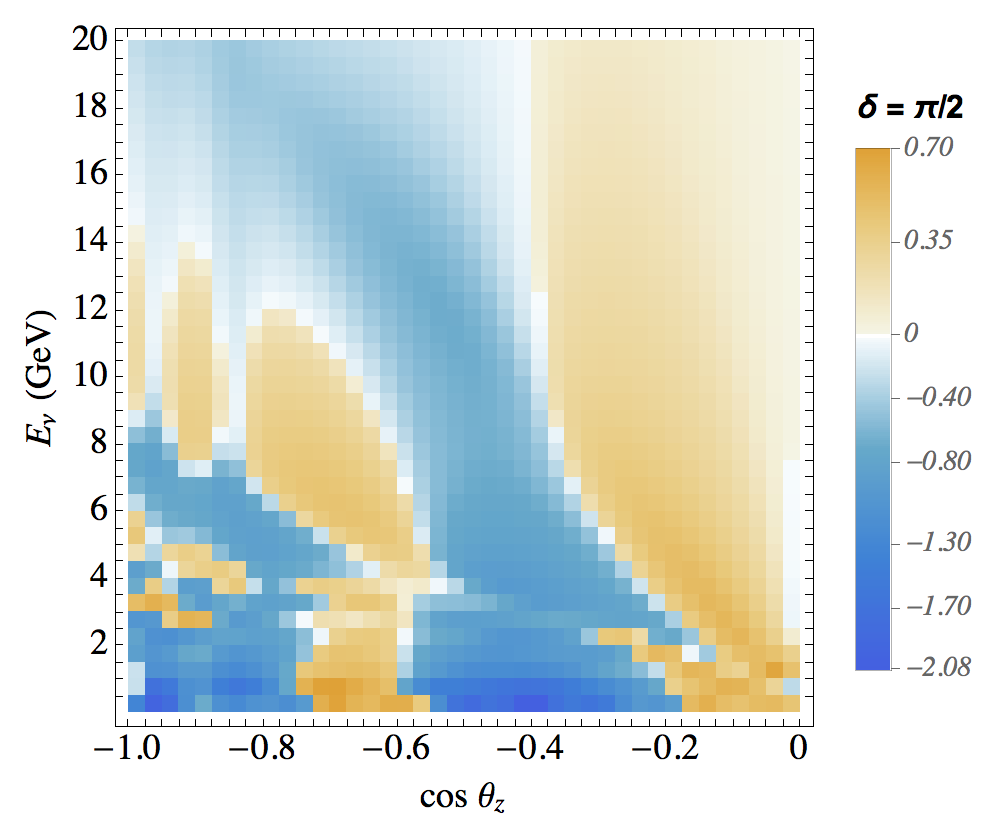}\\ \vskip 0.25in
\includegraphics[width=3in]{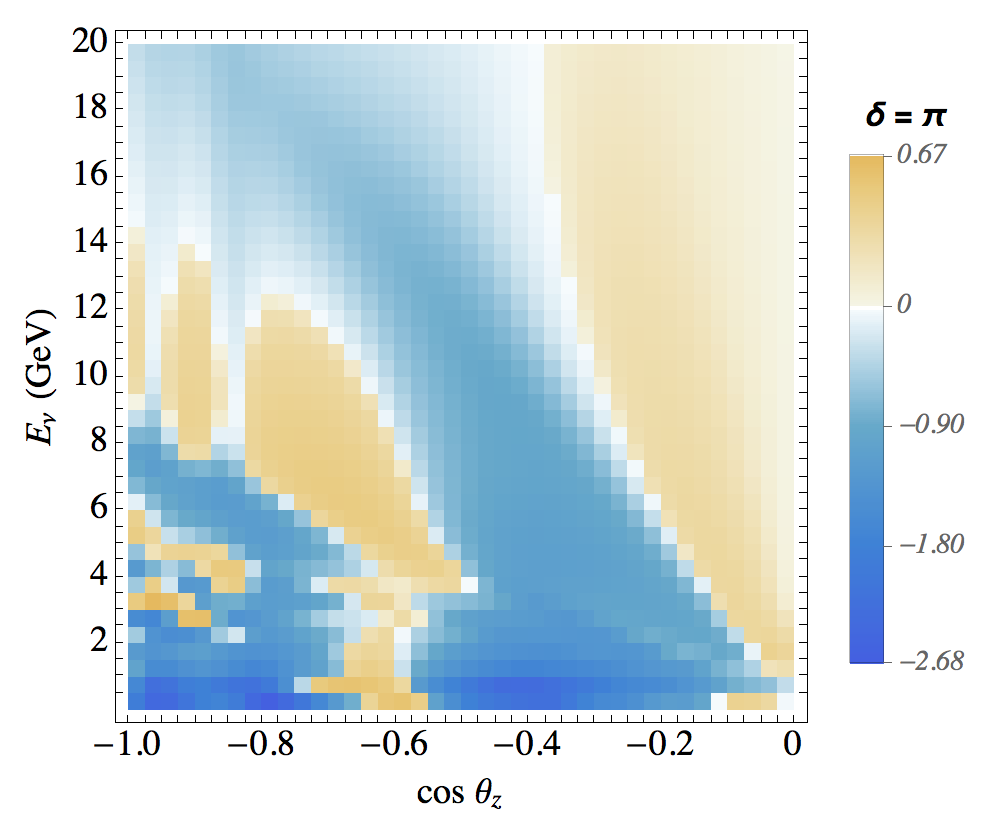} \hskip 0.25in
\includegraphics[width=3in]{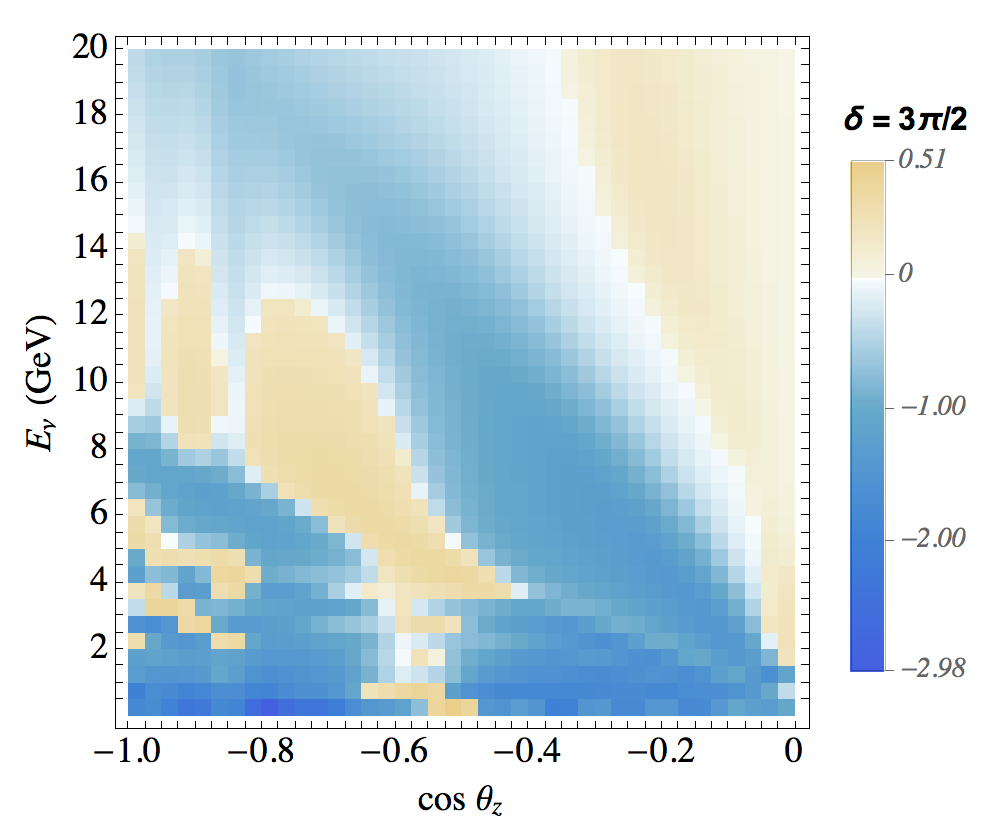}
\caption{ Distributions of the relative CP differences, $S_{ij} (f =
  0) \equiv (N_{ij}^{\delta} - N_{ij}^{0})/ \sqrt{N_{ij}(\delta =
    0)}$, for the $\nu_\mu + {\bar \nu}_\mu$ events in the
  $E_\nu-\cos\theta_z$ plane. 1-year of exposure of Super-PINGU has
  been used. The CP difference is given between $\delta=0$ and
  $\delta=\pi/4$ (top left panel), $\delta=\pi/2$ (top right panel),
  $\delta=\pi$ (bottom left panel) and $\delta=3\pi/2$ (bottom right
  panel). Note that the lowest energy bin is $(0.2-0.5)$ GeV.  Normal
  neutrino mass hierarchy is assumed.}
\label{fig:numu_NH_oscillograms1}
\end{figure}

The fine-binned, $\Delta(\cos \theta_z) = 0.025$ and $\Delta E_\nu =
0.5$ GeV, distribution of CP differences of $\nu_\mu$ events in
Super-PINGU (\ref{eq:nev}) for different values of $\delta$ are shown
in Fig.~\ref{fig:numu_NH_oscillograms1}.  Here contributions from
$\nu$ and $\bar{\nu}$ are summed up.

Let us consider dependence of the distributions of events on $\delta$
given in (\ref{eq:Nmudelta2}) which is explicit and exact.  The first
term in brackets of (\ref{eq:Nmudelta2}) as function of $\delta$ is
symmetric with respect to $\delta = \pi$, whereas the second one is
antisymmetric. The relative contributions of the two terms are
determined by the phase $\phi$.  As we will, see the first term
dominates and the second one produces shift of maximum of $S_{\sigma}$
to $\delta > \pi$.

The patterns of distributions of events are determined by the domain
structure formed by the magic lines.  Namely, the borders of domains
are inscribed in the grid of magic lines with interconnections in the
resonance regions \cite{our3}.  Non-zero value of $D_{23}$ produces
further shift of borders.

Let us consider the magic lines for the difference of densities of
events for a given $\delta$ and $\delta = 0$.  Since $A_{e\tilde{3}}$
and $A_{e\tilde{2}}$ appear as common factors (in the approximation of
$D_{23} = 0$) the solar and atmospheric magic lines are the same as
for the probabilities if neutrinos and antineutrinos are considered
separately.  The exact interference phase condition for number of
events corresponds to zero value of the terms in the brackets of
(\ref{eq:Nmudelta2}) which gives
\be
\tan \phi  \approx - \tan \phi_{31}^m  \approx
\frac{(r - 1)(1 -  \cos \delta) }{\sin \delta}.   
\label{phi-d}
\ee
At high energies the $\nu_\mu$ flux dominates ($r \gg 1$) and the
pattern of the $d_\mu$ distribution follows dependence of the
probability $P_{\mu \mu}$ on $E_\nu$ and $\cos\theta_z$, in particular
the $P_{\mu \mu}$ domain structure.  For $E_\nu \gtrsim 3$ GeV one can
clearly see three solar (vertical) magic lines at $\cos \theta_z$
presented in (\ref{solarlll}).  The interference phase condition is
given in (\ref{eq:magicE}) with $\phi(\delta)$ obtained from
(\ref{phi-d}).  The oblique lines with different values of the slope
parameter $A$ in Fig. \ref{fig:numu_NH_oscillograms1} correspond to
$\phi = \phi_{min}$, $\phi = \phi_{min} + \pi$ , $\phi = \phi_{min} +
2\pi$, {\it etc.}.  E.g., for $\delta = \pi$, we obtain from
(\ref{phi-d}) $\phi_{min} = \pi/2$, and the lines correspond to $A =
25, ~ 8.5$ and $5$ GeV.  The phase $\phi$ and consequently, the
slopes change slightly with $\delta$. In the range below 6 GeV the
pattern of distribution is also determined by the atmospheric magic
lines.  The pattern is also affected by non-zero $D_{23}$ as well as
by contribution from antineutrinos, which have shifted magic lines
with respect to the neutrino lines.

With increase of $\delta$ the domain structure does not change
qualitatively although the domains of the negative CP-difference
(blue) expand, especially the one which is aligned to the magic line
(\ref{eq:magicE}) with $A = 25$ GeV.  The values of CP-differences
increase and asymmetry between negative and positive CP differences
increases with $\delta$, as one can read from numbers at the
explanatory bars.  Maximal CP phase effect is in the lowest energy
bins.

At low energies $r \approx 2$, and the difference of the densities
equals
\be
d_\mu^\delta - d_\mu^{0} ~=  \frac{1}{2}   
\sigma^{CC} \Phi_\mu^0 \sin 2\theta_{23} 
|A_{e\tilde{2}} A_{e\tilde{3}}| 
\left[\cos \phi(1 - \cos\delta) - \sin\phi \sin\delta \right]. 
\label{eq:Nmudelta4}  
\ee    

Effect of averaging over $E_\nu$ and $\theta_z$ can be explored using
the constant density approximation.  From (\ref{pemu-av1}) we obtain
for the neutrino part (the first term in (\ref{eq:Nmudelta1}))
\be
\langle d_{\mu}^\delta \rangle - 
\langle d_{\mu}^{0} \rangle = 
- \sigma^{CC} \Phi_\mu^0  \frac{J_\theta}{2} 
\left[\left(1 - \frac{1}{r} \right) 
(1 -  \cos \delta)  \sin^2 \phi_{21}^m 
- \frac{1}{2r}
\sin \delta \sin 2 \phi_{21}^m  
\right].
\ee 
At low energies, when  $r \approx 2$,  this expression  reduces to 
\be
\langle d_\mu^{\delta}\rangle - \langle d^{0}_\mu \rangle =
- \sigma^{CC} \Phi_\mu^0 \frac{ J_\theta}{4}  \left[(1 -  \cos \delta)  
\sin^2 \phi_{21}^m  - \frac{1}{2}\sin \delta \sin 2 \phi_{21}^m  
\right], 
\label{diffmu-d}
\ee
where the additional factor $1/2$ is due to the flavor suppression.
Comparing (\ref{diffmu-d}) and (\ref{eq:Nmudelta4}) we find that
averaging is reduced to substitution in the CP-factor $\cos\phi
\rightarrow \sin^2 \phi_{21}^m$ and $\sin\phi \rightarrow 0.5 \sin
2\phi_{21}^m$.  Notice that dependence of the differences on $E_\nu$
and $\theta_z$ is in $\phi_{21}^m$, $\Phi_\mu^0$ and $r$.  Expression
in (\ref{diffmu-d}) is a combination of two functions: $\sin^2
\phi_{21}^m$ and $\sin 2 \phi_{21}^m$ with weights determined by the
phase $\delta$.  The first function is even and the second is odd in
$\delta$ and both functions vanish along the magic lines.  For several
specific values of $\delta$ we obtain (in units $\frac{1}{4}
\sigma^{CC} \Phi_\mu^0 J_\theta$)
\ba
\langle d^{\pi/4}_\mu \rangle - \langle d^{0}_\mu \rangle  
& \propto & - (1 -  \frac{1}{\sqrt{2}}) \sin^2 \phi_{21}^m 
+ \frac{1}{2\sqrt{2}} \sin 2 \phi_{21}^m , 
\nonumber\\
\langle d^{\pi/2}\rangle  - \langle d^{0}_\mu \rangle 
& \propto & - \sin^2 \phi_{21}^m +  \frac{1}{2}\sin 2 \phi_{21}^m ,   
\nonumber \\
\langle d^{\pi}\rangle - \langle d^{0}_\mu \rangle 
&  \propto  & - 2 \sin^2 \phi_{21}^m , 
\nonumber \\
\langle d^{3\pi/2} \rangle - \langle d^{0}_\mu \rangle 
& \propto  & -\sin^2 \phi_{21}^m  \frac{1}{2} \sin 2 \phi_{21}^m .  
\nonumber
\ea

Similarly according to (\ref{anpemu-avlim}) and (\ref{anpmumu-avdiff})
the difference for antineutrinos equals
\be
\langle \bar{d}_\mu^\delta \rangle  - \langle \bar{d}_\mu^0 \rangle = 
\sigma^{CC} \bar{\Phi}_\mu^0 \frac{\bar{J}_\theta}{2}  
\left[(1 -  \cos \delta)  \sin^2 \bar{\phi}_{21}^m
\left(1 - \frac{1}{\bar{r}}\right)
- \frac{1}{2 \bar{r}} 
\sin \delta \sin 2 \bar{\phi}_{21}^m \right].
\label{diffmu-d-anti}
\ee
If mixings and phases in neutrino and antineutrino channels are
approximately equal, the effect of inclusion of antineutrinos could be
accounted for by the overall suppression factor $1 - \langle \bar{P}
\rangle / \langle {P}\rangle \kappa_\mu \approx 0.8$ to that of
neutrino only without change of the shape of the distribution.
Differences of the phases and mixing angles in the neutrino and
antineutrino channels lead to distortion of the neutrino distribution
mainly in the regions around the magic lines.

The $\nu_{\tau}-$ flux appears at the detector due to the $\nu_\mu -
\nu_\tau$ oscillations. In turn, the $\nu_{\tau}$ interactions $
\nu_{\tau} + N \rightarrow \tau + h \rightarrow \mu + \nu_\tau +
\nu_\mu + h \nonumber $ will contribute to the sample of $\nu_\mu
-$events with a muon and a hadron cascade in the final state.
However, the number of these events is relatively small due to its
small branching ratio and small cross-section of the $\tau$ production
near the energy threshold.  Also these events have certain features
which can be used to discriminate them from the true $\nu_\mu -$events
\cite{ARS}.  As we will see, the highest sensitivity to $\delta$ is in
the sub-GeV region where $\tau$ leptons are not produced.  In our
simplified study the effect of the $\tau$ decays can be accounted by
adding a systematic error.

\subsection{Smearing}

\begin{figure}[t]
\includegraphics[width=3in]{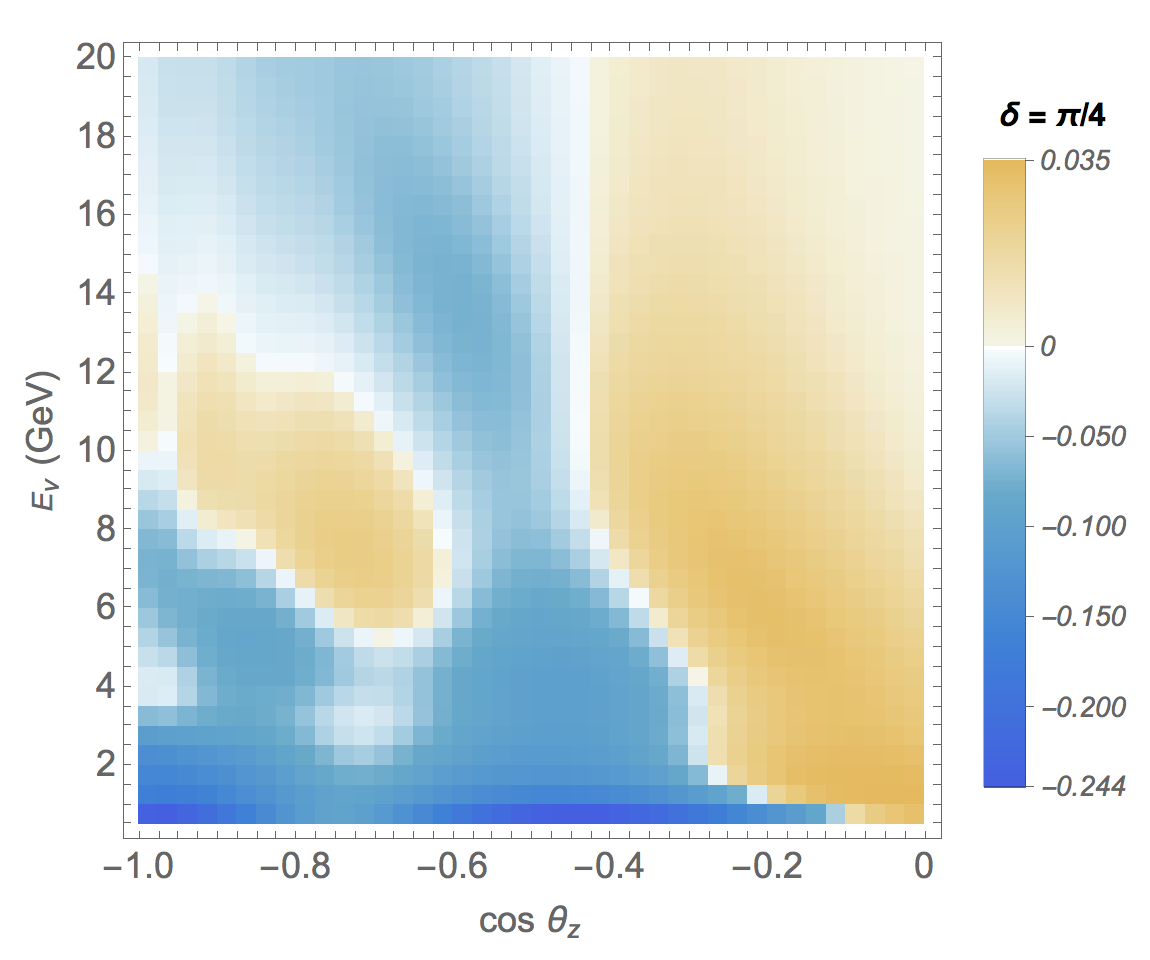} \hskip 0.25in
\includegraphics[width=3in]{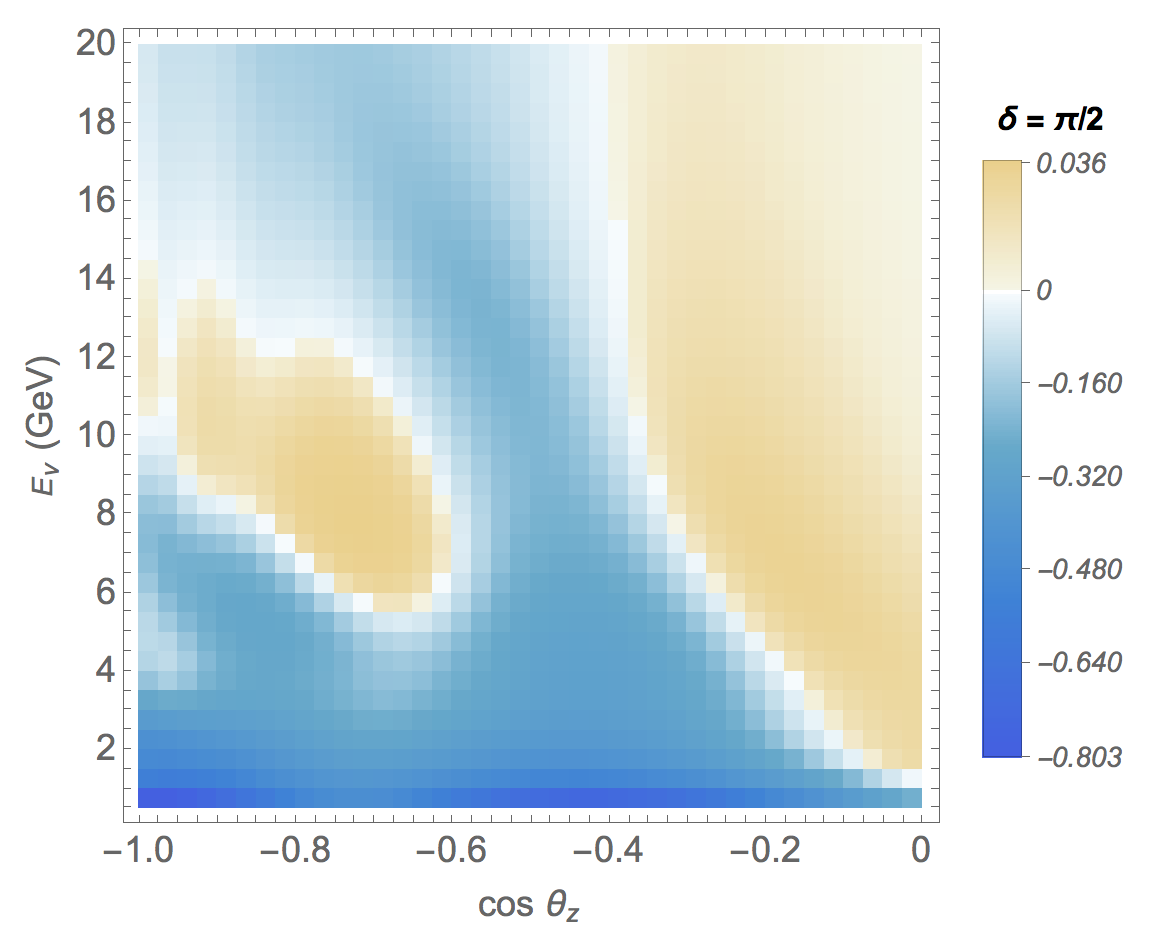}\\ \vskip 0.25in
\includegraphics[width=3in]{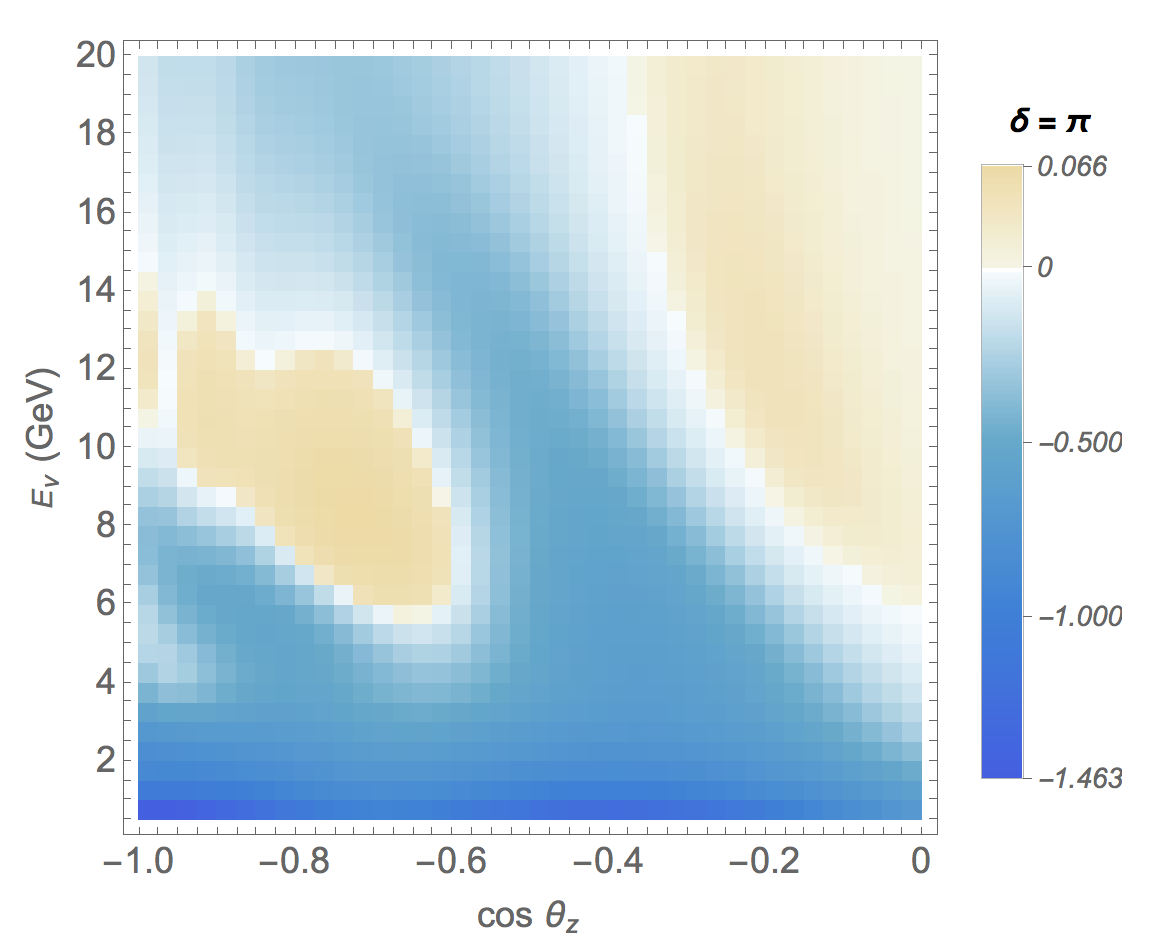} \hskip 0.25in
\includegraphics[width=3in]{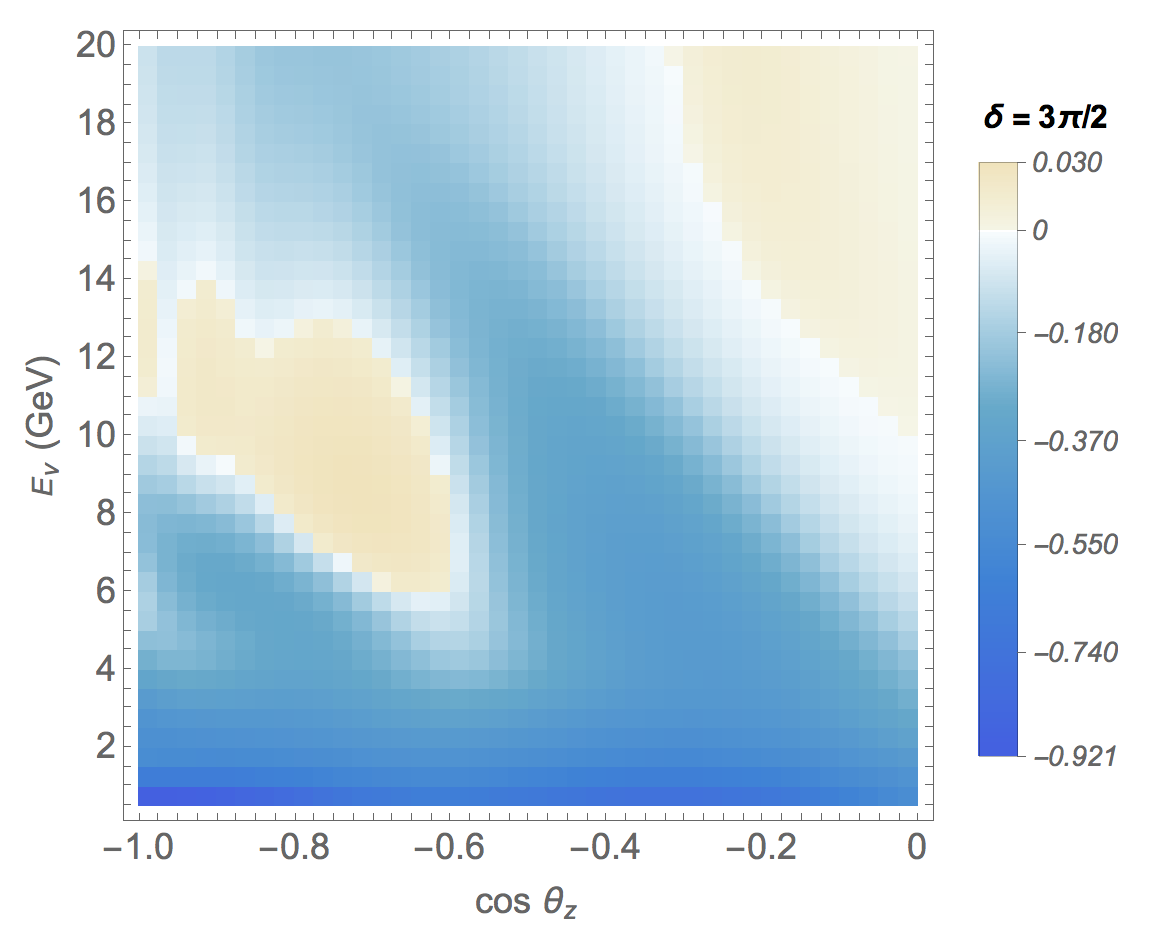}
\caption{
  The same as in Fig.~\ref{fig:numu_NH_oscillograms1}, but after
  smearing of the distributions over the energy and zenith angle of
  neutrinos. The smearing functions have been taken in the form of the
  PINGU reconstruction functions with widths reduced by factor
  $1/\sqrt{3}$.}
\label{fig:numu_NH_oscillograms2}
\end{figure}

In Fig.~\ref{fig:numu_NH_oscillograms2} we show the relative
CP-difference distributions of the $\nu_\mu$ events, $S_{ij}(f = 0)$
in (\ref{sij}), smeared with the Super-PINGU reconstruction functions
as defined in Sec. III B. Smearing leads to disappearance of fine
structures and to merging of regions of the same sign CP-differences.
There is dominant region of the negative CP-differences $S_{ij} < 0$
and two separate regions with $S_{ij} > 0$. The large one is at $|\cos
\theta_z| < 0.4$ and at high energies restricted from below by the
first interference phase line.  The smaller region is at $\cos
\theta_z < - 0.6$ and energies $E = (6 - 14)$ GeV.  Actually, it
consists of three regions situated between the solar (vertical) magic
lines and between two lines (second and third) determined by the
interference phase conditions.

With increase of $\delta$ the region with $S_{ij} < 0$ expands.
Correspondingly, the region of positive $S$ at $|\cos \theta_z| < 0.4$
shifts to higher energies, whereas the smaller region slightly
shrinks. Asymmetry between maximal positive and maximal negative
CP-differences increases.  Below 5 GeV the region of negative $S$
expands to horizontal directions (larger $\cos \theta_z$) and for
$\delta \sim (1 - 1.5)\pi$ we find $S_{ij} < 0$ for all $\cos
\theta_z$.  At low energies and the bins with the highest
CP-difference are in the zenith angle intervals $|\cos \theta_z| = 0.9
- 1.0$ and $|\cos \theta_z| < 0.3 - 0.5$.  For $\delta > \pi/2$ the
dip near vertical directions is larger than in outer regions.  With
increase of $\delta$ the latter shifts from $\cos \theta_z = - 0.43$
to $- 0.3$.  It is these features that should be used to measure
$\delta$.

Smearing leads to a substantial decrease of the sensitivity to
$\delta$. This reduction is a consequence of the integration over
regions with different values and signs of $S_{ij}$.  The decrease of
sensitivity is characterized by factors (2.3 - 1.7) for small $\delta
\sim (0.25 - 0.5)\pi$ and by factor 1.3 in the range $\delta \sim (1 -
1.5)\pi/2$, where the CP phase effect is large (see also Sec.\ VI and
Fig.~\ref{fig:significance_numu_V12}).  The reason is that for small
$\delta$ at low energies the regions with $S_{ij} > 0 $ and $S_{ij} <
0$ are comparable in size and in absolute values of $S_{ij}$ (see
Figs.~\ref{fig:numu_NH_oscillograms1} and
\ref{fig:numu_NH_oscillograms2}). So that smearing (integration over
the energy and zenith angle) leads to partial cancellation.  With
increase of $\delta$ the asymmetry between the regions with positive
and negative $S_{ij}$ increases, thus reducing the cancellation.

Smearing with PINGU reconstruction functions leads to stronger
decrease of sensitivity, mostly in the $\delta < \pi$ region (see
Sec.\ VI).

To evaluate contributions to the total distinguishability $S_{\sigma}$
from different energy regions we have computed $S_{\sigma}(E_{th})$
for 1 year of exposure and fixed $V_{eff}(E_\nu)$ using different
minimal energies of integration, $E_{th}$.  Here $f = 0$, {\it i.e.},
the systematic errors have not been included.

We find that with decrease of $E_{th}$ from 1 GeV down to 0.5 GeV
$S_{\sigma}$ increases by a factor (1.5 - 1.7) depending on the value
of $\delta$. Decrease of $E_{th}$ from 0.5 GeV down to 0.2 GeV leads
to increase of $S_{\sigma}$ by another factor (1.3 - 1.9), with
strongest increase at small values of $\delta$.  It should be noticed
that extrapolation of the results below 0.5 GeV becomes unreliable.

\subsection{Neutrinos and antineutrinos}

\begin{figure}[t]
\includegraphics[width=3in]{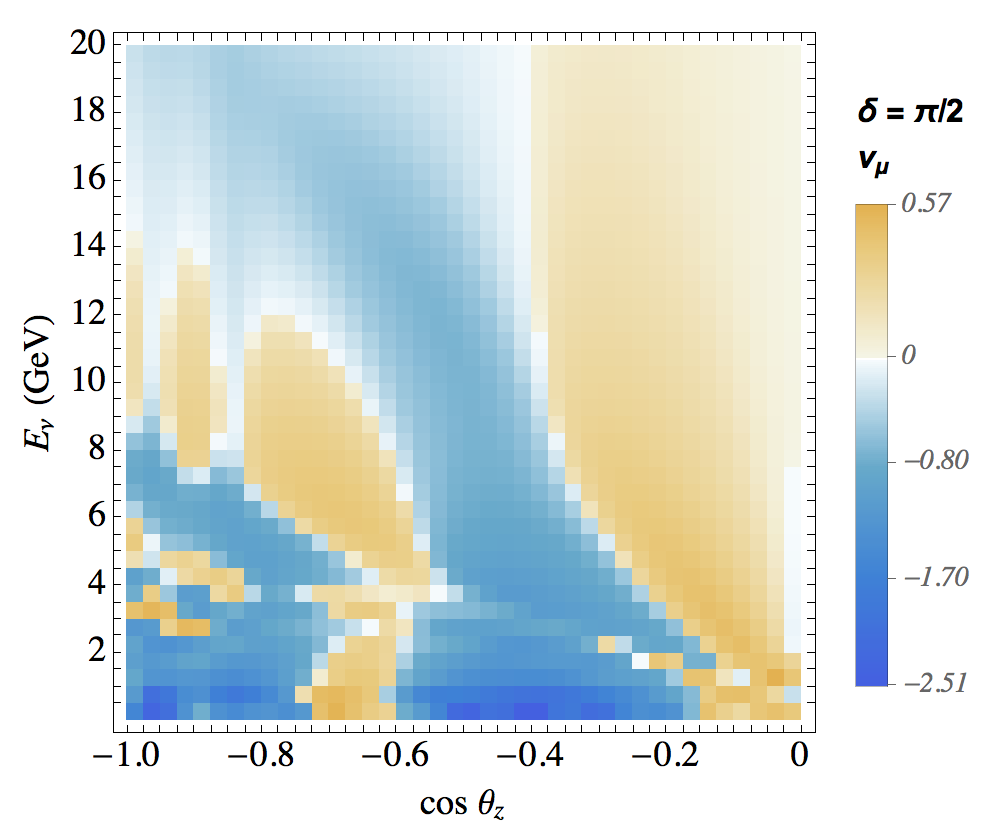} \hskip 0.25in
\includegraphics[width=3in]{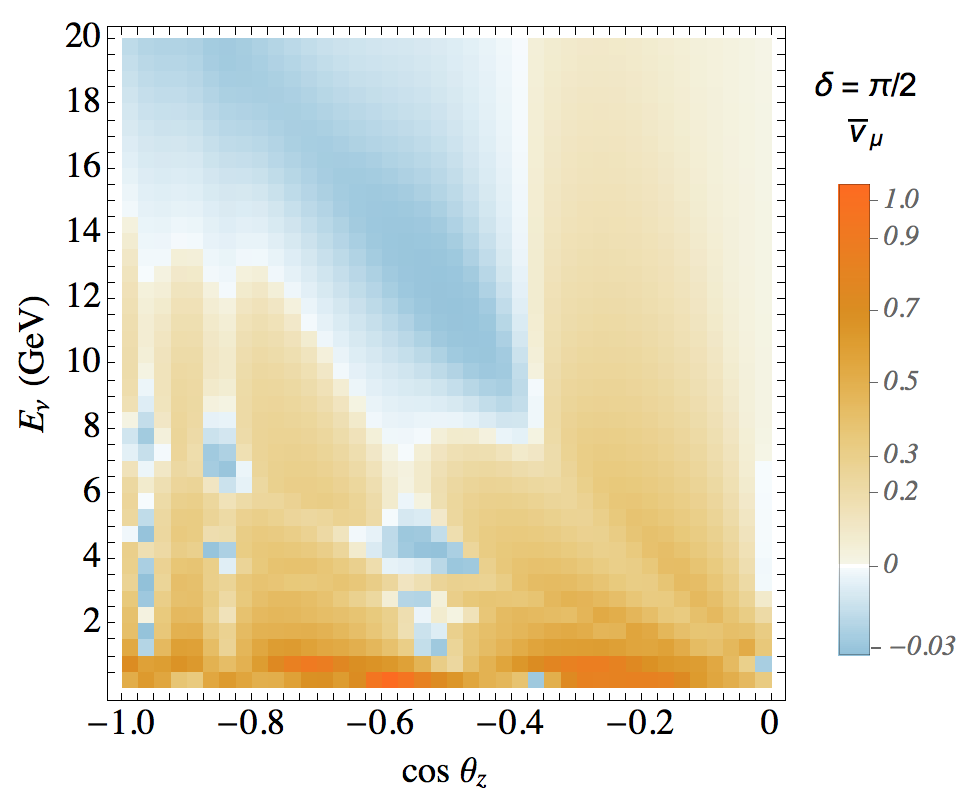}\\ \vskip 0.25in
\includegraphics[width=3in]{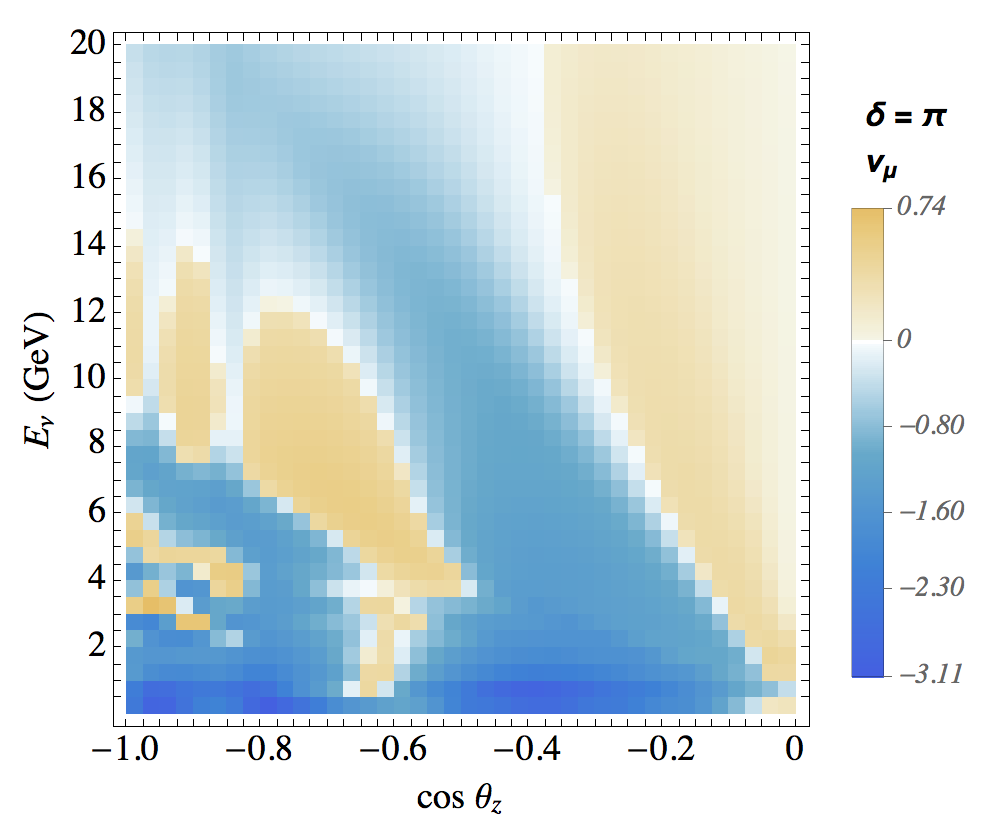} \hskip 0.25in
\includegraphics[width=3in]{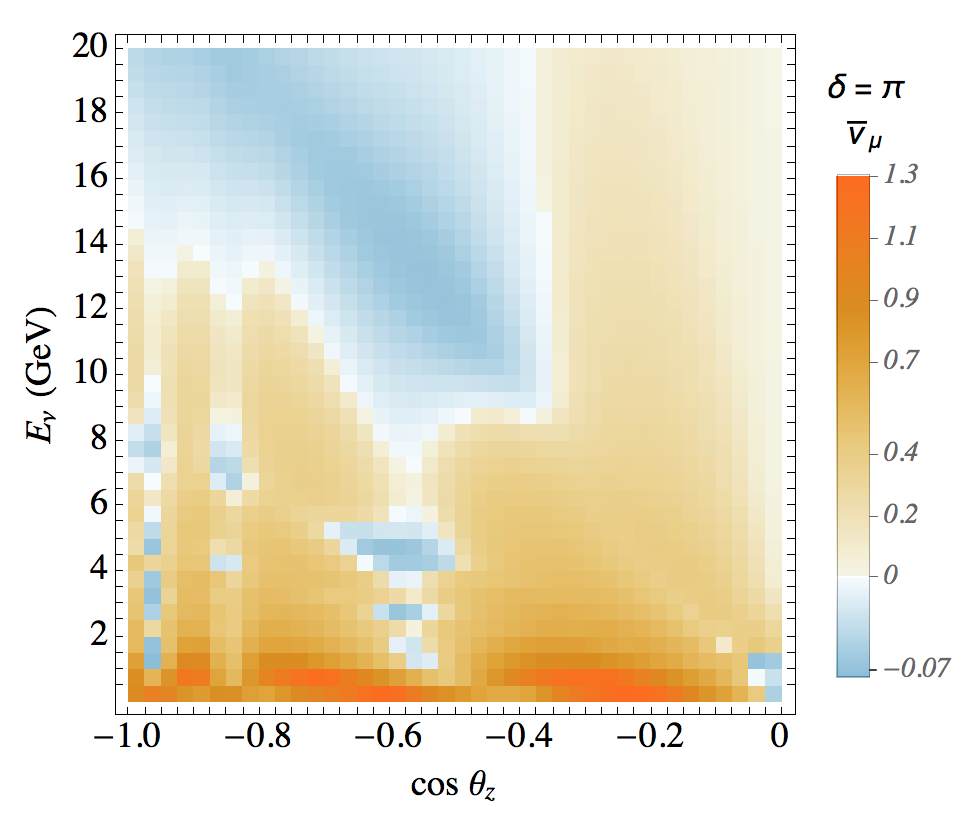}
\caption{The same as in Fig.~\ref{fig:numu_NH_oscillograms1}, but
  separately for $\nu_\mu$ and ${\bar \nu}_\mu$ events and for values
  of the CP phase $\delta=\pi/2$ (upper panels) and $\delta=\pi$
  (lower panels).}
\label{fig:numu_NH_oscillograms3}
\end{figure}

Measurements of the inelasticity ($y-$distribution) allow to make
partial separation of the neutrino and antineutrino signals on
statistical ground~\cite{ribordy}.  Since the CP-differences have
opposite signs for neutrinos and antineutrinos, at least at low
energies, one expects improvement of sensitivity to the CP-phase if
the $\nu$ and $\bar{\nu}$ signals are separated.  To assess the
possible improvement we consider first the ideal situation of complete
separation.  In Fig.~\ref{fig:numu_NH_oscillograms3} we show the
unsmeared CP-difference plots for neutrinos and antineutrinos
separately.  We take $\delta = \pi/2$ and $\delta = \pi$.

For neutrinos (resonance channel) the distribution is rather similar
to that for the sum of $\nu$ and $\bar{\nu}$ signals.  Asymmetry
between positive and negative contributions increases with exclusion
of $\bar{\nu}$.  As a result, the significance for neutrinos alone
increases by a factor $(1.3 - 1.4)$ in comparison with significance
for the sum of the signals.

In contrast, the distribution for antineutrinos shows different
pattern: it has the same sign of the CP difference, $S_{ij}$,
(positive or negative in the same regions) as for neutrinos in high
energy region, $E > 8$ GeV \footnote{This is consequence of the level
  crossing in the neutrino channel.}.  Whereas at low energies the
antineutrino and neutrino distributions have opposite signs, and so
cancellation is strong in the absence of separation.  For
antineutrinos, especially at low energies, the positive asymmetry
dominates.  This can be immediately understood on the basis of our
analytic consideration in Sec.\ II C.

The distinguishability without separation is larger than the
difference of $\nu$ and $\bar{\nu}$ distinguishabilities.  For $E_{th}
= 0.5$ GeV and $\delta = \pi$ we have $S_{\sigma, \nu} = 13.5 $,
$S_{\sigma, \bar{\nu}} = 5.8$ and $S_{\sigma, \nu} - S_{\sigma,
  \bar{\nu}} = 7.7$, which is smaller than total $S_{\sigma } = 9.9$.
This means that cancellation is not complete and reflects the fact
that in the high energy regions the sign of CP-difference is the same
for $\nu$ and $\bar{\nu}$.  There is an asymmetry between the neutrino
and antineutrino contributions related to difference of the
cross-sections and fluxes.  In the case of ideal separation we would
have $S_{\sigma} = 14.7$, instead of 9.9, {\it i.e.} almost 1.5 times
larger than without separation.  Smearing and partial separation will
reduce this enhancement factor substantially.  Separation of neutrinos
and antineutrinos, {\it i.e.}\ reconstruction of $y-$distributions, is
possible at high energies $E > 3$ GeV.  At low energies that becomes
problematic.

\subsection{Inverted mass hierarchy}
 
\begin{figure}[t]
\includegraphics[width=3in]{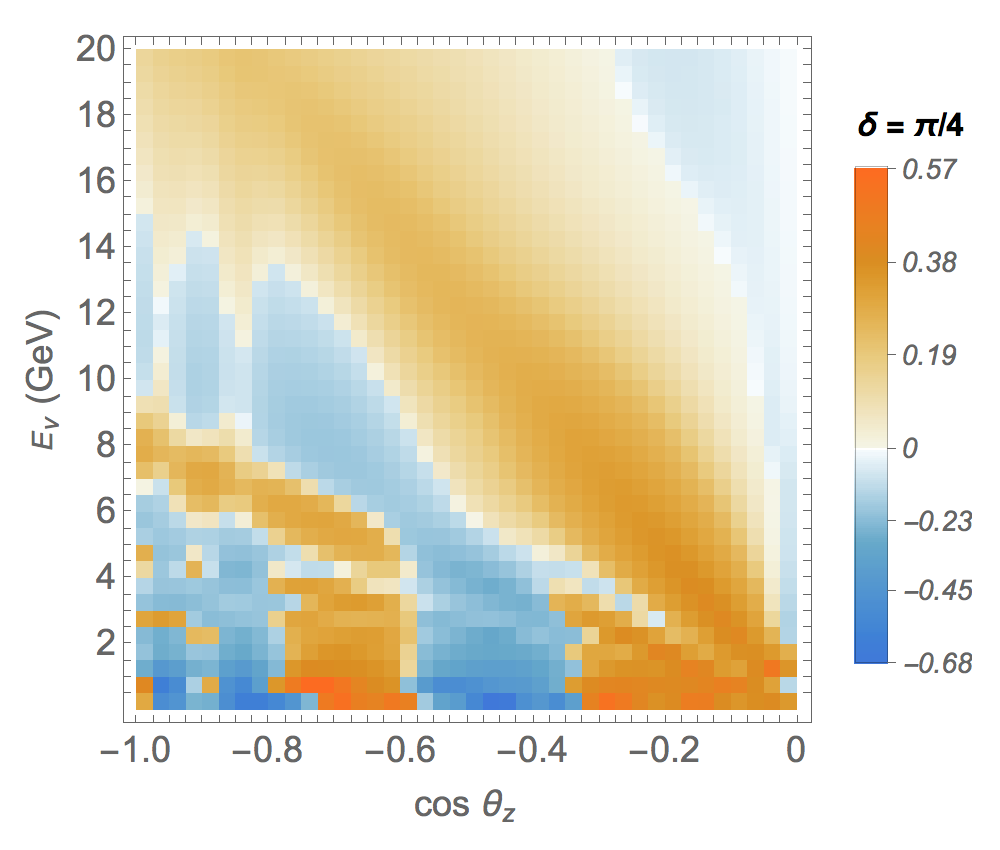} \hskip 0.25in
\includegraphics[width=3in]{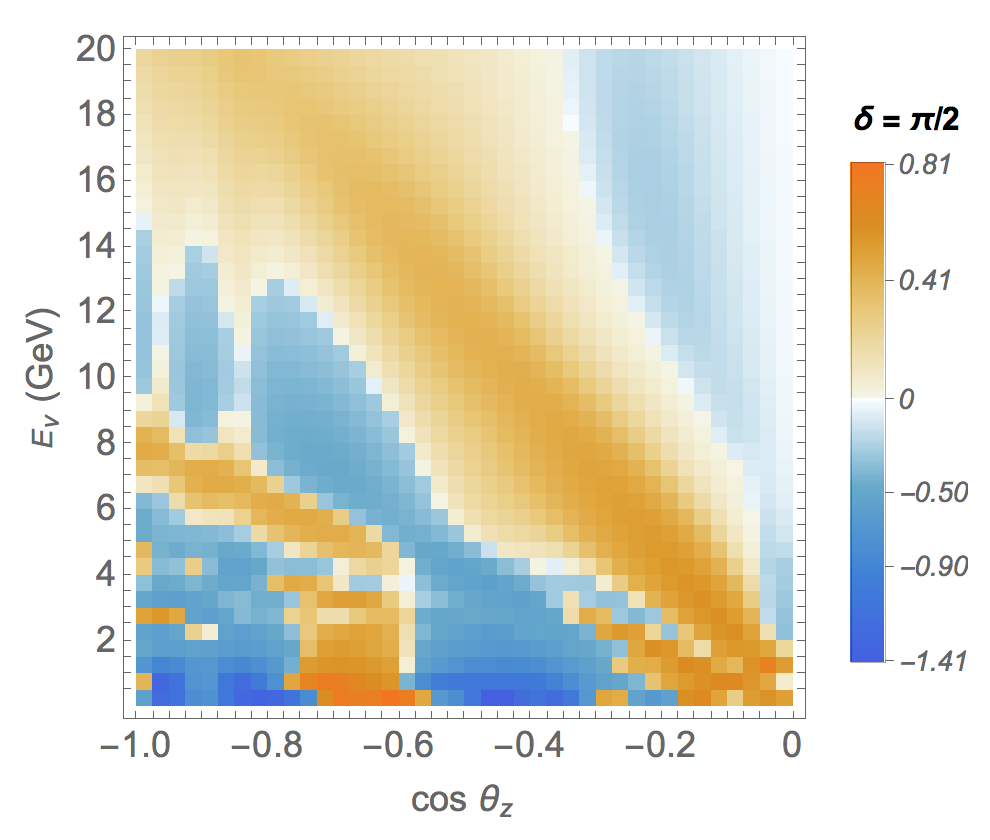}\\ \vskip 0.25in
\includegraphics[width=3in]{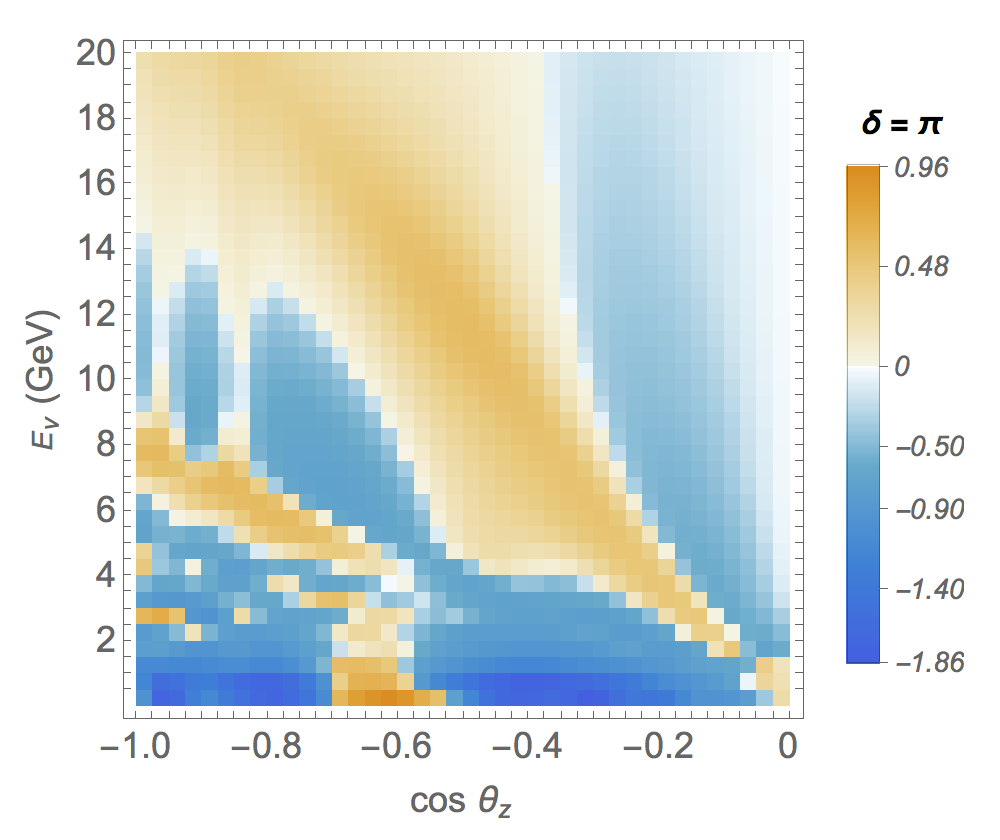} \hskip 0.25in
\includegraphics[width=3in]{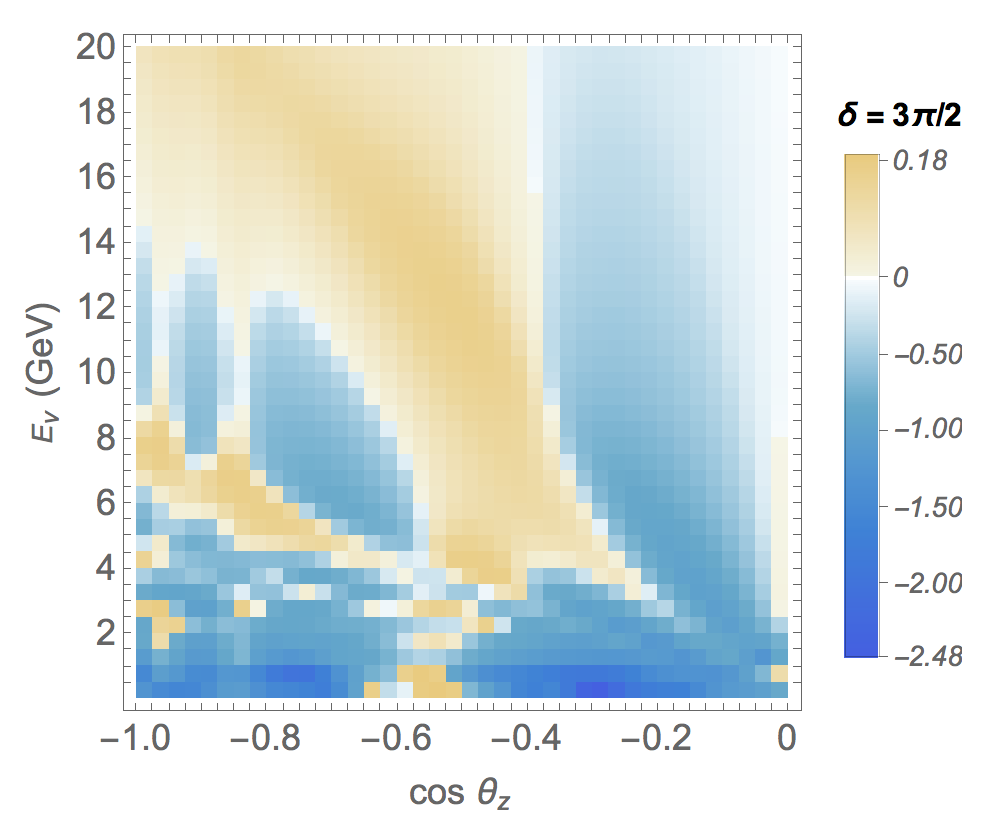}
\caption{ The same as in Fig.~\ref{fig:numu_NH_oscillograms1}, but for
  the inverted neutrino mass hierarchy.}
\label{fig:numu_IH_oscillogram}
\end{figure}

For the inverted neutrino mass hierarchy (IH) the pattern of
distributions is inverted with respect to that for the normal mass
hierarchy at high energies and it is the same for low energies, see
Fig.~\ref{fig:numu_IH_oscillogram}.  The difference is related to the
1-3 resonance whose effect is different for normal and inverted
hierarchies.  At low energies sensitivity to the mass hierarchy
disappears.

Formally all the expressions for probabilities and amplitudes in terms
of mixing angles and phases (eigenvalues) in matter are the same as in
the case of NH but values of the angles and phases change. Also the
signs of the phases $\phi_{31}^m$ and $\phi_{32}^m$ change.  Since the
1-2 mass ordering does not change, averaging over $\phi_{32}^m$ and
$\phi_{31}^m$ will give at low energies the same expression for the
probabilities with the only change $J_\theta \rightarrow
J_\theta^{IH}$, and in $J_\theta^{IH}$ only $\theta_{13}^m$ changes.
Thus, for densities of events we obtain
\be
\frac{d^{IH}}{d^{NH}} \approx \frac{J_\theta^{IH}}{J_\theta^{NH}} 
\approx \frac{\sin 2\theta_{13}^{mIH} 
\cos \theta_{13}^{mIH}}{\sin 2\theta_{13}^{mNH} \cos\theta_{13}^{mNH}}. 
\nonumber
\ee 
Furthermore, for energies much below the 1-3 resonance energy,
$\theta_{13}^{mIH} \approx \theta_{13}^{mNH} \approx \theta_{13}$.  So
that the densities of events in both cases are expected to be
approximately equal.

We find that the integral distinguishabilities of the same phases for
the inverted hierarchy is about $(25 - 30) \%$ lower.  Also the $\nu -
\bar{\nu}$ separation is more important for inverted mass hierarchy
since in this case the difference of signals from neutrinos and
antineutrinos is smaller.

\section{Cascade events}

\subsection{Density of events}

The cascade events are produced by the CC $\nu_e$ interactions $\nu_e
+ N \rightarrow e + X$, $\bar{\nu}_e + N \rightarrow e^+ + X$ and
several other processes (see discussion below).  The density of the CC
$\nu_e$ events is given by
\be
d_e (E, \cos\theta_z) =  
\sigma^{CC} \Phi_e^0
\left[ \left(P_{ee}  + r P_{\mu e}\right)
 +
\kappa_e \left({\bar P}_{ee} + \bar{r} {\bar P}_{\mu e}\right)
\right],  
\nonumber
\ee
where $\kappa_e \equiv (\bar{\sigma}_e /\sigma_e) (\bar{\Phi}_\mu^0 /
\Phi_\mu^0)$.
Its $\delta$-dependent part,
\be
d_e^\delta \equiv \sigma^{CC} \Phi_\mu^0 
\left[P_{\mu e}^\delta 
+\kappa_e \bar{P}_{\mu e}^\delta \right]
\approx 
\sigma^{CC}
\Phi_\mu^0  \left[|A_{e \tilde{3}} A_{e \tilde{2}}| \cos(\phi - \delta)\, 
+ 
\kappa  |\bar{A}_{e \tilde{3}}\bar{A}_{e \tilde{2}}| \cos(\bar{\phi} + \delta)\
\right],  
\label{eq:Nedelta1}
\ee
is determined by the $\nu_\mu \rightarrow \nu_e$ oscillation
probability only, since $P_{ee}$ and $\bar{P}_{ee}$ are
$\delta$-independent. Consequently, there is no flavor suppression of
the CP-violation effects even for low energies.

The difference of densities of the $\nu_e$-events equals
\be
d_e^\delta - d_e^{0}  \approx  - \sigma^{CC} 
\Phi_\mu^0  \sin 2\theta_{23} |A_{e \tilde{3}} A_{e \tilde{2}}|
\left[\cos\phi(1 - \cos\delta) - \sin\phi \sin\delta \right].  
\label{eq:Nedelta2}
\ee
As we discussed before, the antineutrino contribution is suppressed by
a factor 1/4.  Comparing expression (\ref{eq:Nedelta2}) with
(\ref{eq:Nmudelta2}) and (\ref{eq:Nmudelta4}) we find that for $r
\approx 2$ and $D_{23} \approx 0$,
\be
d_e^\delta - d_e^{0} = - 2 (d_\mu^\delta - d_\mu^{0}).  
\label{fact2}
\ee
So, the CP difference of the $\nu_e$ events has an opposite sign with
respect to the CP difference of the $\nu_\mu$ events and its size is
two times larger.  As a result, the cascade events can give even
bigger contribution to distinguishability of different values of
$\delta$ than the $\nu_\mu$ events.  The reason is the flavor
suppression of CP-differences for the $\nu_\mu$ events, which is
absent for the $\nu_e$ events.  With increase of energy the ratio $r$
increases, the flavor suppression becomes weaker. Consequently,
$(d_\mu^\delta - d_\mu^{0})$ increases and the numerical factor in
equation (\ref{fact2}) becomes smaller, approaching 1.  This increase
depends on the value of $\delta$.

Explicit expression for the density of events averaged over
$\phi_{32}^m$ (which is valid at low energies) can be obtained using
the constant density approximation.  Since $P_{\mu e} = P_{e \mu}
(\delta \rightarrow - \delta)$ we have from (\ref{pemu-av})
\be
\langle {P}_{\mu e}^\delta \rangle = \frac{J_\theta}{2} 
\left[\cos \delta \cos 2 \theta_{12}^m \sin^2 \phi_{21}^m
-  \frac{1}{2}\sin \delta \sin 2 \phi_{21}^m  \right].
\label{pmue-av}
\ee
Correspondingly, when $\cos 2 \theta_{12}^m \approx - 1$, the
difference of probabilities for a given $\delta$ and $\delta = 0$
equals
\be
\langle P_{\mu e}^\delta \rangle - \langle P_{\mu e}^{0} \rangle = 
\frac{J_\theta}{2}  
\left[(1 -  \cos \delta) \sin^2 \phi_{21}^m 
- \frac{1}{2}\sin \delta \sin 2 \phi_{21}^m \right].
\label{pmue-av1}
\ee
For antineutrinos such a difference has similar expression but with
overall minus sign and mixing angles and phases in matter taken for
antineutrinos.

For difference of the densities of events we obtain
\ba
\langle d_e^{\delta} \rangle  - \langle d_e^{0}  \rangle &  = & 
\sigma^{CC} \Phi_\mu^0 \frac{J_\theta}{2}  
\left[(1 -  \cos \delta)  \sin^2 \phi_{21}^m
 -  \frac{1}{2}\sin \delta \sin 2 \phi_{21}^m \right] 
\nonumber\\ 
& - & \bar{\sigma}^{CC}\bar{\Phi}_\mu^0 \frac{\bar{J}_\theta}{2}  
\left[(1 -  \cos \delta)  \sin^2 \bar{\phi}_{21}^m
 - \frac{1}{2}\sin \delta \sin 2 \bar{\phi}_{21}^m \right]. 
\ea
In the range where the phases of neutrinos and antineutrinos  
are approximately equal  we have 
\be
\langle d_e^{\delta}\rangle  - \langle d_e^{0} \rangle \approx 
\sigma^{CC} \Phi_\mu^0 \frac{1}{2} J_\theta 
\left(1 - \kappa \frac{\bar{J}_\theta}{J_\theta}\right)
\left[(1 -  \cos \delta)  \sin^2 \phi_{21}^m
 - \frac{1}{2}\sin \delta \sin 2 \phi_{21}^m \right],     
\label{diffe-d}
\ee
with the factor in the first brackets describing the C-suppression.

\begin{figure}[t]
\includegraphics[width=3in]{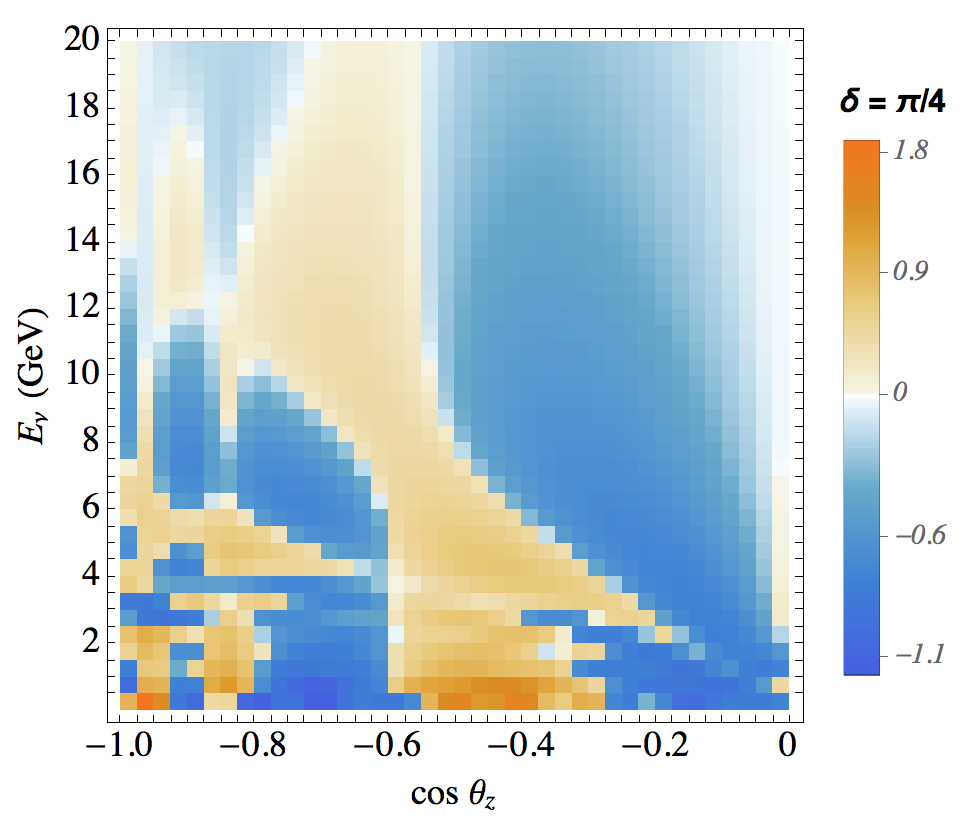} \hskip 0.25in
\includegraphics[width=3in]{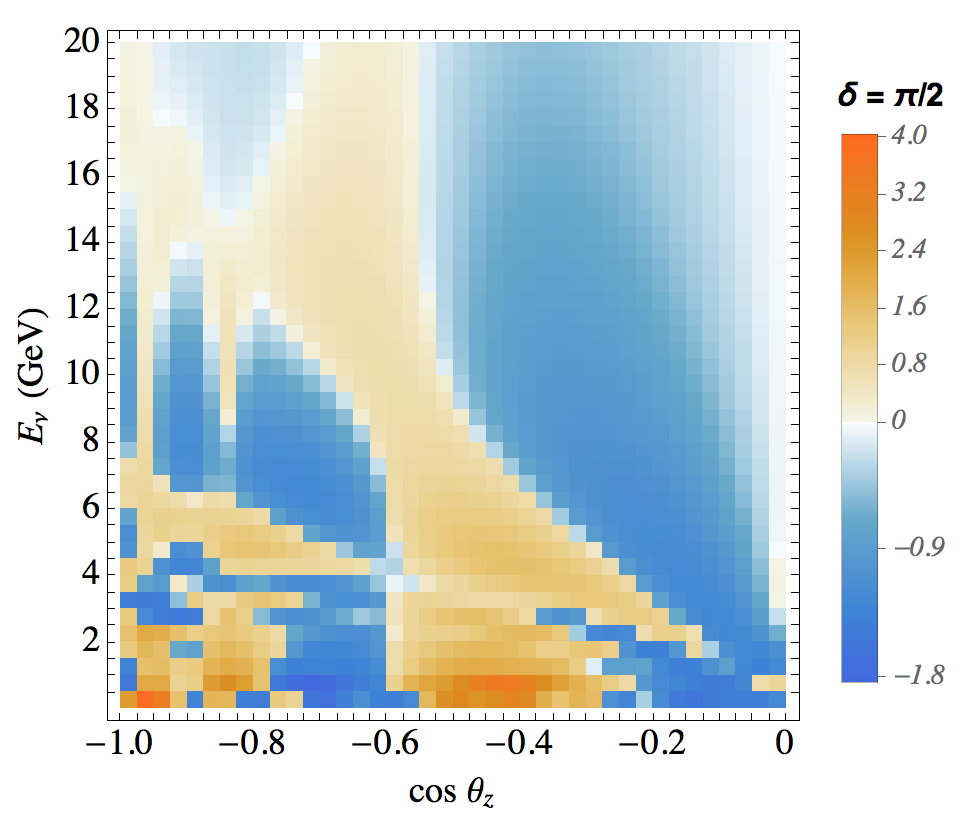}\\ \vskip 0.25in
\includegraphics[width=3in]{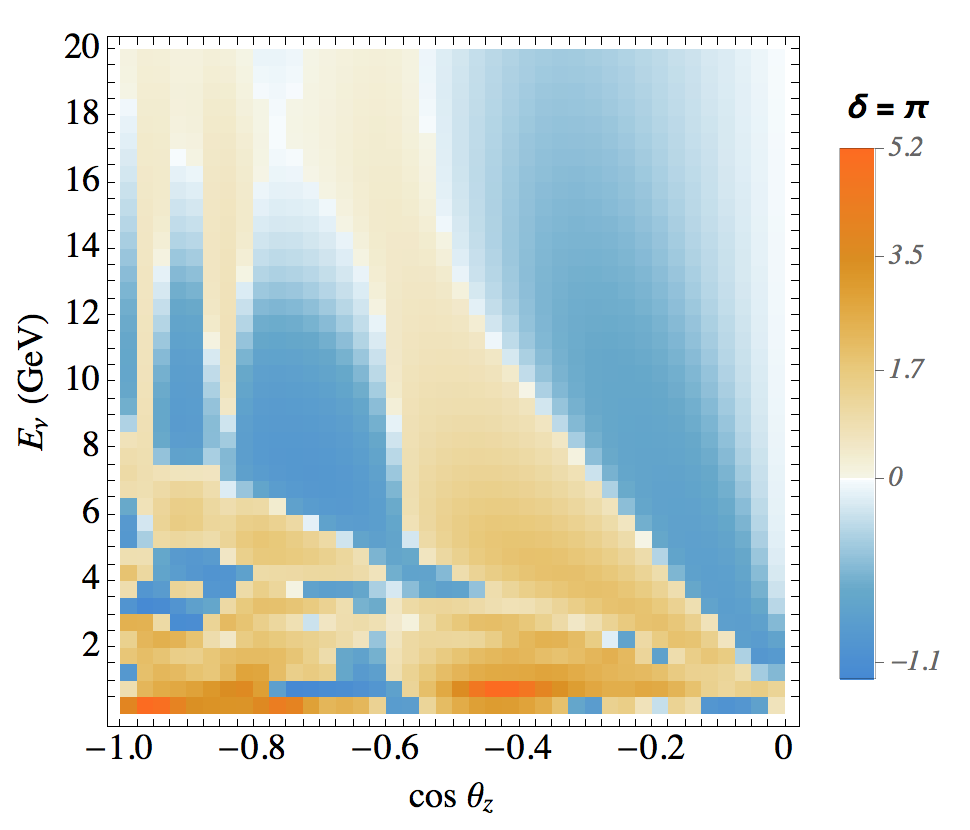} \hskip 0.25in
\includegraphics[width=3in]{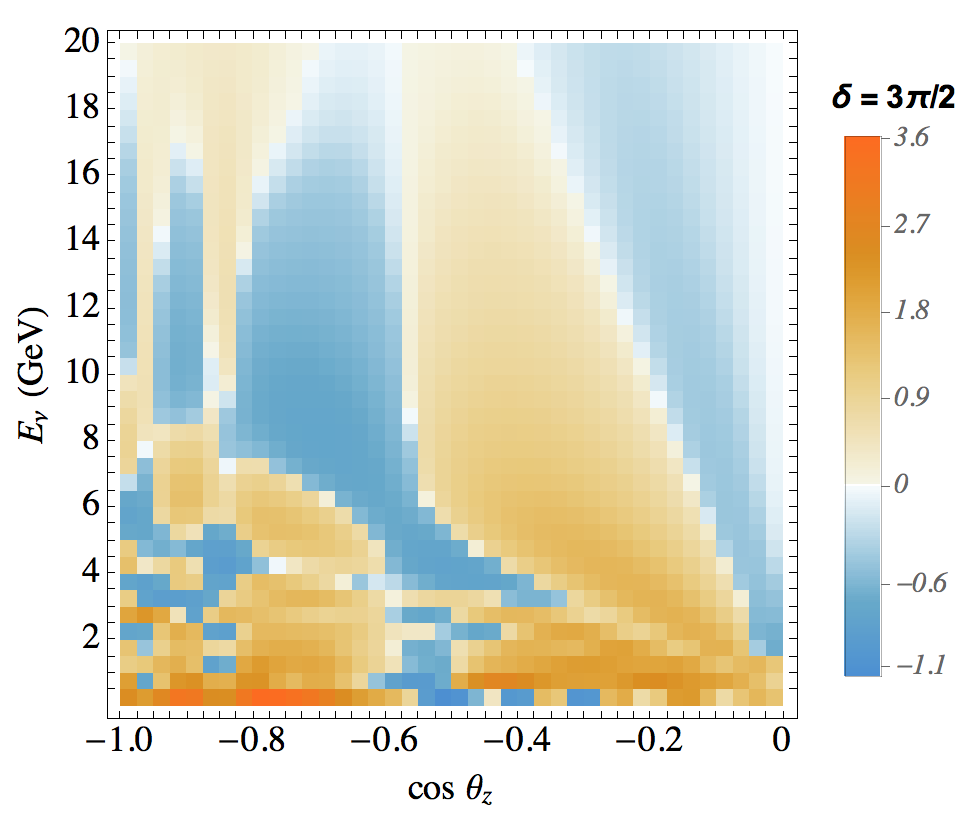}
\caption{ The same as in Fig.~\ref{fig:numu_NH_oscillograms1}, but for
  the $\nu_e + {\bar \nu}_e$ events.}
\label{fig:nue_NH_oscillogram}
\end{figure}

In Fig.~\ref{fig:nue_NH_oscillogram} we show the unsmeared
distribution of the CP differences of $\nu_e$ events $S_{ij}(f = 0)$
for different values of $\delta$.  As we marked, the transition
probability $\nu_\mu \rightarrow \nu_e$ gives unique contribution to
the CP difference, the $\nu_e \rightarrow \nu_e$ contribution and
$D_{23}$ are absent.  As a result, the distributions follow closely
the domain structure of $\nu_\mu \rightarrow \nu_e$ probability
determined by the magic lines.  Now the interference phase lines
depend on $\delta$.  From (\ref{eq:Nedelta2}) we obtain the condition
for the phase, i.e., zero value of the expression in the brackets of
eq (\ref{eq:Nedelta2}), as:
\be
\tan \phi  \approx - \tan \phi_{31}   \approx
\frac{1 -  \cos \delta }{\sin \delta}. 
\label{phi-dem}
\ee
This condition does not depend on $r$ in contrast to (\ref{phi-d}),
since only one transition probability enters.  For $\delta = \pi/2$ we
obtain $\tan \phi = 1$ or $\phi = \pi/4 + n\pi$.  Now the pattern of
distributions changes with $\delta$, since the interference phase and
CP phase dependencies do not factor out.

One can see in Fig.~\ref{fig:nue_NH_oscillogram} three solar magic
lines (\ref{solarlll}).  The oblique structures are determined at high
energies by the interference phase condition (\ref{eq:magicE}) with
$\phi$ determined in (\ref{phi-dem}).  Pattern of the distribution is
especially simple for $\delta = \pi$ when the equation (\ref{phi-dem})
gives $\phi_{31} = \pi/2 + \pi n$ which coincides with the phase
condition for probability.  The oblique lines with the slope factor $A
= 25, ~ 8.2$ and 5 GeV correspond to $\phi_{31} = \pi/2, ~ 3\pi/2, ~
5\pi/2$.  With increase of $\delta$ the slopes of interference phase
lines given by $A$ increase and so domains shift to higher energies.
The upper right domain (high energies and small $|\cos\theta_{z}|$) is
determined by the solar magic line and the first interference phase
line, {\it etc.}.  At energies below the 1-3 resonance also the
atmospheric magic lines determine the structure.

For the $\nu_e$ events, the pattern (regions of positive and negative
CP-difference) is inverted in comparison to the pattern for the
$\nu_\mu$ events in the whole energy range. So, good separation of the
$\nu_e$ and $\nu_\mu$ events, {\it i.e.} flavor identification, is
crucial (see Sec.\ VI C).

For the $\nu_e-$ events the positive CP-difference dominates.  The
total CP-distinguishability from $\nu_e-$ events is higher than from
$\nu_\mu$, e.g., $S_{\sigma}(\nu_e)$ is a factor of (1.5 - 1.8) bigger
than $S_{\sigma}(\nu_\mu)$ for $E_{th} = 0.5$ GeV, with the biggest
difference at small values of $\delta$.  Decrease of the threshold
from 1 GeV down to 0.2 enhances distinguishability by a factor of (1.3
- 1.5).

The problem here is that the cascade events are not only due to
$\nu_e$ interactions but also due to other processes which should be
taken into account:

\begin{enumerate}[(i)]

\item Neutral current (NC) interactions of all types of neutrinos.
  These events are not affected by oscillations, and so do not
  contribute to the CP difference of the events.  Still they increase
  the total number of events and therefore the statistical error in
  the denominator of $S_\sigma$ thus, diluting the significance.  The
  NC contribution could be disentangled, if the hadron cascades are
  distinguished from the EM cascades induced by electrons.  Notice
  that at low energies we deal with just few (1 - 2) individual pions
  and they can be distinguished from electrons.

\item The $\nu_\tau$ CC interactions which produce $\tau$ leptons.
  The latter generate cascades in all the decays of $\tau$ leptons but
  $\mu$.  At low energies contribution of these events is suppressed
  due to high threshold of the $\tau$ lepton production.

\item Contribution of the CC $\nu_{\mu}$ events with faint muons
  (close to threshold of Cherenkov radiation).  Fraction of these
  events is higher at low energies.  A part of the CC $\nu_e$ events
  can be confused with the $\nu_\mu$ events when one of the pions will
  be misidentified with muon.  This problem may be cured at least
  partly by introduction of additional kinematical cuts.

\end{enumerate}

\subsection{Smearing of the cascade events}

\begin{figure}[t]
\includegraphics[width=3in]{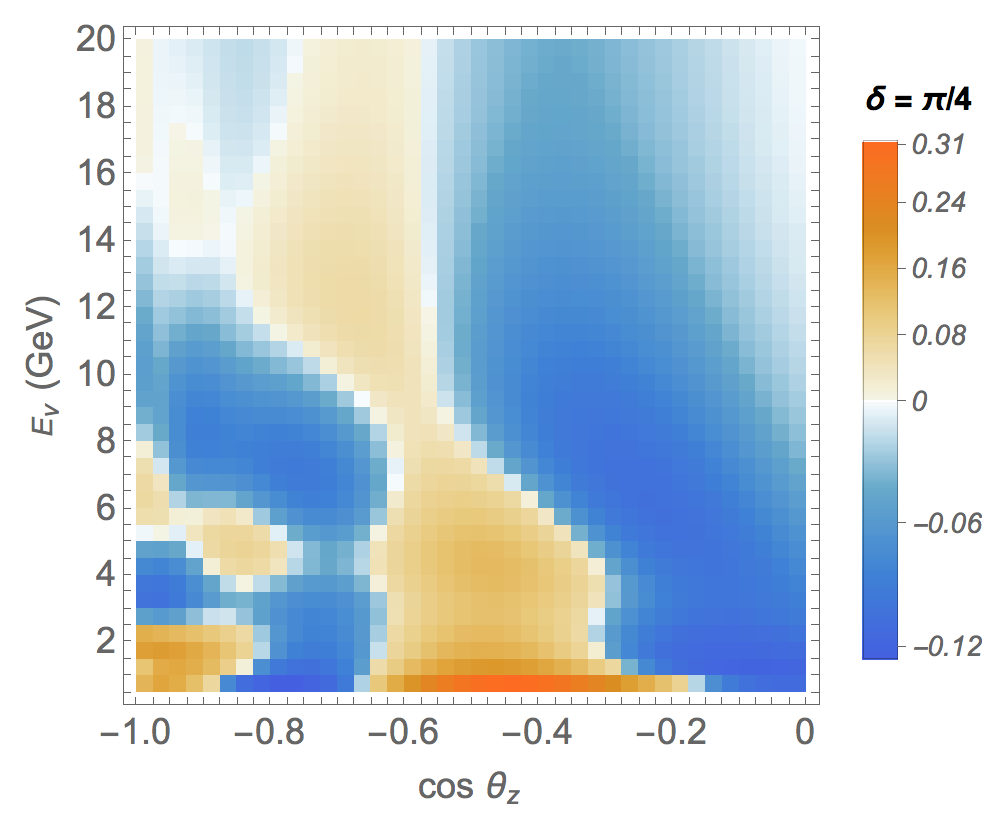} \hskip 0.25in
\includegraphics[width=3in]{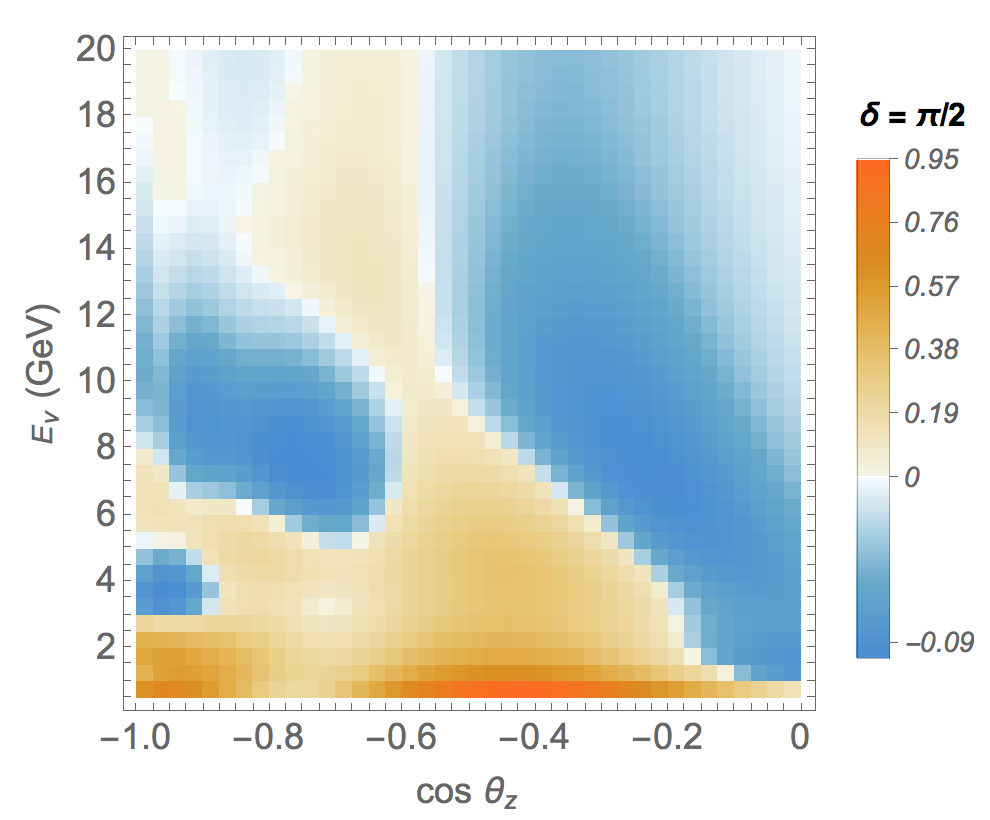}\\ \vskip 0.25in
\includegraphics[width=3in]{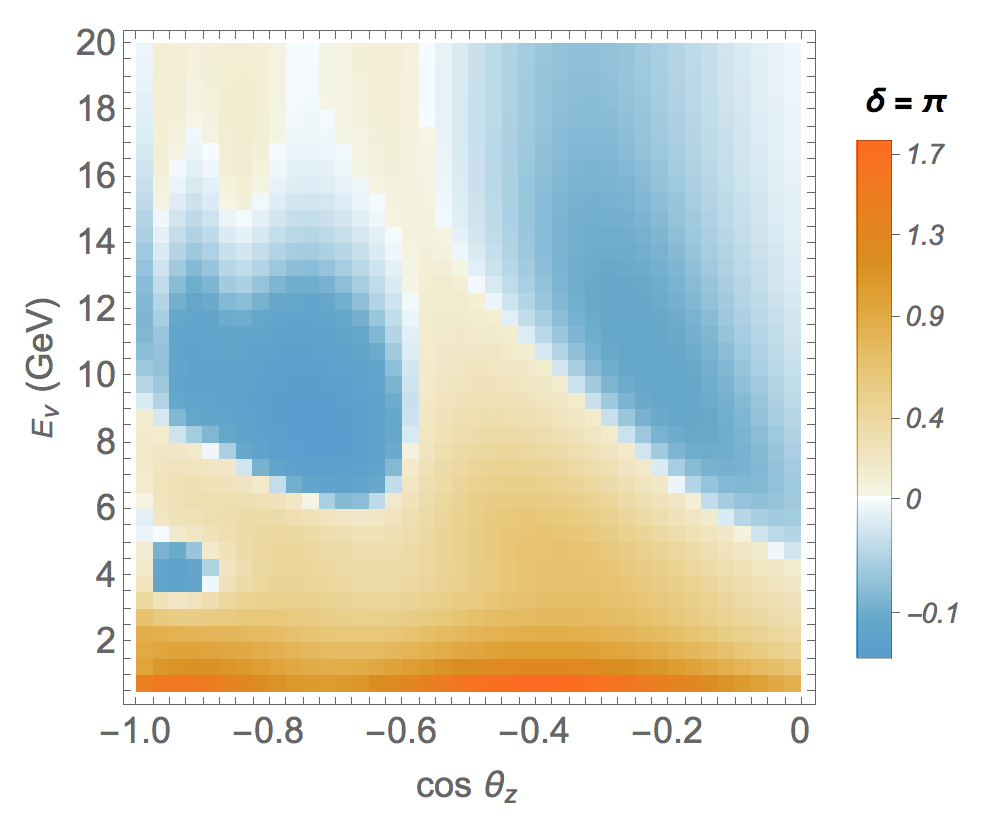} \hskip 0.25in
\includegraphics[width=3in]{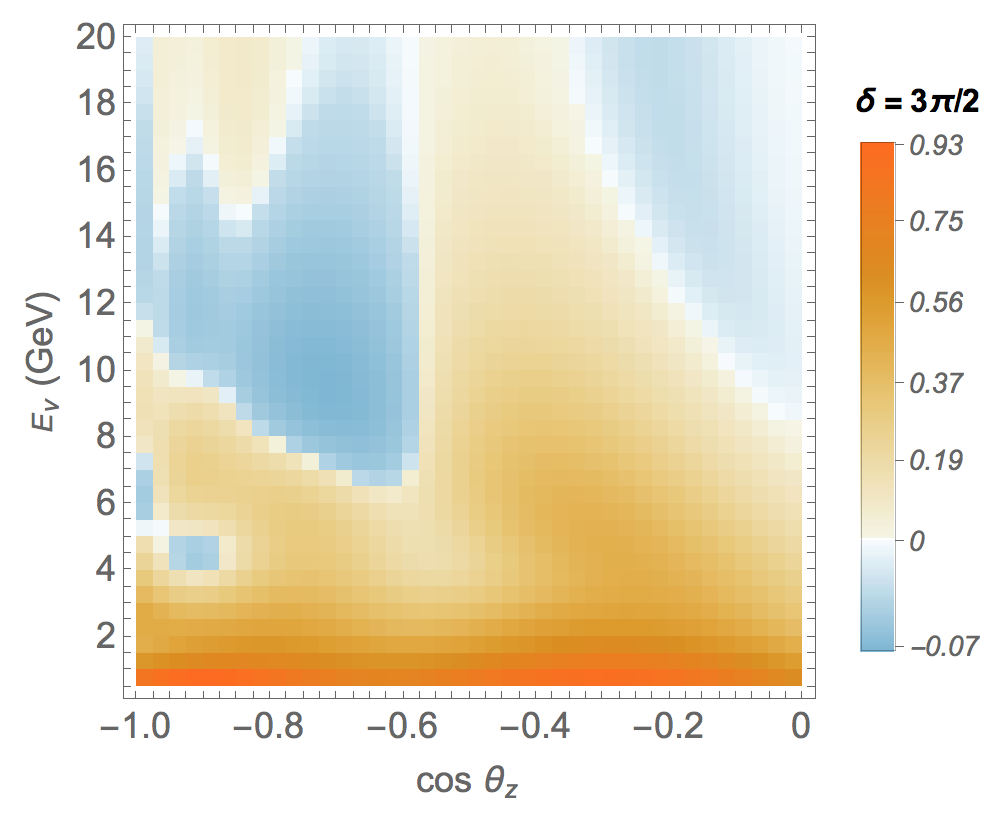}
\caption{Same as in Fig.~\ref{fig:numu_NH_oscillograms2}, but for the
  $\nu_e + {\bar \nu}_e$ events.}
\label{fig:numu_NH_oscillograms_e}
\end{figure}

In Fig.~\ref{fig:numu_NH_oscillograms_e} we show results of smearing
of cascades with the energy and angle reconstruction functions.  We
used the Super-PINGU reconstruction functions defined in Sec.\ III B.
Even after smearing one can see the CP-domain structure determined by
the magic lines.

According to Fig.~\ref{fig:numu_NH_oscillograms_e} the effect of
smearing on distinguishability of the $\nu_e$ events is stronger than
that of the $\nu_\mu$ events.  The suppression is a factor (2.4 - 3.5)
times in the interval from $\delta = (0.5 - 0.25)\pi$ and it is weaker
for large phases: a factor (1.4 - 1.5) in the interval $\delta = (1.5
- 1)\pi$. The pattern of the $S_{ij}$ distribution is inverse to that
for $\nu_\mu$ events in a sense that $(E_\nu - \theta_z)$ regions with
$S_{ij} < 0$ for $\nu_\mu$ become regions with $S_{ij} > 0$ for
$\nu_e$. Below 5 GeV with increase of $\delta$ the region with $S_{ij}
> 0$ expands towards the horizontal direction and for $\delta > \pi$
we find that $S_{ij} > 0$ for all values of zenith angle.  For fixed
energy, two peak in the dependence of $S_{ij}$ on $\cos \theta_z$ are
at $\cos \theta_z \sim - 0.9$ and at $|\cos \theta_z| \sim (0.3 -
0.45)$.

At low energies, due to loss of angular resolution structures become
essentially horizontal.  Asymmetry between the positive and negative
CP-differences increases with $\delta$ and $S_{ij} > 0$ becomes
dominant for $\delta \gtrsim \pi$. In spite of stronger smearing, the
distinguishability of $\nu_e$ events is a factor (1.3 - 1.7) times
larger (depending on value of $\delta$ and $E_{th}$) than the
distinguishability of $\nu_\mu$ events.

\section{Sensitivity to the CP-phase}

In this section we will present the total distinguishability of
different values of the CP-phase and discuss how it can be affected by
various systematic errors as well as flavor misidentification of
events.  This allows us to identify the main challenges of
determination of $\delta$ and evaluate the level of admissible errors.
We will comment on possible ways to reduce the errors and improve
sensitivity to the CP-phase.

\subsection{Total Distinguishability and the energy threshold}

\begin{figure}[t]
\includegraphics[width=4.5in]{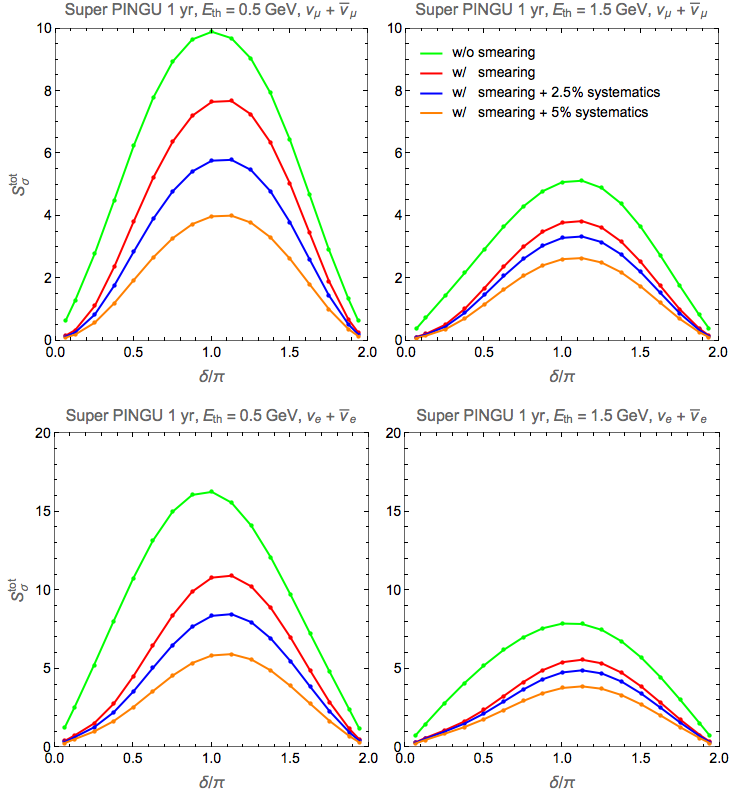}
\caption{Integrated Super-PINGU distinguishabilities between a given
  value of $\delta$ and $\delta=0$ as functions of $\delta$ for the
  $\nu_\mu + {\bar \nu}_\mu$ events (upper panels) and for $\nu_e +
  {\bar \nu}_e$ events (lower panels).  The dependencies have been
  computed for the energy thresholds $E_{th} = 0.5$ GeV (left panels)
  and $E_{th} = 1.5$ GeV (right panels). Different lines show
  distinguishabilities without smearing, with smearing and different
  levels of the uncorrelated systematic errors: $f = 0$, $2.5\%$ and
  $5\%$. Normal mass hierarchy is assumed.}
\label{fig:significance_numu_V12}
\end{figure}

\begin{figure}[t]
\includegraphics[width=4.5in]{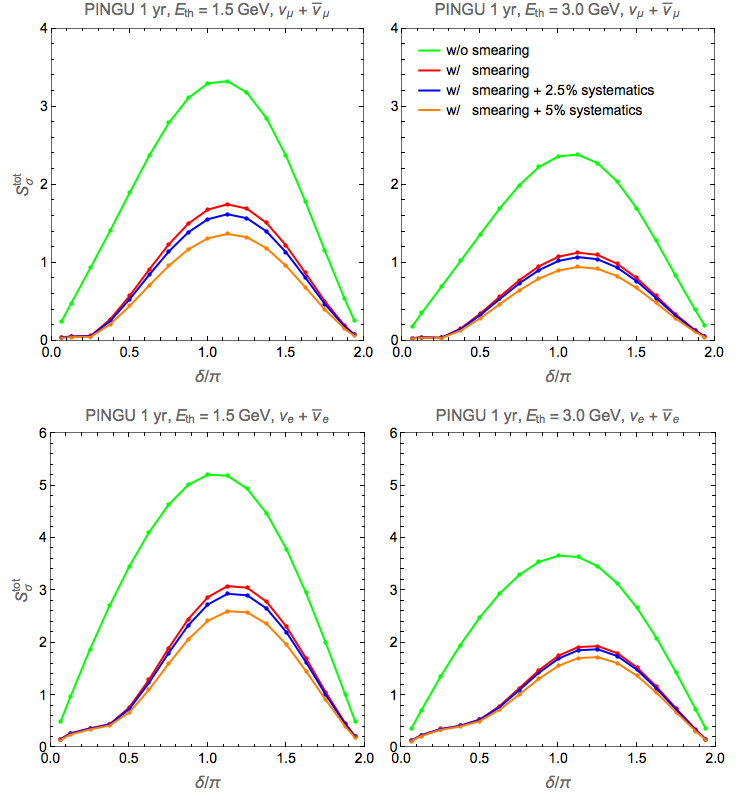}
\caption{The same as in Fig.~\ref{fig:significance_numu_V12}, but for
  the PINGU detector. The energy thresholds  are 1.5 
  GeV and 3.0 GeV in this case.}
\label{fig:significance_numu_VLoI}
\end{figure}

\begin{figure}[t]
\includegraphics[width=4.5in]{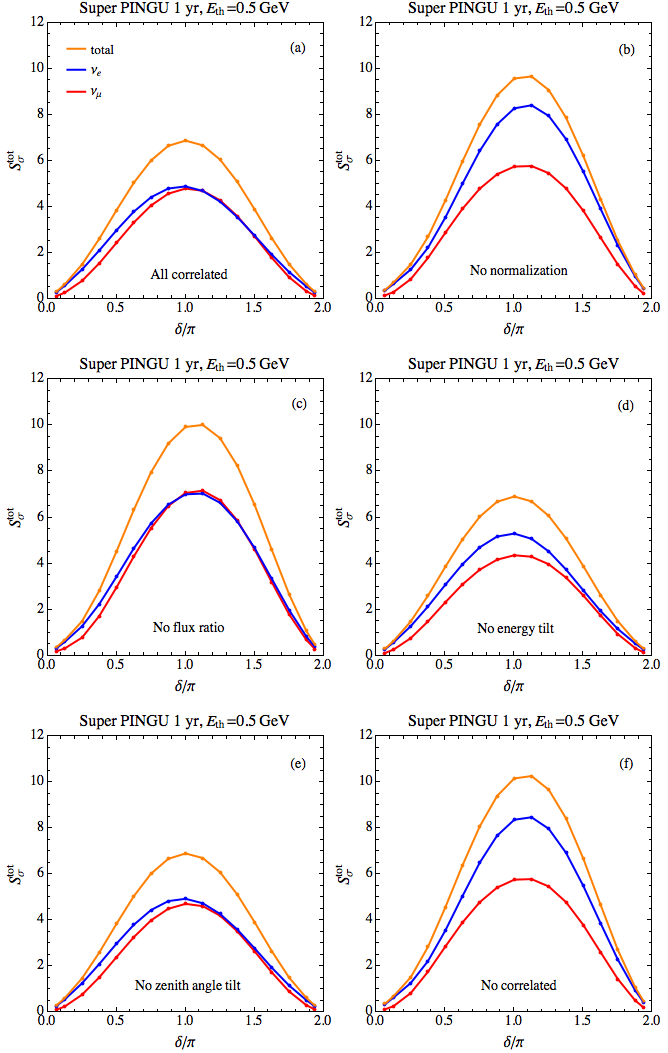}
\caption{Effects of different correlated systematic errors on
  sensitivity to the CP-phase. Shown are the total distinguishability
  as well as integrated Super-PINGU distinguishabilities from
  $\nu_\mu$ and $\nu_e$ events between a given value of $\delta$ and
  $\delta=0$ as functions of $\delta$.  Different panels correspond to
  the cases when (a) all errors are included; (b) normalization
  uncertainty of $20\%$ is removed; (c) flux ratio uncertainty is
  removed; (d) the energy tilt uncertainty is removed; (e) the angular
  tilt uncertainty is removed; (f) all correlated systematic
  uncertainties are removed.  The distinguishabilities have been
  computed after smearing, with $2.5\%$ uncorrelated systematics 1
  year exposure, $E_{th} = 0.5$ GeV and for sum of $\nu$ and $\bar
  \nu$ signals.}
\label{fig:systematics}
\end{figure}

Figs.~\ref{fig:significance_numu_V12} and
\ref{fig:significance_numu_VLoI} show the total distinguishabilities
of a given value of the phase $\delta$ from $\delta = 0$ for the
$\nu_\mu$ and $\nu_e$ events, $S_{\sigma}(\nu_\mu)$ and $S_{\sigma}
(\nu_e)$, after 1 year of exposure.  We use thresholds $E_{th} = 0.5$
GeV and $1.5$ GeV for Super-PINGU and $E_{th} = 1.5$ GeV and $3$ GeV
for PINGU.  Shown are the dependence before smearing and after
smearing over $E_\nu$ and $\theta_z$, with $f = 0$, $2.5\%$, and $5\%$
uncorrelated systematic error.  Let us comment on various features of
the obtained distributions.

\begin{enumerate}[(i)]

  \item Dependence of the distinguishabilities on $\delta$ before
    smearing is nearly symmetric with respect to $\pi$.  Maximum is
    slightly shifted to $\delta < \pi$ for the $\nu_e$ events.
    Smearing diminishes $S_{\sigma}$, the suppression factor depends
    on $\delta$ and is different for the $\nu_\mu$ and $\nu_e$ events.
    As we discussed in Sec.\ IV B, $S_{\sigma}$ is suppressed by
    smearing stronger for $\delta < \pi$.  As a result, smearing
    shifts maximal sensitivity toward $\delta \approx 1.1\, \pi$.

    Smearing suppresses $S_{\sigma} (\nu_e)$ stronger than
    $S_{\sigma}(\nu_\mu)$, especially in the interval $\delta = (0 -
    0.5)\pi$.  For $E_{th} = 0.5$ GeV the factors of suppression are
    (1.4 - 3.5) for $\nu_e$ and (1.3 - 2.3) for $\nu_\mu$. Here in
    brackets, small number corresponds to $\delta = 3\pi/2$ and large
    one to $\pi/4$.  For $\delta = (0.5 - 1.5)\pi$ the suppression is
    given by 1.3 for $\nu_\mu$ and (1.4 - 1.5) for $\nu_e$.

\item The systematic uncorrelated errors suppress $S_{\sigma}$ but do
  not affect significantly the shape of $S_{\sigma}$ dependence on
  $\delta$.  For $E_{th} = 0.5$ GeV, the $f = 2.5\%$ level errors
  diminish $S_{\sigma}(\nu_\mu)$ by a factor (1.3 - 1.5) in the
  interval $\delta = (0.25 - 1.5)\pi$.  In this interval
  $S_{\sigma}(\nu_e)$ is suppressed by a factor $\approx 1.3$.  An
  error $f = 5\%$ gives an additional suppression with respect to $f =
  2.5\%$ case: e.g., for $\delta = \pi$ the suppression factor is 1.4
  for both $S_{\sigma}(\nu_\mu)$ and $S_{\sigma}(\nu_e)$.  With
  increase of $E_{th}$, effect of the uncorrelated systematics
  decreases since the number of events decreases.  For Super-PINGU
  with $E_{th} =1.5$ GeV we obtain about 2 times smaller suppressions
  for $\delta = \pi$: $S_{\sigma}(\nu_\mu)$ decreases by a factor 1.15
  (1.46) for $f = 2.5\%$ ($5\%$) and $S_{\sigma}(\nu_e)$ decreases
  even weaker by a factor 1.13 (1.40).  For PINGU with $E_{th} = 3$
  GeV the reduction is $4\%$ ($12\%$) for $f = 2.5\%$ $(5\%)$.

\item With decrease of threshold from 1.5 GeV down to 0.5 GeV, the
  integral distinguishability of Super-PINGU increases in the interval
  $\delta = (1.0 - 1.5)\pi$ by a factor $\sim 1.7$ for both types of
  events.  For $\delta \leq \pi/2$ the increase is bigger: (1.9 - 2.0)
  for $\nu_\mu$ and smaller: (1.4 - 1.7) for $\nu_e$.

\item Smearing reduces distinguishability for PINGU much stronger than
  for super-PINGU (see Fig.~\ref{fig:significance_numu_VLoI}).
  Moreover the suppression increases with decrease of $\delta$.
  Taking $E_{th} = 1.5$ GeV we obtain the following numbers for
  suppression factor of $S_{\sigma}(\nu_\mu)$ ($S_{\sigma}(\nu_e)$):
  $\delta = 3\pi/2$: 1.64 (1.62), $\delta = \pi$: 2 (1.9), $\delta =
  \pi/2$: 3.2 (4.5), $\delta = \pi/4$: $> 20$ (4.5). Such dependence
  is related to bigger width of the reconstruction function and
  comparable regions of the positive and negative $S_\sigma$ before
  smearing.  Similar factors arise for $E_{th} = 3$ GeV.  In contrast,
  the uncorrelated systematic errors affect PINGU sensitivity much
  weaker (less than $10\%$) than Super-PINGU, which is related to
  smaller number of PINGU events.

  Comparing PINGU and Super-PINGU at the same threshold, $E_{th} =
  1.5$ GeV, we obtain that distinguishability for Super-PINGU is
  bigger by a factor (1.7 - 2.4) in the interval $\delta = (1 -
  1.5)\pi$, mainly due to increase of the effective volume. For
  $\delta \leq \pi/2$ the increase is much bigger: $> 5$ due to both
  increase of $V_{eff}$ and better reconstruction.

  Going from PINGU with $E_{th} = 3$ GeV to Super-PINGU with $E_{th} =
  0.5$ GeV (with smearing and $2.5\%$ systematics) the
  distinguishability increases by a factor (6 - 7) for $\nu_\mu$ and
  by a factor (4 - 5) for $\nu_e$ in the interval $\delta = (1 -
  1.5)\pi$.

\item We have also computed the Super-PINGU distinguishability using
  the PINGU reconstruction functions without rescaling. Improvement of
  the reconstruction ($\sqrt{3}$ decrease of widths) affects very
  weakly the distinguishability in the range $\delta = (1 - 1.5)\pi $.
  For $\delta = \pi/2$ we find $25\%$ increase of $S_{\sigma}(\nu_e)$
  and $10\%$ increase of $S_{\sigma}(\nu_\mu)$. The improvement is
  very strong for small values of $\delta$. Maximal increase of
  $S_{\sigma}(\nu_e)$ given by a factor 2 is at $\delta = \pi/4 $ and
  maximal increase of $S_{\sigma}(\nu_\mu)$ is by a factor 2.2 at
  $\delta = \pi/8$.  The improved resolution for large $\delta$ is
  important for measurements of $\delta$.

\end{enumerate}

We define the total distinguishability (both $\nu_\mu$ and $\nu_e$
channels) as
\be
S_{\sigma}^{tot} = \sqrt{S_{\sigma}^2(\nu_\mu) + S_{\sigma}^2(\nu_e)}. 
\label{eqstot}
\ee As the reference setup we take Super-PINGU with $0.5$ GeV
threshold and $2.5\%$ systematics. We use 4 years exposure and
$\sqrt{t}$ scaling of the total distinguishability with exposure time.
According to Fig.~\ref{fig:significance_numu_V12} and
Fig.~\ref{fig:systematics} (f), the phases $\delta = \pi/4, ~\pi/2,
~\pi, ~ 3\pi/2$ can be distinguished from zero at $S_{\sigma}^{tot} =
3.0, ~ 9,~ 21$ and $13.4$ correspondingly.  The contribution from
$\nu_e$ events is about (1.4 - 1.5) times larger than that from
$\nu_\mu$ events alone. For $E_{th} = 1.5$ GeV, the total
distinguishability $S_{\sigma}^{tot} = 2.1, ~ 5.0,~ 11.5$ and $7.9$
for $\delta = \pi/4, ~\pi/2, ~\pi$, and $3\pi/2$ respectively. So, the
decrease is by a factor (1.5 - 1.7).

\subsection{Correlated systematic error} 

The correlated systematic errors require special consideration in view
of the facts that effects of CP-differences are small, $\sim (1 - 2)
\%$, the $(E_\nu - \cos \theta_z)$ distributions are rather flat
(especially in the region of relatively large CP-violation at low
energies), and at low energies the asymmetry (after smearing) has the
same sign for all zenith angles.  The most important correlated
systematic errors are related to uncertainties in

\begin{itemize} 

\item the atmospheric neutrino fluxes.  In the first approximation
  they can be described by a) the overall normalization factor, b) the
  energy spectrum tilt; c) the flux (flavor) ratio, and d) the zenith
  angle dependence tilt.  At low energies the azimuthal dependence
  also becomes non-trivial.
 
\item the cross-sections of neutrino and antineutrino interactions.
  Uncertainties should be considered separately for different
  reactions.

\item the effective volume $V_{eff}$ and its energy dependence.

\item parameters of the reconstruction functions.

\item neutrino mixing angles and mass squared differences.

\end{itemize}

The cross-sections, fluxes and effective volumes enter expressions for
numbers of events as a product $\sigma \Phi V_{eff}$, therefore their
uncertainties can be described simultaneously: so that the number of
$l-$events in $ij$ bin with the uncertainties included equals
\be 
N_{ij, l}^\delta (\delta, \xi_k ) =  \alpha z_l
\left( \frac{E}{2~ {\rm GeV}} \right)^{\eta} 
[1  + \beta (0.5 + \cos \theta_z)]
N_{ij,l}^{\delta} (\xi_k^{st}), ~~~~~ l = e, \mu.  
\label{corr-syst}
\ee
Here $\alpha$ is the overall normalization factor with error
$\sigma_\alpha = 0.2$, $z_l$ is the flux (flavor) ratio uncertainty
($z_e \equiv 1$ for $\nu_e$ events) with error $\sigma_z = 0.05$,
$\eta$ is the energy tilt parameter with error $\sigma_\eta = 0.1$,
$\beta$ is the zenith angle tilt with error $\sigma_\beta = 0.04$.  We
denote these parameters collectively as
\be
\xi_k \equiv (\alpha, \beta, \eta, z_\mu), 
~~~ \xi_k^{st} \equiv (1, 0, 0, 1). 
\ee

In the distinguishability approach these uncertainties can be
accounted by modifying $S_\sigma^{tot}$ in the following way
\be
S_{\sigma}^{tot}(\xi_k) = 
\sqrt{\sum_{l = e, \mu} \sum_{ij} \frac{[N_{ij, l}(\delta, \xi_k) -
N_{ij}(\delta = 0, \xi_k^{st})]^2}{\sigma^2_{ij,l}}  
+ \sum_k \frac{(\xi_k - \xi_k^{st})^2}{\sigma_{k}^2}}. 
\label{ssigma}
\ee 
This modification is analogous to the pull method in $\chi^2$
analysis.  $S_{\sigma}^{tot}(\xi_k)$ should be then minimized over all
parameters $\xi_k$.  Since $S_{\sigma}^0 (\xi_k^{st}) = 0$ (there is
no fluctuation in our analysis), ${\rm min}
[S_{\sigma}(\xi_k)^{tot}]$ gives the final significance with
systematic errors taken into account:
\be
S_{\sigma}^{tot} = {\rm min} [S_{\sigma}(\xi_k)^{tot}]. 
\ee

Notice that at minimization of $S^{tot}_{\sigma}$, the errors change
the CP-differences in the individual $E_\nu - \theta_z$ bins,
$S_{ij}$, in such a way that positive and negative $S_{ij}$
equilibrate, thus making them smaller in absolute value.  This lead to
decrease of total distinguishability, especially in the regions where
$S_{ij}$ have the same sign.

Effects of different correlated errors on the total distinguishability
are shown in Fig.~\ref{fig:systematics}.  We present also the
integrated distinguishabilities from the $\nu_e$ and $\nu_{\mu}$
events separately without the pull terms (last term of
(\ref{ssigma})).  Fig.~\ref{fig:systematics} (a) shows the
distinguishabilities when all correlated errors are included.  In the
panels (b)-(e) we show the effects of removal of individual errors. In
the panel (f) all correlated systematic errors are removed.  Let us
consider effects of different systematic errors in order

\begin{enumerate}[(i)]

\item The overall normalization of the product $\sigma_\nu \Phi_\nu
  V_{eff}$ modifies the CP-difference as
\be
S_{ij, l} = \frac{z_l\alpha N_{ij, l} (\delta)-N_{l, ij} (\delta = 0)}
{\sigma_\beta} = 
\frac{(z_l\alpha - 1)N_\mu^{ind} + z_l\alpha N_\mu^{\delta} - N_\mu^{0}}
{\sqrt{N_\mu^{ind} + N_\mu^0  + f^2(N_\mu^{ind} + N_\mu^0)^2 }}  
\label{cp-diff3}
\ee
(recall that $z_e \equiv 1$). Here in the lowest (zero) approximation
in $s_{13}$,
\be
N_\mu^{ind} \approx (s_{23}^4 + c_{23}^4), 
~~~~ N_e^{ind} \approx \frac{1}{r}. 
\label{ind-exp}
\ee
(They correspond to the averaged $\nu_\mu - \nu_\tau$ oscillations.)

There are various factors which restrict the renormalization effect.
Although in the case of independent analysis of the $\nu_e$ and
$\nu_\mu$ events $S_\sigma (\nu_\mu)$ and $S_\sigma (\nu_e)$ can be
strongly affected by the normalization uncertainty $\alpha$, the joint
analysis of the $\nu_\mu$ and $\nu_e$ events shows only moderate
reduction of $S^{tot}_\sigma$ (see Fig.~\ref{fig:systematics}).
Indeed, according to (\ref{ind-exp}) $N_\mu^{ind}\approx N_e^{ind}
\approx 1/2$, {\it i.e.,} they have the same sign and size.  At the
same time, as we have established before, the CP-differences of the
$\nu_e$ events and the $\nu_\mu$ events have opposite signs.
Consequently, $(\alpha N_e^{\delta} - N_e^{0})$ and $(z_\mu \alpha
N_\mu^{\delta} - N_\mu^{0})$ in nominator of (\ref{cp-diff3}) have
opposite signs.  Also as we have seen the absolute values of these
differences (after smearing) are of the same size.  Therefore the
overall normalization (the terms $(\alpha - 1)N_e^{ind}$) can suppress
$S(\nu_e)$ but then the term $(z_\mu\alpha - 1)N_\mu^{ind}$ will
enhance $S(\nu_\mu)$, or vice versa. According to
Fig.~\ref{fig:systematics} (b), removal of the normalization
uncertainty increases $S(\nu_\mu)$ by a factor 1.14, and $S(\nu_e)$ by
1.66, whereas the total $S^{tot}$ increases by a factor 1.4.  We find
that reducing the normalization error down to $\sigma_\alpha = 0.1$
practically does not change the result.

\item Essentially the freedom of normalization is restricted by the
  errors in the flavor ratio of fluxes, the ratio of cross sections
  and ratio of the effective volumes for $\nu_e$ and $\nu_\mu$.
  Indeed, according to Fig.~\ref{fig:systematics} switching off the
  uncertainty in the flux ratio increases both $S_\sigma (\nu_\mu)$
  and $S_\sigma (\nu_e)$ by a factor $\approx 1.36$ and
  $S^{tot}_\sigma$ by 1.4, as in the case of normalization.

  Other factors which restrict the normalization uncertainty effect
  (also other tilts uncertainties) include the following

  \begin{itemize}

  \item In the ``core domain'' ($\cos \theta_z < - 0.83$) the sign of
    the CP difference is opposite to that in the ``mantle domain''
    ($\cos \theta_z > - 0.83$).  So, a suppression of the sensitivity
    to $\delta$ in mantle enhances sensitivity in the core domain.

  \item Since at high energies, effect of CP phase becomes negligible
    the overall shift can not be significant.

  \item For the down-going events (although at low energies it is
    difficult to determine the direction) the CP and oscillation
    effects are small. Therefore inclusion of these events will
    further restrict the uncertainty of normalization. In other words,
    one can use the down-going neutrino events to fix the overall
    normalization.

  \item The CP differences for neutrinos and antineutrinos have
    opposite signs, and the overall normalization factor can diminish
    only one difference.

  \end{itemize}

\item The energy tilt uncertainty in the form (\ref{corr-syst}), see
  Fig.~\ref{fig:systematics} (d), produces only $(2 - 3)\%$ decrease
  of sensitivity.

\item Exclusion of the angular tilt uncertainty, as follows from
  Fig.~\ref{fig:systematics} (e), leads to few percent increase of
  $S^{tot}_\sigma$. It is not excluded, howver, that the tilt
  uncertainties in some other form will lead to stronger diminishing.
  However, this is partially accounted for by the uncorrelated
  systematic errors.

\item Removal of all correlated uncertainties produces similar effect
  on $S^{tot}_\sigma$ as removal of the normalization or the flux
  ratio uncertainties see, Fig.~\ref{fig:systematics} (f).

  The energy scale uncertainty, $E \rightarrow E + \epsilon$, gives
  $N^\delta \rightarrow N^\delta (1 + \epsilon/E)^{- \gamma + 1}$,
  where $\gamma$ is the spectral index of the atmospheric neutrinos.
  In the case of linear uncertainty, $\epsilon \propto E_\nu$, the
  effect is reduced to renormalization.  Here we can extrapolate
  result on the absolute energy scale uncertainty from the PINGU
  simulation \cite{pingu2}, which is small.

  Uncertainties of the reconstruction functions (width, shape) can be
  estimated after the corresponding simulations will be done in
  future.  We expect that by the time of operation of Super-PINGU, the
  values of neutrino oscillation parameters will be measured with high
  enough accuracy and we do not include their errors in the present
  estimations.

  For simplicity we fixed all the oscillation parameters to their
  present best fit values, assuming that they will be known well
  enough by the time of Super-PINGU measurements.  Notice that the 2-3
  mixing is degenerate with the CP-violating phase, especially for the
  beam experiments (see for example \cite{Coloma:2014kca}).  In
  contrast to the beam experiments, there is no substantial degeneracy
  of the CP phase and the 2-3 mixing in the atmospheric neutrino
  experiments.  The key point is that with atmospheric neutrinos one
  measures two dimensional distributions in wide range of energy and
  wide range of baseline (zenith angle). Furthermore, both appearance
  and disappearance channels contribute.  Change of pattern of
  distribution with value of the 2-3 mixing and $\delta$ are
  different. In particular, the highest sensitivity to the 2-3 mixing
  is at high energies and large $|\cos \theta_z|$, as can be seen from
  Fig.~8 of \cite{ARS} and Fig.~\ref{fig:s23dep}.  In
  Fig.~\ref{fig:s23dep} we show the distribution of the relative
  CP-differences for fixed $\delta = \pi$ and different values of
  $\theta_{23}$.  According to Fig.~\ref{fig:s23dep} distribution of
  the $\nu_\mu$ events modifies substantially with $\theta_{23}$,
  which is due to the term $D_{23}$ in (\ref{eq:Nmudelta2}), whereas
  the distribution of the $\nu_e$ events practically does not change
  (see (\ref{diffe-d})).  The highest sensitivity to $\delta$ is at
  low energies. Variations of $\theta_{23}$ can not mimic effect of
  the CP phase.
  
\begin{figure}[t]
\includegraphics[width=3in]{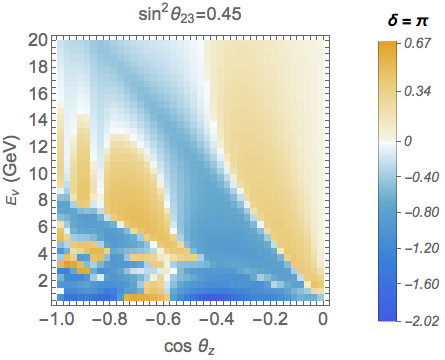}\hskip 0.25in
\includegraphics[width=3in]{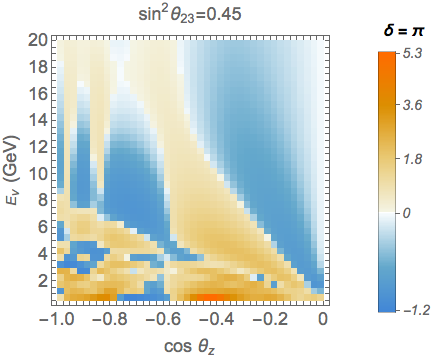}\\ \vskip 0.25in
\includegraphics[width=3in]{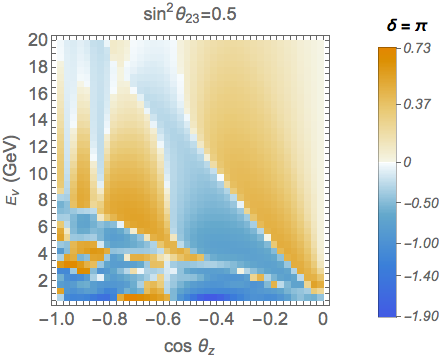}\hskip 0.25in
\includegraphics[width=3in]{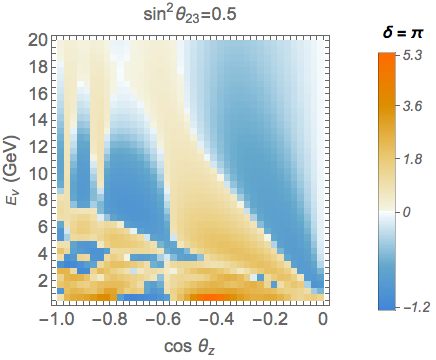}\\ \vskip 0.25in
\includegraphics[width=3in]{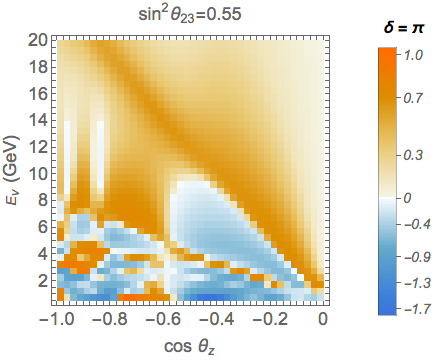}\hskip 0.25in
\includegraphics[width=3in]{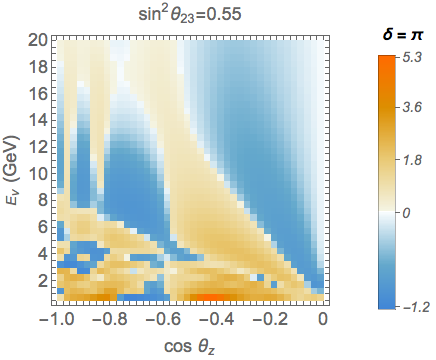}
\caption{ The distributions of the relative CP-differences for $\delta
  = \pi$ and different values of the 2-3 mixing.  Left row for
  $\nu_\mu$ events, right row for $\nu_e$ events.  }
\label{fig:s23dep}
\end{figure}

To further illustrate weak correlation between $\delta$ and $\theta_{23}$ in Super-PINGU we show in Fig.~\ref{fig:s23dep2} (a) dependence on $\sin^2 \theta_{23}$ of distinguishabilities $S_e$, $S_\mu$ and $S_{tot}$ between $\delta = 0$ and $\delta = 3\pi/2$.  As follows from the figure in the range $\sin^2 \theta_{23} = 0.45 - 0.55$ which can be achieved by PINGU, $S_e$ slightly increases; $S_\mu$ decreases from 2.3 down to 2.0, and the total distinguishability decreases from 3.7 to 3.6. So, variations of  $\theta_{23}$ in this range can reduce the distinguishability by about $3\%$. This shows that there is no significant degeneracy of $\theta_{23}$ and $\delta$.

We consider also the distinguishability between the distribution of events for phase $\delta = 0$ and fixed 2-3 mixing $\sin^2 \theta_{23} = 0.40$, $N (0, 0.40)$, and the distributions of events
for value of phase $3\pi/2$ and various values of $\sin^2 \theta_{23}$, $N(3\pi/2, \sin^2 \theta_{23})$:
\be
S_{\sigma} (\sin^2\theta_{23}) =
\frac{N(0, 0.40) - N(3\pi/2, \sin^2\theta_{23})}
{\sqrt{N (0, 0.40)}}.
\label{eq:new-s}
\ee
In Fig. \ref{fig:s23dep2} (b) we show dependences of these $S_{\sigma}$  on $\sin^2\theta_{23}$. According to this figure variations of $\sin^2\theta_{23}$ do not change minimum of $S_{\sigma}$  which is at  the same true value $\sin^2\theta_{23} = 0.40$  and equals 3.9. This means that variations of $\sin^2\theta_{23}$ can compensate difference of phases. Furthermore,  $S_{\sigma}$ increase fast with deviations of $\sin^2\theta_{23}$  from the true value, which shows high sensitivity of superPINGU to 2-3 mixing. Thus, Fig.~\ref{fig:s23dep2} (b) gives  an idea on result of fit when both the phase and 2-3 mixing are varied simultaneously.


\begin{figure}[t]
\includegraphics[width=3in]{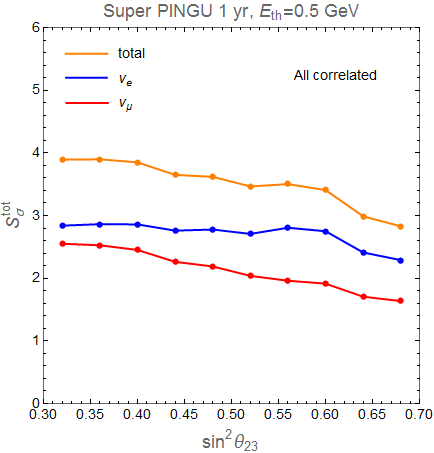}\hskip 0.25in
\includegraphics[width=3in]{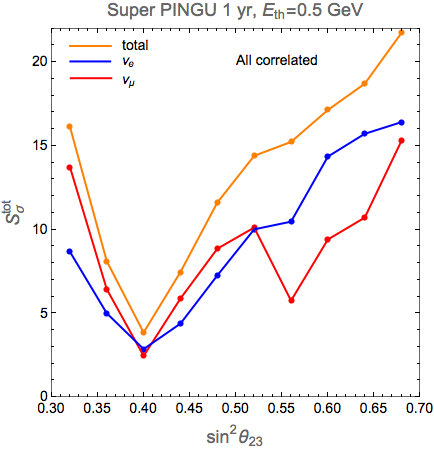}
\caption{Dependence of distinguishabilities $S_{\sigma}(\nu_e) $ (blue), $S_{\sigma}(\nu_\mu$) (red)
and $S_{\sigma}^{tot}$ (orange) between $\delta = 0$ and $\delta = 3\pi/2$
on $\sin^2 \theta_{23}$. The distinguishabilities have been computed for super-PINGU with $E_{th} = 0.5$ GeV and 1 year exposure. Panel (a): the same values of $\sin^2 \theta_{23}$ are taken for both distributions for both values of the phase; panel (b): the distribution for $\delta = 0$ is taken for fixed value $\sin^2 \theta_{23} = 0.40$. }
\label{fig:s23dep2}
\end{figure}

  PINGU will be capable to improve determination of the 2-3 mixing
  substantially without knowledge of $\delta$. Both 2-3 mixing and
  $\delta$ can be determined from Super-PINGU.  To minimize effects of
  $\theta_{23}$ one needs to analyze first the high energy data (say,
  above 10 GeV), where dependence on delta is very low (negligible),
  and analyze low energy data for sensitivity to $\delta$.

\end{enumerate}

\subsection{Effect of flavor misidentification}

Let us consider first the $\nu_\mu$ (track) events.  Due to
misidentification they get contributions from $\nu_e$ CC, NC of all
three neutrino types and $\nu_\tau$ CC interactions. We will describe
this by the misidentification parameters, $g^{\beta}_{\mu}$, which
give fractions of events of type $\beta$ that are identified as the
$\nu_\mu$ events. In turn, part of the true $\nu_\mu$ events will be
identified as events of other types, e.g., the NC interactions.  We
will describe this by the fraction of $\nu_\mu$ events identified as
tracks, $g^{\mu}_{\mu}$.  Using these parameters the number of events
identified as tracks (including misidentification) can be written as
\be
\tilde{N}_\mu (\delta) = g_{\mu}^{\mu}(N_\mu^{ind} + N_\mu^{\delta}) + 
g_{\mu}^e (N_e^{ind} + N_e^{\delta}) + 
g_{\mu}^{NC} N_{NC} +  g_{\mu}^\tau N_\tau (\delta),    
\label{misiden}
\ee
where $N_\beta $ is the number of $\nu_\beta$ events without
misidentification; recall that $N_\mu^{ind} + N_\mu^{\delta} \equiv
N_\mu (\delta)$, etc., and we have omitted the bin indices $ij$.

According to Fig.~9 of \cite{pingu2}, the fractions of events
$g^{\alpha}_\beta$ ($\alpha \neq \beta$) increase with decrease of
energy reaching (for PINGU) $\approx 0.5$ at about (1 - 2) GeV.  In
this case the suppression of the sensitivity to CP phase can be very
strong. For super-PINGU misidentification is expected to be lower. For
estimations we assume that $g^{\beta}_{\mu}$ and $1 - g^{\mu}_{\mu}$
are $\sqrt{3}$ times smaller in Super-PINGU than in PINGU.

Although the fraction of the CC $\nu_\tau$ events misidentified as
tracks is large ($\sim 30 \%$) (in PINGU), at energies below 3 GeV the
cross-section of the CC $\nu_\tau$ interactions, and consequently
$N_\tau$, is very small.  So, we will neglect this contribution.
 
The NC events at low energies contribute to the track events when,
e.g., charged $\pi$ is produced and misidentified with muon (CC
$\nu_{\mu}$ events). These events have, however, smaller cross-section
than the elastic scattering events. They contribute mostly in the
region of $\Delta$ resonance.

For the CP-difference with misidentification taken into account we
obtain
\be
\tilde{S}_\mu = \frac{g_{\mu}^{\mu} (N_\mu^{\delta} - N_\mu^{0}) +  
g_{\mu}^e (N_e^{\delta}  - N_e^0)}{\sqrt{\tilde{N}(\delta = 0) + 
[f \tilde{N}(\delta = 0)]^2}},  
\label{difftilde}
\ee
where according to (\ref{misiden}) 
$$
\tilde{N}(\delta = 0) \approx g_{\mu}^{\mu} N_\mu(\delta = 0) +
g_{\mu}^e N_e(\delta =  0) + g_{\mu}^{NC} N_{NC}.  
$$
NC (as well as $N_\mu^{ind}$) do not contribute to nominator since
$N_{NC}$ does not depend on phase $\delta$ and there is no
fluctuations in our approach.

As we discussed before, effects of change of $\delta$ have opposite
signs for the $\nu_e$ and $\nu_\mu$ events, which leads to flavor
suppression of the CP-differences.  In general,
\be  
N^{\delta}_e - N^{0}_e = 
- \zeta (N^{\delta}_\mu - N^{0}_\mu), 
\ee  
where $\zeta = \zeta (r, \phi, \delta) = \zeta (E_\nu, \theta_z,
\delta) > 0$, see (\ref{eq:Nmudelta2}) and (\ref{eq:Nedelta2}).
Using this relation the expression for $\tilde{S}_\mu$
(\ref{difftilde}) can be written as
\be
\tilde{S}_\mu (f) = \kappa_\mu {S}_\mu (f^\prime), 
\ee
where $f^{\prime} = f \sqrt{g^{\mu}_{\mu} + R^e_\mu g^{e}_{\mu} +
  R^{NC}_\mu g^{NC}_{\mu}}$ is close to 1 and the suppression factor
due to the misidentification equals
\be
\kappa_{\mu} =
\frac{g^{\mu}_{\mu} -
\zeta g^{e}_{\mu}}{\sqrt{g^{\mu}_{\mu} +
R^e_\mu g^{e}_{\mu} + R^{NC}_\mu g^{NC}_{\mu}}}.  
\label{kappadef}
\ee
Here 
$$
R^e_\mu \equiv \frac{N_e(\delta = 0)}{N_\mu (\delta = 0)}, ~~~~ 
R^{NC}_\mu \equiv  \frac{N_{NC}}{N_\mu (\delta = 0)} =
\frac{\sigma_{NC}}{\sigma_{CC}} \frac{\Phi_{tot}}{\Phi_{\mu}},   
$$
and $\Phi_{tot}$ is the total neutrino flux at the detector which is
not affected by oscillations.  Taking in the first approximation, the
$\delta-$independent parts of the probabilities we have according to
(\ref{ind-exp}) is $R^e_\mu \approx 2 r^{-1} $.

For estimations we take $\Phi_{tot} = \Phi_\mu^0 (1 + r^{-1}) \approx
\Phi_\mu (c_{23}^4 + s_{23}^4)^{-1} (1 + r^{-1}) \approx 2 \Phi_\mu (1
+ r^{-1})$.  Here $\Phi_\mu^0$ is the muon flux at the production.
Using $\sigma_{NC}/\sigma_{CC} \approx 1/3$ we obtain
\be
R^{NC}_\mu \approx \frac{2}{3}\left(1 + \frac{1}{r}\right).  
\ee  
At low energies ($r = 2$), $R^{NC}_\mu \approx 1$ and with increase of
energy it decreases down to 2/3.  In the first approximation we will
also neglect difference between $f$ and $f^\prime$.

The minus sign in the nominator of (\ref{kappadef}) ($\zeta > 0 $) is
the main origin of suppression.  For low energies, $E_\nu < 2$ GeV,
when $r = 2$ we have $\zeta = 2$ independently of other parameters.
For higher energies $\zeta$ can be found using the constant density
approximation. From (\ref{diffmu-d}) and (\ref{diffe-d}) we obtain
after averaging out the second terms in both expressions:
\be  
\zeta \approx \frac{r}{r - 1}.  
\ee  
With increase of $r$ (increase of energy) $\zeta \rightarrow 1$, and
flavor suppression becomes weaker.
 
Similarly for $\nu_e$ events we can introduce the misidentification
parameters $g_e^e$ and $g_e^{\beta}$.  NC interactions contribute to
$\tilde{N}_e$ via the $\pi^{0}$ production. The suppression factor for
the CP-difference equals
\be
\kappa_{e} \approx 
\frac{g^{e}_{e} -
\zeta^{-1} g_{e}^{\mu}}{\sqrt{g^{e}_{e} +
(R_\mu^e)^{-1}  g_{e}^{\mu} + R^{NC}_e g^{NC}_{e}}}.
\label{kappadefe}
\ee
Here using similar consideration as for $\nu_\mu$ events we find
$$
R^{NC}_e \equiv \frac{N^{NC}}{N^{CC}_e(\delta = 0)} \approx 
\frac{1}{3}(r + 1), ~~~~~ R^\mu_e = (R_\mu^e)^{-1}. 
$$
At low energies, $R^{NC}_e \approx 1$.

We have introduced the misidentification factors for individual
$(E_\nu - \cos \theta_z)$ bins. This allows us to estimate suppression
of distinguishability from certain energy regions. In particular, for
low energies the factor $\zeta$ does not depend on the energy and
angle, and if $g_\beta^{\alpha}$ are nearly constant we have for the
integrated distinguishability: $\tilde{S}_{int} = \kappa S_{int}$.  In
general, one needs to take into account the energy and zenith angle
dependence of $g^\alpha_\beta$ and perform integration with $\kappa_e$ and $\kappa_\mu$.

To perform estimations of the misidentification factors $\kappa_\mu$
and $\kappa_e$ we use the following relations between the
misidentification parameters extracted from Fig.~9 of \cite{pingu2}:
$$
(1 - g^{\mu}_{\mu})  = 0.7 g^{e}_{\mu}, ~~~~ 
g^{NC}_{\mu} \approx 2 g^{e}_{\mu}. 
$$
No results on $\nu_e$ misidentification is available, so we assume
that $g_{e}^{\mu} \leq (1 - g^{\mu}_{\mu})$, $(1 - g^{e}_{e}) \geq
g^{e}_{\mu}$, and $g_e^{NC} \approx 2 g_\mu^e$.  All the quantities
here are expressed in terms of $g_\mu^e$ and for different values of
this parameter we obtain, varying $\zeta$, the following:
for $g_\mu^e = 0.05$: 
$\kappa_\mu = (0.82 - 0.87)$ and  $\kappa_e = (0.87 - 0.89)$;  
for $g_\mu^e = 0.1$:  
$\kappa_\mu = (0.66 - 0.75)$ and $\kappa_e = (0.75 - 0.80)$; 
for $g_\mu^e = 0.2$:
$\kappa_\mu = (0.38 - 0.55)$ and $\kappa_e = (0.55 - 0.63)$; 
for $g_\mu^e = 0.3$:
$\kappa_\mu = (0.15 - 0.38)$ and $\kappa_e = (0.38 - 0.46)$. 

The distinguishability of the $\nu_\mu$ events is reduced more
strongly than that for the $\nu_e$ events.  According to Fig.~9 of
\cite{pingu2}, $g_\mu^e = 0.1$ can be achieved by PINGU for $E_\nu >
30$ GeV. If the parameters for Super-PINGU are scaled by a factor
$\sqrt{3}$ we obtain instead $E_\nu > 14$ GeV. For $g_\mu^e = 0.2$ the
corresponding energies are 12 GeV (PINGU) and 5 GeV (Super-PINGU) and
for $g_\mu^e = 0.3$ we have 7 GeV and 1 GeV.  At the same time, at low
energies where quasi-elastic scattering dominates the flavor
identification can be better. We conclude that flavor identification
at low energies is crucial for measurements of CP-phase with
Super-PINGU.

A consistent way to treat misidentification is to include its effects
in simulations or use $g^{\alpha}_{\beta}$ as nuisance parameters and
to perform marginalization over them.  Unfortunately, only few
parameters $g^{\alpha}_{\beta}$ are known even at high energies and
their accuracy (say $1 \sigma$ intervals) are not clear. In this
circumstance (in view of absence of information about
$g^{\alpha}_{\beta}$) the correct question to address is what should
be the size of $g$ (level of misidentification) which would allow to
make measurement of the CP-phase.  We answered this question by
estimating effect on sensitivity of different values of $g$. We find
that the level of misidentification should not be higher than $20\%$.

Not only values of $g_{\beta}^{\alpha}$  but also accuracy with which $g_{\beta}^{\alpha}$ will be known are important for determination of $\delta$. Indeed, the uncertainties  of misidentification parameters,  $\delta g_{\beta}^{\alpha}$, propagate to the uncertainties of suppression factors  $\kappa_\mu$ and $\kappa_e$,  and consequently, will further reduce the distinguishability $S_{\alpha}$. If the errors of different $g_{\beta}^{\alpha}$ are uncorrelated the uncertainty in $\kappa_\beta$  can be written as 
\be
\delta \kappa_\beta =
\sqrt{
\sum_i 
\left( \frac{d \kappa_\beta}{d g_{\beta}^i} \right)^2     
(\delta g_{\beta}^i)^2, 
}
\, \, \, \, 
\beta = e, \mu, 
\, \, \, \, 
i = e, \, \mu, \,  NC.  
\label{eq:g-var}
\ee
According to Fig.~9 of \cite{pingu2}, in wide energy range the errors are rather small: $\delta g_{\beta}^i = 0.01 - 0.03$  and they are  about the same size for all the parameters:  $g_{\beta}^{\alpha} \approx \delta g$. We assume that similar uncertainties will be for Super-PINGU. Then using expression  (\ref{kappadef}) we obtain from (\ref{eq:g-var}) $\delta \kappa_\mu \approx 2.2 \delta g$  and $1.7 \delta g$ for $g_\mu^e = 0.1$ and $0.2$ correspondingly. The uncertainty in $\kappa_e$ is smaller: according to (\ref{kappadefe}) and (\ref{eq:g-var}), $\delta \kappa_e \approx \delta g$ for both values of $g_\mu^e$. Taking  $\delta g = 0.02$ we have for $g_\mu^e = 0.1$ that  $\kappa_\mu = 0.66 \pm 0.04$ and $\kappa_e = 0.80 \pm 0.02$. The corresponding numbers for $g_\mu^e = 0.2$ are $\kappa_\mu = 0.380 \pm 0.035$ and $\kappa_e = 0.63 \pm 0.02$. Consequently, uncertainty in $g_{\beta}^{\alpha}$ can reduce $S_\mu$ events by $(7 - 10)\%$ and $S_e$ -- by $\sim 3\%$, if $\delta g = 0.02$.

\subsection{Sensitivity to CP-phase and its possible improvements}

Let us first summarize estimations of the total distinguishability of
Super-PINGU in the case of 4 years of exposure, with $E_{th} = 0.5$
GeV, uncorrelated systematic errors at the level $f = 2.5\%$ and all
correlated errors included.  We have obtained that the values of
phases $\delta = \pi/4, \pi/2, \pi, 3\pi/2$ can be distinguished from
$\delta = 0$ with $S^{tot}_\sigma = 3.0,~ 7.6,~ 13.6,~ 7.6$.  Taking
that the flavor misidentification averaged over energies reduces the
distinguishability by a factor 1.3, we have $S^{tot}_\sigma = 2.1,
~5.3,~ 9.5, ~5.3$ for the same values of the phase.  That is, in the
interval $\delta = (0.5 - 1.5)\pi$ the phase $\delta$ can be
distinguished from 0 with $S^{tot}_\delta > 5$.  For $E_{th} = 1.5$
GeV we would have about a factor 1.7 lower distinguishability:
$S^{tot} = 1.2,~ 3.1,~ 5.6,~ 3.1$, so that in the interval $\delta =
(0.5 - 1.5)\pi$ the phase can be distinguished from 0 at
$S^{tot}_\sigma > 3$.  Lower level of the uncorrelated systematic
error can increase $S^{tot}_\sigma$ by a factor 1.3.

According to Figs.~\ref{fig:significance_numu_V12} and
\ref{fig:significance_numu_VLoI} correlated and uncorrelated
systematic errors (even without misidentification) reduce significance
by a factor of 1.5 - 2. Misidentification can further reduce it by a
factor 0.5 - 0.7. So, total effect of systematics is reduction of
distinguishability by a factor of 2 - 4. Therefore, systematics
dominate in Super-PINGU as in beam experiments.

There are several possibilities to improve the sensitivity.

\begin{itemize}

\item Further decrease of the energy threshold.  The increase of
  $S_{\sigma}^{tot}$ could be about $30\%$ for $E_{th} = 0.2$ GeV in
  comparison with the case of $E_{th} = 0.5$.  This would probably
  require a denser instrumentation of the detector.

\item Stringent kinematical cuts: Selection of subset of events with
  high quality of reconstruction of neutrino flavor, energy and
  direction. That will lead, however, to reduction of numbers of
  events (efficiency).  In any case an optimization of quality of
  reconstruction and statistics is needed.

\item Increase of the exposure time or/and increase of the effective
  volume.  E.g., after 9 years the distinguishability for $\delta =
  \pi/4$ can reach $S_{\sigma}^{tot} \approx 2 - 3$.

\item Improvements of the flavor identification at low energies.

\item Increase the photocathode area using denser array of DOM's or
  photosensors of new type, e.g., as considered for MICA.  This will
  improve reconstruction of energy and direction of neutrino as well
  as flavor identification of events.

\item Partial (statistical) separation of the neutrino and
  antineutrino signals can enhance distinguishability by up to $30\%$.

\end{itemize}

\begin{figure}[t]
\includegraphics[width=6in]{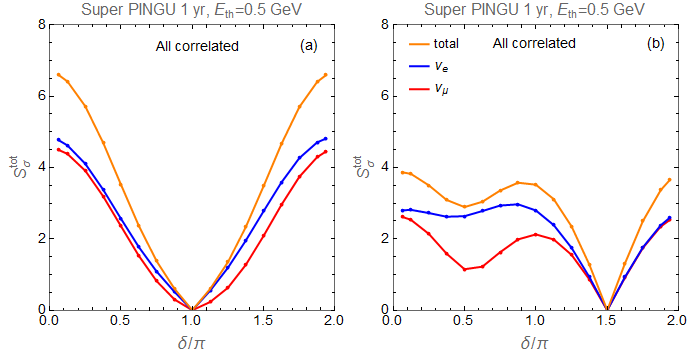}
\caption{The distinguishabilities of a given value of $\delta$ from
  (a) $\delta_0 = \pi$ and (b) $\delta_0 = 3\pi/2$.  Shown are the
  total distinguishability (orange), as well as distinguishabilities
  from $\nu_e$ (blue) and $\nu_\mu$ (red) events as functions of the
  fit value of phase $\delta$.  }
\label{fig:cp180and270}
\end{figure}

We have computed the distinguishability from different values
$\delta_{0}$ (apart from $\delta_{0} = 0$).  For $\delta_0 = \pi$ the
distinguishability can be obtained from results for $\delta_0 = 0$, see Fig.~\ref{fig:cp180and270} (a).
In the first approximation the dependence is simply inverted, i.e.,
substituting maxima by minima and vice versa in
Figs.~\ref{fig:significance_numu_V12}-\ref{fig:systematics}.
Equivalently, it is just shifted by about $\pi$. Situation is more
complicated for other values of $\delta_0$ due to non-linear
dependence of results on the phase.  For distinguishability of
$\delta$ from $\delta_{0} = 3\pi/2$ (favored now) we find according to Fig.~\ref{fig:cp180and270} (b)  
that for $\delta = \pi$ the distiguishability 
equals $S_\sigma^{tot} = 3.5$. This    
is comparable with distinguishability 
between $3\pi/2$ and 0 discussed before. The largest 
$S_\sigma^{tot} = 3.9$ is for $\delta = 0$ and there 
is local minimum, $S_\sigma^{tot} = 3$, for 
$\delta = \pi/2$.

Our results show that Super-PINGU may be potentially competitive with
other proposals \cite{lbne,hyperk,ess,lbno}.  Notice that accelerator
experiments have good sensitivity at small values of $\delta$ but
typically show degeneracy at $\delta = 0$ and $\pi$ ~\footnote{This
  degeneracy is due to the fact that neutrino beam determinations of
  CP are based on measurements of the CP asymmetry (or $\nu -
  \bar{\nu}$ asymmetry) which is proportional to $\sin \delta$.
  Clearly, fit of all available data also sensitive to $\cos \delta$
  will remove this degeneracy.}.  In contrast, Super-PINGU has
relatively low sensitivity at $\delta< \pi/2$, but distinguishability
of $\delta = 0$ and $\pi$ is nearly maximal.  Interestingly, the
strongest distinguishability is for $\delta = 0$ and $\sim 3\pi/2$
(both for beams and Super-PINGU), consequently, if the present
indications of $\delta \sim 1.5\pi$ are true, it will be easier to
establish the CP-violation in lepton sector.

\subsection{Towards realistic estimation of sensitivity}

In view of the fact that characteristics of detection and
reconstruction of neutrino parameters are largely unknown at low
energies, we introduced a number of simplifications, assumptions and
extrapolations of results from high energies.  Therefore the emphasis
is on identification of the main factors which affect the sensitivity
rather than on final numbers, which should be considered as tentative
and very preliminary. The crucial factors include:
\begin{itemize}
\item withds of the energy and angle reconstruction functions; 
\item flavor misidentification parameters;
\item level of uncorrelated systematic errors. 
\end{itemize}
Further progress can be achieved once PINGU (and ORCA) update their
proposals and dedicated study of volume detection of the low energy
events are performed.

We could have overestimated the sensitivity somehow, maybe by a factor
of 2 but certainly not by an order of magnitude.  This can be
understood comparing our results with sensitivity of the
HyperKamiokande atmospheric neutrino studies.  According to
\cite{hyperk} HK will be able to disentangle maximal CP violation from
zero with $\sim 99\%$ confidence level.  These numbers are obtained
after 10 years exposure and 0.56 Mton fiducial volume. Taking 10 times
bigger effective volume we would get after 4 years with the same
reconstruction capacities $5\sigma$ confidence, which agrees with our
estimations ($3 - 8 \sigma$).  Clearly Super-PINGU can not reach
HyperKamiokande resolution but at the same time, since events we
consider are at higher energies (0.5 - 1) GeV instead of 0.01 GeV,
such a high quality instrumentation is not needed. Furthermore, volume
detection may have advantages for higher energies. So, one would
expect somehow lower than $5 \sigma$ significance.

Configuration of the Super-PINGU experiment is taken as an
illustration.  If needed, it is possible to consider denser array of
photomultiplier tubes which will improve capacities of the detector.
Furthermore, methods for volume detection of neutrinos are under fast
development now. The progress is both in the directions of
improvements of characteristics of the optical modules (DOM) and
improvement of analysis of events (which differ from events in SK or
HK).  For instance, recent simulations by PINGU and ORCA show that
reconstruction of the neutrino direction from the cascade events can
be as good as from the tracking events.

Even if we have overestimated the sensitivity and real one is lower,
this can be compensated by future developments of techniques and/or an
increased density of DOMs in the detector. Results we present
constitute a kind of reference point, which can be updated in
different directions: one can improve quality of evaluation of
sensitivity for a given configuration when more information will be
obtained or one can change the assumed configuration of the detector.

Concerning the analysis, eventually distinguishability approach (which
does not contain fluctuations) should be substituted by the $\chi^2$
or maximal likelihood analyses.  For this, one needs to perform Monte
Carlo (MC) simulations of events at Super-PINGU. This should be done
by experimental collaborations and may take few years.

\section{Conclusions} 

Assuming that the neutrino mass hierarchy is identified, we have
explored a possibility to measure the CP-phase with future
multi-megaton scale and low energy threshold atmospheric neutrino
detectors.  The method consists of comparison of the $(E_\nu - \cos
\theta_z)$ distributions of events produced by $\nu_e$ and $\nu_\mu$
for different values of $\delta$.  We use the relative CP-difference
of the distributions to quantify distinguishability and sensitivity.

We have presented simple analytic formalism which allows us to
describe properties of the distributions and gives their exact and
explicit dependence on $\delta$.  The pattern of distributions is
determined to a large extent by the grid of magic lines of three
types, which in turn determine the borders of the CP domains - the
areas with the same sign of the CP-difference.  At low energies the
distributions are averaged over fast oscillations driven by the 1-3
mass splitting.  Using the quasi-constant density approximation we
have derived analytic expressions for the averaged distributions.  In
spite of this averaging the CP-effects are not washed out, and
furthermore, increase with decrease of energy.

In this connection we considered, as illustration, Super-PINGU which
is a further possible upgrade of PINGU detector with multi-megaton
effective volume in the sub-GeV range. Similarly, one can explore
extensions of ORCA. Super-PINGU with large volume at $(0.1 - 0.2)$ GeV
can be used also for proton decay searches.

We have computed distributions of events and the relative
CP-differences in the $(E_\nu - \cos \theta_z)$ plane and studied
their properties.  There are various factors which suppress the
observable CP-effects.  In particular, the flavor suppression related
to the presence of both $\nu_\mu$ and $\nu_e$ original fluxes, and the
fact that CP-asymmetries of the $\nu_\mu - \nu_\mu$ and $\nu_e -
\nu_\mu$ probabilities have opposite signs.  This leads to partial
cancellation of CP-phase effect in the $\nu_e$ and $\nu_\mu$ fluxes at
the detector.  The C-suppression is related to summation of the
neutrino and antineutrino events, since $\nu$ and $\bar{\nu}$ have
CP-asymmetries of opposite sign.  This suppression could be reduced by
partial separation of the $\nu$ and $\bar{\nu}$ signals.

Smearing of the distributions over the neutrino energy and direction
reconstruction functions washes out fine structures of the
distributions and leads to decrease of distinguishability by a factor
(1.5 - 3), depending on the values of $\delta$.  Distinguishability
from $\delta = 0$ is rather low for $\delta < \pi/2$ and maximal in
the interval $\delta = (1 - 1.5)\pi$.

Flavor misidentification of events at the detector produces strong
decrease of sensitivity to the CP phase. Mainly, this is related to
the fact that the $\nu_e$ and $\nu_\mu$ events have CP-differences of
opposite sign, and numbers of these events are comparable at low
energies.  Their misidentification leads to significant cancellation
of the CP violation effects and suppression of $S_{\sigma}^{tot}$ can
be by a factor $(4 - 5)$ at low energies. So, good flavor
identification ($g_{\alpha}^\beta < 0.2$) is crucial for the CP phase
measurement.

We find that inclusion of $f = 2.5\%$ uncorrelated errors (in the case
of our binning and $E_{th} = 0.5$ GeV) reduces the distinguishability
by a factor $(1.3 - 1.4)$.  The correlated systematic errors can
further reduce $S_{\sigma}^{tot}$ by about $30\% - 50\%$.  The total
normalization of fluxes, and cross-sections uncertainties as well as
uncertainties in the flux ratios give the main contribution to this
reduction.  Simultaneous analysis of $\nu_\mu$ and $\nu_e$ events
allows to reduce effect of the correlated systematic errors.

Using Fig.~\ref{fig:systematics} (a) (line which corresponds to total
distinguishability), we obtain that after 4 years of exposure
(approximately double the numbers) and $2.5\%$ uncorrelated
systematics the total distinguishabilities equal $S_{\sigma}^{tot} =
3, ~8, ~14, ~8$ for the phases $\pi/4$, $\pi/2$, $\pi$, $3\pi/2$
correspondingly. As follows from our discussion in Sec.\ III C, flavor
misidentification at $20\%$ level can reduce these numbers by a factor
$(0.3 - 0.5)$ (with stronger effect at smaller values of the phase).
This will give minimal values: $S_{\sigma}^{tot} = 1, ~3, ~6, ~3$ for
the same phases in discussion.  So, the value $\pi/4$ can be
distinguished from zero with $S_{\sigma}^{tot} = (1 - 3)$.  For other
phases we obtain $S_{\sigma}^{tot} (\pi/2) = (3 - 8)$,
$S_{\sigma}^{tot} (\pi) = (6 - 14) $, $S_{\sigma}^{tot} (3\pi/2) = (3
- 8)$.  There are various reasons in addition which can modify these
numbers (which in any case should be considered as tentative) in both
directions. Conditionally $S_{\sigma}$ can be interpreted as
significance and the corresponding quantities as numbers of sigmas.
These estimations show that Super-PINGU may potentially be competitive
with neutrino beam projects.

Going from PINGU with $E_{th} = 3$ GeV to Super-PINGU with $E_{th} =
0.5$ GeV increases the total distinguishability by a factor $(4 - 6)$.
The contributions of the $\nu_\mu$ and $\nu_e$ events to the total
distinguishability can be comparable.

Most of computations have been made assuming that by the time of
Super-PINGU experiment the neutrino mass hierarchy will be established
and for definiteness we took the normal mass hierarchy.  We estimated
that significance of measuring $\delta$ in the case of inverted mass
hierarchy is about $30\%$ lower.

The presented study of sensitivity to CP phase should be considered as
very preliminary since various experimental features are not known
yet.  There is a number of issues related to detection of low energy
events and determination of their characteristics (flavor, energy,
direction, {\it etc.}).  At the same time one can expect that new
experimental developments will further improve the sensitivity.  In
any case, the results obtained here look very promising and
encouraging and certainly show that ``Super-PINGU for CP violation and
proton decay'' deserves further detailed study.

\section*{Acknowledgments}

We are grateful to D.~F.~Cowen and J.~P.~A.~ Marcondes de Andr\'e,
E.~Resconi and J.~Koskinen for correspondence on the design and
performance of the PINGU.  S.R. thanks the Abdus Salam International
Centre for Theoretical Physics for hospitality, where this work was
initiated. S.R. acknowledges support from the National Research
Foundation (South Africa) grant CPRR 2014 number 87823.

\end{document}